\documentclass[letterpaper,12pt]{article} 

\usepackage{amsmath}
\usepackage{amssymb}
\usepackage[round]{natbib}
\usepackage[dvips]{epsfig}
\usepackage{dcolumn}
\usepackage{enumerate}
\usepackage{hhline}
\usepackage{dsfont}
\usepackage{afterpage}
\usepackage{arydshln}
\usepackage{graphicx}
\usepackage{color}
\usepackage[usenames,dvipsnames]{xcolor}
\usepackage{rotating}
\usepackage[breaklinks]{hyperref}
\usepackage{breakurl} 
\usepackage{xr}
\usepackage[percent]{overpic}
\usepackage{subfig}
\usepackage[]{caption}

\usepackage{algorithmicx}
\usepackage[noend]{algpseudocode}
\usepackage{algorithm}
\usepackage{diagbox}
\usepackage{graphicx}
\usepackage{wrapfig}
\usepackage{lscape}
\input epsf
\usepackage{fontenc}
\usepackage{setspace}
\usepackage{bm}
\usepackage{slashbox}
\usepackage{lscape}
\usepackage{breakurl} 
\usepackage{multirow}
\usepackage{eurosym}
\epsfverbosetrue
\def\theequation{\thesection.\arabic{equation}}  
\def\abstract{\if@twocolumn
\section*{Abstract}
\else \normalsize 
\begin{center}
{\bf Summary\vspace{-.5em}\vspace{0pt}} 
\end{center}
\quotation 
\fi}
\def\endabstract{\if@twocolumn\else\endquotation\fi}

\newcommand{\bfZ}{\mbox{\boldmath $Z$}}

\makeatletter
\newcommand{\myappendix}[1]{
	\setcounter{section}{1}
        \renewcommand{\thesection}{A\arabic{section}}}

\def \dsE {\text{$\mathds{E}$}}
\def \dsR {\text{$\mathds{R}$}}

\DeclareMathOperator{\diag}{diag}

\DeclareMathOperator{\ND}{N}

\def \bvec {\text{\boldmath$b$}}    
    
\def \dvec {\text{\boldmath$d$}}

\def \kvec {\text{\boldmath$k$}}

\def \rvec {\text{\boldmath$r$}}

\def \uvec {\text{\boldmath$u$}}    
\def \vvec {\text{\boldmath$v$}}    
\def \wvec {\text{\boldmath$w$}}    
\def \xvec {\text{\boldmath$x$}}    
\def \yvec {\text{\boldmath$y$}}    
\def \zvec {\text{\boldmath$z$}}    \def \mZ {\text{\boldmath$Z$}}

\def \alphavec        {\text{\boldmath$\alpha$}}
\def \betavec         {\text{\boldmath$\beta$}}

\def \deltavec        {\text{\boldmath$\delta$}}

\def \varepsilonvec   {\text{\boldmath$\varepsilon$}}
\def \zetavec         {\text{\boldmath$\zeta$}}
\def \etavec          {\text{\boldmath$\eta$}}
\def \thetavec        {\text{\boldmath$\theta$}}
\def \varthetavec     {\text{\boldmath$\vartheta$}}

\def \lambdavec       {\text{\boldmath$\lambda$}}
\def \muvec           {\text{\boldmath$\mu$}}

\def \xivec           {\text{\boldmath$\xi$}}

\def \rhovec          {\text{\boldmath$\rho$}}

\def \psivec          {\text{\boldmath$\psi$}}

\def \varthetahatvec     {\text{\boldmath$\hat \vartheta$}}

\def \nullvec {\mathbf{0}}

\usepackage{color}
\usepackage{colordvi}
\newlength{\breite}
\breite\textwidth
\newcounter{aufg}[section]
  {\refstepcounter{aufg}\noindent\textbf{Exercise \arabic{aufg}:}
   \\*[1ex]\noindent}{\vspace{.5cm}}
   
 \newcounter{notes}[section]
  {\refstepcounter{aufg}\noindent\textbf{}
   \\*[1ex]\noindent}{\vspace{.5cm}}
   
\usepackage{amsthm}  % Theoreme
\theoremstyle{definition}

\newtheorem*{beisp*}{Example}
\newtheorem{Proof}{Proof}


\newtheoremstyle{break}% name
  {}%         Space above, empty = `usual value'
  {}%         Space below
  {}% Body font
  {}%         Indent amount (empty = no indent, \parindent = para indent)
  {\bfseries}% Thm head font
  {.}%        Punctuation after thm head
  {\newline}% Space after thm head: \newline = linebreak
  {}%         Thm head spec
  
\theoremstyle{break}

\newcommand{\head}[2]%
 {\hrule \vspace{.15cm} {\sfbold Advanced Statistical Inference, Summer Term 2012, Georg-August-University G\"ottingen}\hfill
{\sfbold Sheet #1}\\
{\sfbold Prof. Dr. Thomas Kneib, Nadja Klein}\hfill {\sfbold #2}

\vspace{.2cm}
\hrule

\vspace{1cm}

}

\newcounter{auf}
{\refstepcounter{auf}
\begin{center}
\fcolorbox[gray]{0}{.95}{
\makebox[\breite]{
\textbf{Exercise \arabic{auf}}
}}\\*[1ex]\noindent
\end{center}
}{\vspace{.5cm}}

\newcounter{loes}[section]
{\stepcounter{loes}
\begin{center}
\fcolorbox[gray]{0}{.95}{
\makebox[\breite]{
\textbf{L"osung \arabic{loes}}
}}\\*[1ex]\noindent
\end{center}
}{}

{\begin{center}
\fcolorbox[gray]{0}{.95}{
\makebox[\breite]{
\textbf{Zu Aufgabe #1}
}}\\*[1ex]\noindent
\end{center}\vspace{1cm}
}{\vspace{1cm}}

\newcounter{ka}
{\refstepcounter{ka}
\begin{center}
\framebox[\textwidth]{
\textbf{Aufgabe \arabic{ka}} \hfill #1 Punkte
}\\*[1ex]\noindent
\end{center}
}{\vspace{1cm}}

\newcounter{lka}
{\refstepcounter{lka}
\begin{center}
\framebox[\textwidth]{
\textbf{L\"osung \arabic{lka}} \hfill #1 Punkte
}\\*[1ex]\noindent
\end{center}
}{\vspace{1cm}}

\newcounter{myremark}

\newcounter{mynotation}

\usepackage{paralist}

\renewenvironment{itemize}[1]{\begin{compactitem}#1}{\end{compactitem}}

\newtheorem{theorem}{Theorem}

\makeatletter
\def\@seccntformat#1{\@ifundefined{#1@cntformat}%
	{\csname the#1\endcsname\quad}  % default
	{\csname #1@cntformat\endcsname}% enable individual control
}
\let\oldappendix\appendix %% save current definition of \appendix
\renewcommand\appendix{%
	\oldappendix
	\newcommand{\section@cntformat}{\appendixname~\thesection\quad}
}
\makeatother

\usepackage{titlesec}
\titlespacing*{\section}{0pt}{0.5\baselineskip}{0.5\baselineskip}
\titlespacing*{\subsection}{0pt}{0.5\baselineskip}{0.3\baselineskip}
\titlespacing*{\subsubsection}{0pt}{0.5\baselineskip}{0.3\baselineskip}

\usepackage{scalerel,stackengine}
\stackMath
\newcommand\reallywidehat[1]{%
\savestack{\tmpbox}{\stretchto{%
  \scaleto{%
    \scalerel*[\widthof{\ensuremath{#1}}]{\kern-.6pt\bigwedge\kern-.6pt}%
    {\rule[-\textheight/2]{1ex}{\textheight}}%WIDTH-LIMITED BIG WEDGE
  }{\textheight}% 
}{0.5ex}}%
\stackon[1pt]{#1}{\tmpbox}%
}

\begin{document}

\pagestyle{empty}
%\singlespacing
\begin{titlepage}
%\title{Bayesian Inference for the Implicit Copula from a Heteroscedastic P-Spline Regression}
\title{Bayesian Inference for Regression Copulas}
\author{Authors}
\author{Michael Stanley Smith$^{1}$ and Nadja Klein$^{2,\star}$\footnote{To cite this article: Michael Stanley Smith \& Nadja Klein (2020): BayesianInference for Regression Copulas, Journal of Business \& Economic Statistics, DOI:10.1080/07350015.2020.1721295}\\ \\
$^1$University of Melbourne, $^2$Humboldt  Universit\"at zu Berlin}
%\date{}
\maketitle
\vspace{3in}

\noindent
{\small $\mbox{}^\star$Correspondence should be directed to Nadja Klein
	at Humboldt  Universit\"at zu Berlin, Unter den Linden 6, 10099 Berlin. Email: nadja.klein@hu-berlin.de. 
	Michael Stanley Smith is Professor of Management (Econometrics) at Melbourne Business School, University of Melbourne. Nadja Klein is an Assistant Professor of Applied Statistics at Humboldt  Universit\"at zu Berlin.
	Nadja Klein gratefully acknowledges funding from the Alexander von Humboldt foundation and  the German research
foundation (DFG) through the Emmy Noether grant KL 3037/1-1. The authors thank the Editor, Associate Editor and three referees whose comments improved the paper.}

%\normalsize
\newpage
\begin{center}
\mbox{}\vspace{2cm}\\
{\LARGE Bayesian Inference for Regression Copulas}\\
\vspace{1cm}
{\Large Abstract}
\end{center}
\vspace{-1pt}
\onehalfspacing
\noindent
We propose a new semi-parametric distributional regression smoother that is
based on a copula decomposition of
the joint distribution of the vector of response values. The copula
is high-dimensional and constructed by inversion of a pseudo regression, where the
conditional mean and variance are
semi-parametric functions of covariates modeled using regularized
basis functions.
By integrating out the 
basis coefficients, an implicit copula process on the
covariate space is obtained, which we call a `regression copula'. We combine this with a non-parametric margin to define a copula model, where
the entire
distribution---including the mean and variance---of the response
is a smooth semi-parametric function of the covariates. The copula is estimated using both Hamiltonian
Monte Carlo and variational Bayes; the latter of which is scalable to 
high dimensions.
Using real data examples and a simulation study
we illustrate the efficacy of these estimators and the copula model. 
In a substantive example, we  
estimate the distribution of half-hourly electricity spot prices 
as a function
of demand and two time covariates using radial bases and horseshoe regularization. The copula model produces distributional estimates that 
are locally adaptive with respect to the covariates, and predictions that are more accurate than those
from benchmark models.
\vspace{2.5in}

\noindent
{\bf Keywords}: Distributional regression; Hamiltonian Monte Carlo; Implicit copula, P-splines; Radial basis functions, Variational Bayes.

\end{titlepage}
%\doublespacing

\newpage
\pagestyle{plain}
\setcounter{equation}{0}
\renewcommand{\theequation}{\arabic{equation}}
%\linespread{1.5}\selectfont
%\doublespacing %%%EDIT
\setlength{\abovedisplayskip}{0.1cm}
\setlength{\belowdisplayskip}{0.1cm}

\vspace{-15pt}
\section{Introduction}%\label{sec:intro}
\vspace{-10pt}
Non- or semi-parametric regression methods typically estimate only
the mean of a response variable as an unknown smooth function of covariates.
Yet in many applications, other features of the response distributions---such as higher
moments and quantiles---also vary with the covariates.
% For example,
%\cite{YauKoh2003}, \cite{LazTitsias2011} and others consider heteroscedastic regressions, 
%where both the mean and variance of the response are unknown smooth functions of the covariates.
For example, to address this \cite{RigSta2005} and~\cite{KleKneLanSoh2015}
make all the parameters of a response
distribution unknown smooth functions of the covariates.
However, these authors assume
a specific parametric distribution for the response, conditional on the functions. 
%Alternatively, \cite{HotKneBue2014} estimate
%conditional transformations as smooth multivariate functions of the covariates, which [NK: WE NEED TO DISCUSS
%THE PROS AND CONS OF THIS??]
In this paper, we propose a novel class of semi-parametric distributional regression
models for continuous data that avoids such an assumption. It uses a copula decomposition
of the joint distribution of a vector of values from a single response variable. 
To do so, we employ a new  copula 
with a dependence structure that is an unknown smooth function of the covariate values, and model
the marginal distribution of the response variable non-parametrically. The distributional regression 
is therefore flexible in two ways: non-parametric in a distributional sense with respect to the margin of the
response, and semi-parametric in a functional sense with respect to the covariates via the copula. 
It allows the entire distribution
of the response
%---including higher moments and 
%quantiles---
to be a smooth
unknown function of the covariates. 

Copula models~\citep{McNFreEmb2005,Nel2006} are popular
because the marginal distributions can be modeled arbitrarily and separately from the dependence structure. In
this paper, the copula has dimension equal to the length of the vector 
of response values, which can be high ($87,648$ 
in one of our examples).  Few existing copulas can be used in such a situation, although copulas
constructed by the inversion of a parametric distribution~\citep[Sec. 3.1]{Nel2006} can. Such copulas are called either `inversion' or `implicit' copulas, and those constructed by the inversion of Gaussian~\citep{Song2000}, t~\citep{Demarta2005} and skew t~\citep{SmiGanKoh2012} distributions are popular. 
More flexible implicit copulas can be constructed
by inverting the distribution of values of one or more response variables from 
parametric statistical models. We label these response variables  `pseudo-responses' because they are not observed directly.
Examples include implicit copulas constructed from
factor models~\citep{Mur2013,OhPat2017}, vector
autoregressions~\citep{SmiVah2015}, nonlinear state space models~\citep{SmiMan2015}, Gaussian processes~\citep{wauthier2010,Wilson2010} and regularized regression~\citep{KleSmi2017}. 
These implicit copulas  
reproduce the dependence structure
of the pseudo-response variables,
and combining them with arbitrary margins
produces a more flexible model that allows for a wide range of data distributions.

In this paper we show how to construct an implicit copula from a heteroscedastic 
semi-parametric regression. Both the mean and variance of the pseudo-response
are unknown smooth functions of covariates, each modeled using 
function bases with regularized coefficients.
Because implicit copulas do not retain any information about the marginal 
(i.e.~unconditional on the covariates) location
and scale of the pseudo-response, we normalize the pseudo-response
to have zero mean and unit variance marginally.
By integrating out the basis coefficients of the functions, 
we derive a copula that is a smooth function of the covariate values
and regularization parameters only. 
We call this a `regression copula', because when used in a copula
model for the vector of response values, it captures the effect of the covariates. The
regularization parameters become
the copula parameters, and these require estimation.

There are two main challenges
when estimating the copula parameters: (i)~the copula function and density
are unavailable in closed form, and~(ii) the copula has dimension equal to the sample size, 
which may be high. We outline two Bayesian approaches to overcome these challenges.
The first is a Markov chain Monte Carlo (MCMC) sampler with a Hamiltonian Monte Carlo step \citep{Nea2011,HofGel2014} to evaluate
the posterior distribution exactly. 
The second is a variational Bayes (VB) estimator~\citep{JorGhaJaaSau1999,OrmWan2010}
to compute approximate posterior inference quickly when the
sample size and dimension are high. The VB estimator is based on a Gaussian approximation with a sparse factor representation of its covariance 
matrix~\citep{OngNotSmi2018}. We calibrate this using
stochastic gradient ascent~\citep{HonRaiKuuTorKar2010,SalKno2013} with
gradient estimates computed efficiently~\citep{KinWel2014}. 
The result is a VB estimator for the 
regression copula that is applicable to large datasets and is accurate in our empirical 
work.

We derive  properties of the regression copula, including dependence 
metrics, and show that the independence copula is a limiting case.
The entire (Bayesian posterior)
predictive distribution of the observed response variable 
can be computed from the copula model. This distribution is a smooth function
of the covariates, and its
first and second moments
are estimates of the regression and variance functions. 
The inclusion of
a heteroscedastic term for the pseudo-response produces 
a regression copula that is a more flexible function of the covariates
than the implicit copula of a homoscedastic regression discussed by~\cite{KleSmi2017}.
This results in predictive density and 
regression mean and variance function estimates for the observed response with levels
of smoothing
that are `locally adaptive' 
with respect to the covariates. Such local adaptivity is difficult to achieve in
alternative approaches to distributional regression.
  
We first demonstrate the efficacy of our approach using
four real univariate datasets.
Each has a response with a margin that is non-Gaussian that we estimate non-parametrically.
The
unknown smooth functions are modeled using B-spline bases and autoregressive
priors for the coefficients. The estimated regression and variance functions of the response from the copula model
are nonlinearly related to the covariates.
 Their estimates using the exact and approximate posteriors prove very similar,
 yet the latter are faster to evaluate using VB. 
 %The inclusion of
 %a heteroscedastic term for the pseudo-response allows for a much richer dependence structure
 %in the regression copula,
 %compared to the implicit copula of a homoscedastic regression (which is a Gaussian copula, as discussed %by~\cite{KleSmi2017}). 
 A simulation study based on fitting distributional 
 regressions to these four datasets, shows that the proposed regression copula model produces more 
 accurate density forecasts than that proposed by~\cite{KleSmi2017}, 
 a P-spline regression with Gaussian disturbances,
 a heteroscedastic P-spline regression, and the most likely transformation
  estimator of~\citet{MoeHotBue2017}.
 
Distributional regression can be used to estimate the relationship between 
intraday electricity prices and exogenous drivers~\citep{gianfreda2018}. 
We apply our approach to a distributional regression for $n=87,648$
half-hourly electricity spot prices in the Australian National Electricity Market (NEM)
between 2014 and 2018. There are three covariates (demand, time of day
and day) and the
unknown smooth functions are
modeled using trivariate radial bases,
along with horseshoe priors to regularize the coefficients. 
The resulting regression copula model links the entire distribution of prices 
to the three covariates. By adjusting the radial bases so that they 
are periodic in the time of day covariate (only), 
the price distribution
is also periodic in this covariate.
The fitted regression copula model captures the changing
impact of demand on the price distribution
at different times of the day and over the four year period. Using cross-validated
density forecasting metrics and the quantile score~\citep{GneRan2011}, we show
the copula model is more accurate than two benchmark 
distributional regression methods.
 
% \NK{Maybe add a paragraph like: The main contributions of our paper are:
% \begin{itemize}
% \item We construct the implicit copula of a heteroscedastic smoother model of a pseudo-response and investigate its properties.
% \item Combined with arbitrary margins the copulas can be extended to a novel class of effective heteroscedastic smoothing models that to not require a parametric assumptions for the error terms and which we call copula regression.
% \item We develop both, exact and approximate posterior estimation which are highly efficient applicable to high dimensional copulas.
% \end{itemize}}
Finally, we note here that copulas have been used extensively in multivariate
regression frameworks, although
our approach is very different in two ways. First, previous approaches use a low-dimensional
copula
to capture the dependence between multiple response variables with regression margins, which is 
often called a `copula regression' \citep{PitChaKoh2006},
whereas we use a copula to capture the dependence
between different observations on a single response variable. Second, most previous methods employ 
elliptical or vine~\citep{AasCzaFriBak2009} copulas with closed form densities. In contrast, while our copula
does not have a closed form density, it is tractable and scalable to
higher dimensions, as illustrated in our empirical work.

The  paper is structured as follows. Sec.~\ref{subsec:copmod} shows how to
construct a distributional regression model using a regression copula and an 
arbitrary margin. Our regression copula is outlined in
Sec.~\ref{sec:implicit:copula}, along with some of its properties in Sec.~\ref{sec:propertiesC2}. Sec.~\ref{sec:estimation}
outlines exact and approximate Bayesian posterior estimators, along with distribution
and functional prediction. Sec.~\ref{sec:application} 
discusses the four univariate real data examples and the comparison with benchmark alternatives via simulation.
Sec.~\ref{sec:electricity} contains the application to electricity prices, and Sec.~\ref{sec:discussion} concludes.
\setlength{\abovedisplayskip}{0.1cm}
\setlength{\belowdisplayskip}{0.1cm}

\vspace{-15pt}
\section{Distributional Regression using Implicit Copulas}\label{sec:druc}
\vspace{-10pt}
In this section we first introduce the copula model used for distributional regression. 
Then we outline our proposed implicit copula, along with 
some of its key properties. 
\vspace{-12pt}
\subsection{Copula model}\label{subsec:copmod}
\vspace{-7pt}
Consider $N$ realizations $\bm{Y}_{(N)}=(Y_1,\ldots,Y_{N})'$ of a continuous-valued response,
with corresponding covariate values $\tilde{\xvec}_{(N)}=\{\tilde{\xvec}_1,\ldots,\tilde{\xvec}_{N}\}$. 
Following~%\cite{Skl1959} 
Sklar's Theorem, the joint density of $\bm{Y}_{(N)}|\tilde{\xvec}_{(N)}$ can always
be written as
\[
p(\yvec_{(N)}|\tilde{\xvec}_{(N)})=c^\dagger(F(y_1|\tilde{\xvec}_1),\ldots,F(y_{N}|\tilde{\xvec}_{N})|
\tilde{\xvec}_{(N)})\prod_{i=1}^{N} p(y_i|\tilde{\xvec}_i)\,,\;\; \mbox{ for }N\geq 2\,.
\]
Here, $c^\dagger$ is the density of an $N$-dimensional copula process,
and $F(y_i|\tilde{\xvec}_i)$ is the distribution
function of $Y_i|\tilde{\xvec}_i$; both of which are unknown. In this paper we 
approximate this 
joint distribution, also conditional on copula parameters $\thetavec$, with the copula model
\begin{equation}
p(\yvec_{(N)}|\tilde{\xvec}_{(N)},\thetavec) =
c_H\left(F_Y(y_1),\ldots,F_Y(y_{N})|\tilde{\xvec}_{(N)},\thetavec \right)\prod_{i=1}^{N} p_Y(y_i)\,.
\label{eq:copreg}
\end{equation}
The distribution $Y_i|\tilde{\xvec}_i$ is assumed to be invariant with
respect to $\tilde{\xvec}_i$,
%\footnote{However, $Y_i$ is {\em not} marginally independent
%	of $\tilde{\xvec}_i$ when also conditioning on the unknown mean and 
%	variance functions of the pseudo-response, as shown in Part~A.1 of the Web Appendix.}
and has density 
$p_Y$ and distribution function $F_Y$. However, the impact of the covariate
values on $\bm{Y}_{(N)}$ is captured by the copula 
with density $c_H(\uvec_{(N)}|\tilde{\xvec}_{(N)},\thetavec)$, where $\uvec_{(N)}=(u_1,\ldots,u_N)'$
and  $u_i=F_Y(y_i)$. We call this a `regression copula'
 because it is
a function of $\tilde \xvec_{(N)}$.
%\footnote{This is not to be confused with the term
%`copula regression' which is sometimes used to refer to a copula model for a multivariate
%response vector with regression margins.} 
It is a
copula process on the covariate space
with parameters $\thetavec$ that do not vary with $N$. 
We use the implicit copula proposed in the sub-section below for $c_H$, 
and a major aim of 
this paper is to show that by doing so, adopting Eq.~(\ref{eq:copreg}) provides a very flexible, but tractable, approach to distributional regression.

Before specifying $c_H$, we stress that even though 
$Y_i|\tilde{\xvec}_i$ is assumed invariant with 
respect to 
$\tilde{\xvec}_i$,
$Y_i$ is {\em not} marginally independent
of $\tilde{\xvec}_i$ when also conditioning on the unknown mean and 
variance functions of the pseudo-response, 
as shown in Part~A.1 of the Web Appendix.
Moreover, to see how 
the response
is affected by the covariates in the distributional regression at~Eq.(\ref{eq:copreg}), 
consider a sample of size $n$ with $\yvec=(y_1,\ldots,y_n)'$,
covariate values $\tilde{\xvec}=\{\tilde{\xvec}_1,\ldots,\tilde{\xvec}_n\}$ and
$\uvec=(u_1,\ldots,u_n)'$. Then a 
new response $Y_{n+1}$ with corresponding covariate values $\tilde{\xvec}_{n+1}$
has predictive density
\begin{equation}
p(y_{n+1}|\tilde{\xvec}_{(n+1)},\thetavec) =
\int p(\yvec,y_{n+1}|\tilde{\xvec}_{(n+1)},\thetavec)\mbox{d}\yvec = 
\int c_H(\uvec_{(n+1)}|\tilde{\xvec}_{(n+1)},\thetavec)\mbox{d}\uvec\, p_Y(y_{n+1})\,.  
\label{eq:pred1}
\end{equation}
This density is a function of all the covariate values
$\tilde{\xvec}_{(n+1)}=\{\tilde{\xvec},\tilde{\xvec}_{n+1}\}$, which includes
$\tilde{\xvec}_{n+1}$. Moreover, integrating over the posterior of $\thetavec$ gives the posterior predictive density of $Y_{n+1}$ from the regression model as
\begin{equation}
p(y_{n+1}|\tilde{\xvec}_{(n+1)},\bm{y}) =\int p(y_{n+1}|\tilde{\xvec}_{(n+1)},\thetavec) p(\thetavec|\yvec)\mbox{d}\thetavec\,.\label{eq:pred2}
\end{equation}
Eq.~(\ref{eq:pred2}) forms the basis for our distributional 
regression predictions as a function of $\tilde{\xvec}_{n+1}$, and its first two moments are estimates
of the regression mean and variance functions.
In Sec.~\ref{subsec:pred:inf} we show how to compute Eq.~(\ref{eq:pred1}) and Eq.~(\ref{eq:pred2}) efficiently 
for our proposed copula.
\vspace{-12pt}
\subsection{Implicit regression copula}\label{sec:implicit:copula}
\vspace{-7pt}
Key to our approach is the  
regression copula with density $c_H$, which is
derived from a semi-parametric heteroscedastic regression model for a pseudo-response. 
To do so, we first outline the regression and then construct its implicit copula with only 
the basis coefficients of the mean function integrated out, which is 
a Gaussian copula. Next, to derive the copula with the basis coefficients of the variance function also integrated out, it is
represented as an integral of the Gaussian copula.
We show that such a representation is computationally
efficient.
\vspace{-10pt}
\subsubsection{Pseudo-response regression model}\label{subsec:mod:constr}
\vspace{-5pt}
Consider a regression model for a pseudo-response $\tilde Z_i$ with covariates $\tilde{\xvec}_i=\{\bm{x}_i,\bm{w}_i\}$ given by
\begin{eqnarray}\label{eq:latent:model:z}
  \tilde Z_i &= &\tilde m(\bm{x}_i)+\varepsilon_i\,,  \;\; \varepsilon_i \sim N(0,\sigma^2\sigma^2_i) \,,
  \nonumber \\
  \sigma^2_i &= &\exp(g(\bm{w}_i))\,,\mbox{ for }i=1,\ldots,n\,,
\end{eqnarray}
where the first and second moments are smooth unknown functions $\tilde m$ and $g$
of the two covariate vectors. We model these using 
 linear combinations of basis functions $b_1,\ldots,b_{p_1}$ and $v_1,\ldots,v_{p_2}$, such that 
$\tilde m(\bm{x})=\sum_{j=1}^{p_1} \beta_j b_j(\bm{x})$ and $g(\bm{w})=\sum_{j=1}^{p_2}\alpha_j v_j(\bm{w})$. Typical choices for the bases include polynomial or B-spline bases %~\citep{DeBoor1978}
for a scalar covariate, 
and additive or radial bases for multiple covariates. %~\citep{Pow1987}. 
With these approximations, the regression 
model is usually called semi-parametric. 

For $n$ pseudo-response values 
$\tilde{\bm{Z}}=(\tilde Z_1,\ldots,\tilde Z_n)'$
 the regression at Eq.~(\ref{eq:latent:model:z})
can be written as
 \begin{equation}\label{eq:latent:model:z12}
  \begin{aligned}
  \tilde{\bm{Z}} &= B\betavec+\varepsilonvec\,,\;\;  \varepsilonvec=(\varepsilon_1,\ldots,\varepsilon_n)' \sim N(0,\sigma^2\Sigma) \,,\\
  \Sigma &=\diag(\sigma_1^2,\ldots,\sigma_n^2)\,,\mbox{ }\sigma_i^2=\exp(\vvec_i'\alphavec),\mbox{ for }i=1,\ldots,n\,,
  \end{aligned}
\end{equation}
where $\betavec=(\beta_1,\ldots,\beta_{p_1})'$, $\alphavec=(\alpha_1,\ldots,\alpha_{p_2})'$, 
and
the design matrices $B\in\dsR^{n\times p_1}$ and $V\in\dsR^{n\times p_2}$ have $i$th rows
$\bvec_i'=(b_1(\bm{x}_i),\ldots,b_{p_1}(\bm{x}_i))$ and $\vvec_i'=(v_1(\bm{w}_i),\ldots,v_{p_2}(\bm{w}_i))$, respectively.
To produce smooth
and efficient function estimates it is usual to regularize the basis coefficients $\betavec$ and $\alphavec$. In a conjugate Bayesian context, this corresponds to adopting
the conditionally Gaussian priors 
\begin{equation}\label{eq:prior:beta}
\betavec|\thetavec_{\beta},\sigma^2  \sim N\left(\nullvec,\sigma^2 P_\beta(\thetavec_{\beta})^{-1}\right)\,,\;\;\;
\alphavec|\thetavec_{\alpha} \sim N\left(\nullvec,P_\alpha(\thetavec_{\alpha})^{-1}\right)\,,
\end{equation}
%\begin{equation}\label{eq:prior:beta}
%  \begin{aligned}
%  \betavec|\thetavec_{\beta},\alphavec,\sigma^2 & \sim N\left(\nullvec,\sigma^2 P_\beta(\thetavec_{\beta})^{-1}\right)\\
%   \alphavec|\thetavec_{\alpha} & \sim N\left(\nullvec,P_\alpha(\thetavec_{\alpha})^{-1}\right),
%     \end{aligned}
%\end{equation}
 with smoothing (or `hyper') parameters $\thetavec_{\beta}$ and $\thetavec_{\alpha}$.
 The forms of the precision matrices $P_\beta,P_\alpha$ are typically  matched with the 
 choice of bases for $\tilde m$ and $g$, for which we give two examples later. 

\vspace{-12pt} 
\subsubsection{Regression copula construction}
\vspace{-7pt} We extract two
copulas from the regression model defined at Eq.~(\ref{eq:latent:model:z})--(\ref{eq:prior:beta}).
They are called `implicit' \citep[p.190]{McNFreEmb2005}
 or `inversion' \citep[p.51]{Nel2006} copulas because they are constructed 
 by inverting Sklar's theorem. The copulas
 are $n$-dimensional with dependence structures that are (smooth)
 functions of  $\tilde{\xvec}=\{\xvec,\wvec\}$, with 
 $\xvec=(\bm{x}_1',\ldots,\bm{x}_n')'$ and $\wvec=(\bm{w}_1',\ldots,\bm{w}_n')'$. 
% so that we also call them `regression copulas' in this paper.

The first regression copula derived is the implicit copula of the distribution
$\tilde{\bm{Z}}|\xvec,\wvec,\sigma^2,\alphavec,\thetavec_\beta,\thetavec_\alpha$, which we label $C_1$. 
To construct $C_1$, note that the prior for $\betavec$ is conjugate and can be integrated out of the distribution for $\tilde{\bm{Z}}$ analytically, giving
  \begin{equation}\label{eq:marg:var:cond}
\tilde\mZ|\xvec,\wvec,\sigma^2,\alphavec,\thetavec_{\beta},\thetavec_{\alpha}
\sim\ND(\nullvec,\sigma^2\lbrack\Sigma^{-1}-\Sigma^{-1}B\Omega B'\Sigma^{-1}\rbrack^{-1})\,,
\end{equation}
where $\Omega =\left(B'\Sigma^{-1}B+P_\beta(\thetavec_{\beta})\right)^{-1}$, and by applying the Woodbury formula 
% ($A=\Sigma^{-1}$,$U=-\Sigma^{-1}B$, $V=B'\Sigma^{-1}$)
  \begin{equation*}
\lbrack\Sigma^{-1}-\Sigma^{-1}B\Omega B'\Sigma^{-1}\rbrack^{-1}=
\Sigma+BP_\beta(\thetavec_{\beta})^{-1}B'.
\end{equation*}
It is straightforward to show that
the copula of a normal distribution is the widely employed Gaussian copula~\citep{Song2000}. It is obtained by 
standardizing the marginal means to zero and the variances to one. 
The margin in $\tilde Z_i$ at Eq.~(\ref{eq:marg:var:cond}) is $\ND(0,\sigma^2\lbrack\exp(\vvec'_i\alphavec)+\bvec'_i P_\beta(\thetavec_{\beta})^{-1}\bvec_i\rbrack)$,
so that we normalize $\tilde{\mZ}$ by the diagonal matrix $\sigma^{-1}S(\xvec,\wvec,\alphavec,\thetavec_{\beta})=\sigma^{-1}\diag(s_1,\ldots,s_n)$ with $s_i=\lbrack\exp(\vvec'_i\alphavec)+\bvec'_i P_\beta(\thetavec_{\beta})^{-1}\bvec_i\rbrack^{-1/2}$, to get
$\mZ=\sigma^{-1}S(\xvec,\wvec,\alphavec,\thetavec_{\beta})\tilde\mZ$ . With this, the regression at Eq.~\eqref{eq:latent:model:z} can be re-written for the standardized pseudo-response as
\begin{equation}
	Z_i = m(\bm{x}_i,\bm{w}_i) + \frac{s_i}{\sigma}\varepsilon_i\,,\;\;\varepsilon_i \sim N(0,\sigma^2 \sigma_i^2)\,,
\label{eq:sreg}
\end{equation}
where $m(\bm{x}_i,\bm{w}_i)=(s_i/\sigma)\tilde{m}(\bm{x}_i)=(s_i/\sigma)\bm{b}_i'\betavec$
is a function of both $\bm{x}_i$ and $\bm{w}_i$, because $s_i$ is also.

Denoting $S\equiv S(\xvec,\wvec,\alphavec,\thetavec_{\beta})$ for 
conciseness, 
the distribution of the normalized vector with $\betavec$ integrated out is
\begin{eqnarray}
\mZ|\xvec,\wvec,\sigma^2,\alphavec,\thetavec_{\beta},\thetavec_{\alpha} &\sim & N(\bm{0},R)\,, \mbox{ with}\nonumber \\
R \equiv R(\xvec,\wvec,\alphavec,\thetavec_{\beta}) &= &S(\Sigma+B P_\beta(\thetavec_{\beta})^{-1} B')S\,,\label{eq:correlationmatrix}
\end{eqnarray}
and $N(0,1)$ margins for all elements $Z_1,\ldots,Z_n$.
It is straightforward to show \citep{Song2000} that the random vectors
 $\tilde{\bm{Z}}$ and $\bm{Z}$ (conditional on $\xvec,\wvec,\alphavec,\thetavec_\beta$) 
 have the same Gaussian copula function
\[
C_1(\bm{u}|\bm{x},\wvec,\alphavec,\thetavec_{\beta})=\Phi\left(\Phi_1^{-1}(u_1),\ldots,
\Phi_1^{-1}(u_n);\bm{0},R\right)\,,
\]
where $\bm{u}=(u_1,\ldots,u_n)'$, and $\Phi(\cdot;\bm{0},R)$ and $\Phi_1$ are
the distribution functions of $N(\bm{0},R)$ and $N(0,1)$ distributions,
respectively.  This is a regression copula because $R$ is a function of $\tilde{\xvec}$.

We make a number of observations on
$C_1$. First, the parameter $\sigma^2$ does not feature in the expression 
for $R$, and is unidentified in the copula, so that we
set $\sigma^2=1$ throughout the paper. Second, if the density
of the distribution for $\bfZ$ at Eq.~(\ref{eq:correlationmatrix}) is denoted 
as $p_Z$, with marginal densities  
$p_{Z_i}$ for $i=1,\ldots,n$, then the copula density
$c_1 = \frac{\partial^n}{\partial u_1\ldots\partial u_n} C_1$ is  
\begin{equation}
c_1(\bm{u}|\bm{x},\wvec,\alphavec,\thetavec_{\beta})=
\frac{p_{Z}(\zvec|\xvec,\wvec,\alphavec,\thetavec_{\beta})}{\prod_{i=1}^n p_{Z_i}(z_i|\xvec,\wvec,\alphavec,\thetavec_{\beta})}
= \frac{\phi(\zvec;\bm{0},R)}{\prod_{i=1}^n \phi_1(z_i)}, \label{eq:c1}
\end{equation}
where $z_i=\Phi^{-1}_1(u_i)$, $\zvec=(z_1,\ldots,z_n)'$, and $\phi(\cdot;\bm{0},R)$ and $\phi_1$
are the densities of $N(\bm{0},R)$ and $N(0,1)$ distributions,
respectively.
Third, if a non-conjugate prior is used for $\bm{\beta}$,
then $C_1$
is not a Gaussian copula (something we do not consider in this paper). Last, because $R$ is a function of $\alphavec$, so is the dependence structure of $C_1$. If $\alphavec=\bm{0}$, 
then $C_1$ corresponds to the copula of a homoscedastic regression,
as discussed by~\citet{KleSmi2017}. 
%We label the corresponding copula
%`PSC' for a (homoscedastic) P-spline regression copula. 
%However, in Sec.~\ref{sec:application}
%we show that allowing $\alphavec$ to vary makes a substantial difference
%to the flexibility of the dependence structure of $C_1$ (and the subsequent 
%copulas we consider), improving the accuracy of the fitted copula models greatly.

The second regression copula derived is the implicit copula of
 $\bm{Z}$ with both $\betavec$ and $\alphavec$ integrated out. We label this $C_H$ (for heteroscedastic regression copula), and
 it is this copula with density $c_H$ that is used to model the observed data at Eq.~(\ref{eq:copreg}).
  
\begin{theorem}[\underline{Definition of $C_H$ and $c_H$}]\label{theo:c2}
{\em If $\tilde{Z}_i$ follows the heteroscedastic regression for the pseudo-response at
Eq.~(\ref{eq:latent:model:z})--(\ref{eq:prior:beta}),  $Z_i=s_i\tilde{Z}_i$ is the 
normalized response at Eq.~(\ref{eq:sreg}) with $\sigma=1$, $\tilde{\xvec}=\{\xvec,\wvec\}$ are the covariate values and $\thetavec=\{\thetavec_\beta,\thetavec_\alpha\}$, 
then the $n$-dimensional implicit copula of the distribution $\bm{Z}|\tilde{\xvec},\thetavec$ has density
\[
c_H(\bm{u}|\tilde{\xvec},\thetavec)=\int c_1(\bm{u}|\bm{x},\bm{w},\alphavec,\bm{\theta}_{\beta})p(\alphavec|\thetavec_{\alpha})\mathrm{d}\alphavec=\frac{\int \phi(\zvec;\bm{0},R)p(\alphavec|\thetavec_\alpha)\mathrm{d}\alphavec}{\prod_{i=1}^n \phi_1(z_i)}\,,
\]
and copula function
\[
C_H(\bm{u}|\tilde{\xvec},\thetavec)=\int C_1(\bm{u}|\bm{x},\bm{w},\alphavec,\bm{\theta}_{\beta})p(\alphavec|\thetavec_{\alpha})\mathrm{d}\alphavec=
\int \Phi(\bm{z};0,R)p(\alphavec|\thetavec_\alpha)\mathrm{d}\alphavec\,,
\]
where $\uvec=(u_1,\ldots,u_n)'$, the 
marginal 
$u_i=F(z_i|\tilde{\xvec},\thetavec)=\Phi_1(z_i)$, so that 
$z_i=\Phi_1^{-1}(u_i)$.
}\\
{\bf Proof}: See Part~A of the Web Appendix.
\end{theorem}
We make three observations on $C_H$ defined in Theorem~\ref{theo:c2}. 
First, integration over $\alphavec$ is required to compute $C_H$ and $c_H$.
In Sec.~\ref{sec:estimation} we show how to 
do this integration exactly using Hamiltonian Monte Carlo (HMC), and approximately
using variational Bayes (VB) methods, when computing posterior inference. Second,
the dependence parameters of $C_H$ are the 
the smoothing parameters $\thetavec=\{\thetavec_\alpha,\thetavec_\beta\}$ of $\tilde{m}, g$ 
in the regression  for the pseudo-response at Eq.~\eqref{eq:latent:model:z}.
Last, it is much simpler to construct the implicit
copula of $\bm{Z}$, rather than $\tilde{\bm{Z}}$ here. This is because 
constructing the latter copula would involve evaluating (and inverting) the $n$ marginal distribution functions
 \[
 \tilde{F}_i(\tilde{z}_i|\tilde{\xvec},\thetavec)=\int 
 \Phi_1(\tilde{z}_i/s_i)p(\alphavec|\thetavec_\alpha) \mbox{d}\alphavec\,,\;i=1,\ldots,n\,.
 \]
 Each of these involves computing a
 $p_2$-dimensional integral using
 numerical methods. In contrast, the margin of  $Z_i|\tilde{\xvec},\thetavec$ 
 is simply a standard normal, which greatly simplifies evaluation of $C_H$.
 
\vspace{-12pt}
\subsection{Properties of $C_H$}\label{sec:propertiesC2}
\vspace{-7pt}
Here, we state some properties of the regression copula $C_H$. First, 
the independence copula
is a limiting case of this copula, as outlined in Theorem~\ref{theo:Icop} below: 
\begin{theorem}\label{theo:Icop}
{\em Let $\Pi(\uvec)=\prod_{i=1}^n u_i$ be the 
	independence copula function~\citep[p.11]{Nel2006}, 
	and $\gamma_\beta(\thetavec_\beta)<\infty$ 
	be the maximum marginal variance of the prior for $\betavec$ at Eq.~(\ref{eq:prior:beta}),
	then
	\[
	\lim\limits_{ \gamma_\beta \rightarrow 0} C_H(\uvec|\tilde{\xvec},\thetavec) 
	=\Pi(\uvec)\,.
	\]
}
{\bf Proof}: See Part~A of the Web Appendix.
\end{theorem}
\noindent 
%Note that for any given values $(\psi_{\beta,1},\psi_{\beta,2})$ of the prior
%partial correlations, $\gamma_\beta \rightarrow 0$ if and
%only if $\tau^2_\beta \rightarrow 0$, so that $\tau^2_\beta$ can be viewed as the copula
%parameter that determines the overall level of dependence in $C_H$.
An implication of Theorem~\ref{theo:Icop} is that the relationship
between the response and covariates is weak when the posterior of $\gamma_\beta$ is
close to zero.

Below we give expressions for some common dependence metrics of the 
bivariate sub-copula $C_H^{ij}$ of $C_H$ in  
elements $1\leq i < j \leq n$. The
 derivations are given in Part~A of the Web Appendix.
\begin{itemize}
\item[(i)] 
For $q\in(0,1)$, if $(U_i,U_j)\sim C^{ij}_H$, the lower and upper quantile dependence are
\begin{eqnarray*}
\lambda^L_{ij}(q|\tilde{\xvec},\thetavec) &\equiv &\mbox{Pr}(U_i<q|U_j<q)=\int \lambda^L_{1,ij}(q|\xvec,\wvec,\alphavec,\thetavec_\beta)p(\alphavec|\thetavec_\alpha) \mbox{d}\alphavec\,,\mbox{ and}\\
\lambda^U_{ij}(q|\tilde{\xvec},\thetavec) &\equiv &\mbox{Pr}(U_i>q|U_j>q)=\int \lambda^U_{1,ij}(q|\xvec,\wvec,\alphavec,\thetavec_\beta)p(\alphavec|\thetavec_\alpha) \mbox{d}\alphavec\,, 
\end{eqnarray*}
where $\lambda^L_{1,ij}$ and $\lambda^U_{1,ij}$ are the lower and upper pairwise
quantile dependences of a bivariate Gaussian copula with correlation parameter
$r_{ij}$ given by the $(i,j)$th element of $R$ in Eq.~(\ref{eq:correlationmatrix}).
\item[(ii)] The lower and upper extremal tail dependence  
\[
\lambda_{ij}^L=\lim_{q \downarrow 0}\lambda^L_{ij}(q|\tilde{\xvec},\thetavec)=0\,,\mbox{ and }
\lambda_{ij}^U=\lim_{q\uparrow 1}\lambda^U_{ij}(q|\tilde{\xvec},\thetavec)=0\,.
\]
\item[(iii)] Spearman's rho and Kendall's tau are
\[
\rho^S_{ij}(\tilde{\xvec},\thetavec) =
 \frac{6}{\pi} \int \arcsin(r_{ij}/2)p(\alphavec|\thetavec_{\alpha})\mbox{d}\alphavec\,,\;\;
\tau^K_{ij}(\tilde{\xvec},\thetavec) =
\frac{2}{\pi}\int \arcsin(r_{ij})p(\alphavec|\thetavec_{\alpha})\mbox{d}\alphavec\,,
\]
where $r_{ij}$ is as defined above and is a function of
$\xvec,\wvec,\alphavec,\thetavec_\beta$.
%\item[(iv)] Kendall's tau is
%\[
%\tau^K_{ij}(\tilde{\xvec},\thetavec) =
%\frac{2}{\pi}\int \arcsin(r_{ij})p(\alphavec|\thetavec_{\alpha})\mbox{d}\alphavec\,.
%\]
\end{itemize}
These  metrics are functions of the 
copula parameters $\thetavec$, and also  
all $n$ covariate values $\tilde{\xvec}=\{\xvec,\wvec\}$, rather than just 
$\bm{x}_i,\bm{x}_j,\bm{w}_i,\bm{w}_j$.
%  of the $i$th and $j$th observations. 
(We return to this 
feature in Sec.~\ref{sec:application}, where we show it corresponds to  local adaptivity of the distributional estimates from the copula model).
The metrics are computed  with respect to the posterior
of $\thetavec$ for the examples in Sec.~\ref{sec:application}.

\setlength{\abovedisplayskip}{0.1cm}
\setlength{\belowdisplayskip}{0.1cm}

\vspace{-15pt}
\section{Estimation}\label{sec:estimation}
\vspace{-10pt}
Estimation of the copula model at Eq.~(\ref{eq:copreg}) requires estimation
of both the marginal $F_Y$ and parameters $\thetavec$. 
It is popular to use two stage estimators, where $F_Y$ is estimated first, followed by 
$\thetavec$,  because they are simpler to implement and only involve a minor loss of efficiency~\citep{joe2005}. For $F_Y$ we use the
adaptive kernel density estimator (labeled `KDE') of~\cite{shimazaki2010} and a Dirichlet 
process mixture estimator~\citep{Neal2000} (labeled `DPhat'). For the latter,
when estimating $\thetavec$ using MCMC,  uncertainty 
with respect to the estimate of $F_Y$ can also be integrated out by following 
\cite{GraLis2017} and using the draws of $F_Y$ at each sweep, instead of conditioning
on its posterior point estimate. We find in our empirical work that 
this has only a minor
effect on the copula and distributional estimates. Last, in some examples
we transform the response variable---e.g. by taking its logarithm---before applying the
KDE. In this case the marginal density of the response on the original scale is easily obtained
by multiplying the KDE and Jacobean of the transformation in the usual
fashion, although we present results on the logarithmic scale for
clarity.

\vspace{-12pt}
\subsection{Likelihood}
\vspace{-7pt}
Estimation of $\thetavec$ 
based on Eq.~(\ref{eq:copreg}) with $N=n$ observations
is difficult because
$c_H$ at Theorem~\ref{theo:c2} is expressed as an integral over $\alphavec$.
Nevertheless, the likelihood can
be evaluated by expressing it conditional
on the coefficients $\betavec$ and $\alphavec$, 
and then integrating them out using Bayesian methods, which 
is the approach we employ. The Jacobian of the transformation from $\bm{Z}$ to $\bm{Y}$ is $J_{Z\rightarrow Y}=\prod_{i=1}^n p_Y(y_i)/\phi_1(z_i)$, and by a change of variables and Eq.~(\ref{eq:sreg}), the conditional likelihood is
%\begin{eqnarray}
\begin{equation}
p(\yvec|\xvec,\wvec,\betavec,\alphavec,\thetavec_{\beta},\thetavec_{\alpha})
= p(\zvec|\xvec,\wvec,\betavec,\alphavec,\thetavec_{\beta},\thetavec_{\alpha})J_{Z\rightarrow Y} 
 = \phi(\zvec;SB\betavec,S\Sigma S)\prod_{i=1}^n\frac{p_{Y}(y_i)}{\phi_1(z_i)}\,,\label{eq:likelihood:y:cond:beta}
\end{equation}%\end{eqnarray}
%with $z_i=\phi_1^{-1}(F_Y(y_i))$, 
which can be evaluated in $O(n)$ operations because $S$ and $\Sigma$ are diagonal.
%\citet{KleSmi2017} exploited a similar observation for the case where $\alphavec=\bm{0}$ (so that the regression model for the pseudo-response at Eq.~(\ref{eq:latent:model:z}) is homoscedastic
%and $C_2=C_1$), for which they propose a MCMC scheme.
%We extend their approach below to generate $\alphavec$ using a Hamiltonian Monte Carlo (HMC) step.
Below we show how to evaluate the posterior of $\thetavec$ exactly by 
generating $\alphavec$ using a Hamiltonian Monte Carlo (HMC) step within
a MCMC scheme.  However, for large $n$ and some choices of $P_\beta,P_\alpha$
exact samplers can be sticky and/or slow, so that 
we also develop a variational Bayes (VB)
estimator for approximate inference requiring less computation. Both approaches estimate the posterior of the parameters augmented with the basis coefficients, denoted
as $\varthetavec=\lbrace\betavec,\alphavec,\thetavec_{\beta},\thetavec_{\alpha}\rbrace$
with dimension $p_{\vartheta}$.

\vspace{-12pt}
\subsection{Exact estimation using MCMC}\label{subsec:exactmcmc}
\vspace{-7pt}
%A MCMC sampler is used compute the augmented posterior. 
Each scalar
element of $\thetavec$ (or of a re-parameterization) is generated
using a normal approximation based on analytical derivatives
of the logarithm of its conditional posterior.
%Through experimentation, we found generating transformations of the partial correlations of the AR(2) priors 
%at Eq.~(\ref{eq:prior:beta})---as opposed to the autoregressive parameters---improves the convergence,
%stability and efficiency of the sampler.
The coefficients $\betavec$ are generated from a multivariate normal. 
Details on these two steps
are given in Part~B.1 of the Web Appendix for the copulas  in Sec.~\ref{sec:application}.
%for the two choices 
%of $P_\alpha$ and $P_\beta$ in our empirical work.

The most challenging aspect of this sampler is 
generating from the conditional posterior of $\alphavec$. We found Gaussian or
random walk proposals result in poor mixing of the Markov chain, so that
a HMC~\citep{Nea2011} step is employed instead. 
This augments $\alphavec$ by momentum variables,
and draws from an extended target distribution that is proportional to the exponential
of the Hamiltonian function. 
Dynamics specify how
the Hamiltonian function evolves, and its volume-conserving 
property results in  high acceptance rates of the proposed iterates.

We use the leapfrog integrator \citep{Nea2011}, which employs the logarithm of the
target density  
\begin{equation*}%\label{eq:cpost:alpha}
\begin{aligned}
l_{\alpha}\equiv \log(p(\alphavec|\xvec,\zvec,\lbrace\varthetavec\setminus\alphavec\rbrace))&\propto  
-\frac{1}{2}\sum_{i=1}^{n}\left(\log(s_i^2)+\log(\sigma_i^2)\right)
-\frac{1}{2}\left(\zvec'(S\Sigma S)^{-1}\zvec-2\betavec'B'\Sigma^{-1} S^{-1}\zvec\right)\\\quad&-\frac{1}{2}\betavec'\Sigma^{-1}\betavec-\frac{1}{2}\alphavec'P_\alpha(\thetavec_{\alpha})\alphavec
\end{aligned}
\end{equation*}
and its gradient
\begin{equation*}%\label{eq:grad:alpha}
\begin{aligned}
&\nabla_{\alpha} l_{\alpha}=-P_\alpha(\thetavec_{\alpha})\alphavec-\frac{1}{2}V'\left\lbrack\left(\frac{\partial s_1^2}{\partial\etavec_{\alpha}}s_1^{-2},\ldots,\frac{\partial s_n^2}{\partial\etavec_{\alpha}}s_n^{-2}\right)'+\left(\frac{\partial \sigma_1^2}{\partial\etavec_{\alpha}}\sigma_1^{-2},\ldots,\frac{\partial \sigma_n^2}{\partial\etavec_{\alpha}}\sigma_n^{-2}\right)'\right\rbrack\\\quad& +\frac{1}{2}V'\left\lbrack(B\betavec)\circ(B\betavec)\circ\left(\frac{1}{\sigma_1^2},\ldots,\frac{1}{\sigma_n^2}\right)'-\left(\frac{\partial \kappa_{1,1}^2}{\partial\etavec_{\alpha}}z_1^2 ,\ldots,\frac{\partial \kappa_{1,n}^2}{\partial\etavec_{\alpha}}z_n^2 \right)'\right\rbrack\\\quad&+V'\left\lbrack\zvec\circ\left( \frac{\partial \kappa_{2,1}^2}{\partial\etavec_{\alpha}},\ldots,\frac{\partial \kappa_{2,n}^2}{\partial\etavec_{\alpha}}\right)'\circ (B\betavec)       \right\rbrack,
\end{aligned}
\end{equation*}
where  $\circ$ is the Hadamard product, $\etavec_{\alpha}=V\alphavec$, $\kappa_{1,i}=(\sigma_i^2 s_i^2)^{-1}$, $(\kappa_{2,i}=\sigma_i^2 s_i)^{-1}$,
a closed form expression for 
$\frac{\partial s_i^2}{\partial\etavec_{\alpha}}$ is given in the Web Appendix and $\frac{\partial \sigma_i^2}{\partial\etavec_{\alpha}}\sigma_i^{-2}=1$.
%\textbf{Note that the derivatives of $s_i^2$ with respect to $\etavec_{\alpha}$ (as well as $\thetavec_{\beta},\thetavec_{\alpha}$ are available in closed form, see the Web Appendix for derivation. }
The step size $\epsilon$ and the number of leapfrog steps $L$ at each sweep
are set using the dual averaging approach of~\citet{HofGel2014} as follows.
 A trajectory length $\iota=\epsilon L=1$ is obtained
by preliminary runs of the sampler
with small $\epsilon$ (to ensure a small discretization error)
and large $L$ (to move far).
The dual averaging algorithm uses this trajectory length and adaptively changes $\epsilon, L$ during $M_{\mbox{\scriptsize adapt}}\leq M$ iterations of the complete sampler with $M$ sweeps, in order to achieve a desired rate of acceptance $\delta$.
In our examples $\delta=0.75$, while the
starting value for $\epsilon$ is given by Algorithm~4 of~\citep{HofGel2014}. 
Algorithm~\ref{hmcstep} gives the HMC step at sweep $m$ of the sampler.  
\begin{algorithm}
\caption{Hamiltonian Monte Carlo with Dual Averaging}\label{hmcstep}
\vspace{0.2cm}
Given $\varthetavec^{(m-1)},\bar{\epsilon}_{m-1},\epsilon_{m-1},\delta,\iota,\mu=\log(\epsilon_0),\bar H_{m-1},M,M_{\mbox{\scriptsize adapt}}$:
\vspace{0.2cm}
\begin{algorithmic}[1]
\State Set $\gamma=0.05$, $t_0=10$, $\kappa=0.75$ as in~\citet{HofGel2014}.
\State Sample $\rvec\sim\ND_{p_2}(\nullvec,I)$.
\State Set $\alphavec^{(m)}\gets\alphavec^{(m-1)}$,$\tilde \alphavec\gets\alphavec^{(m-1)}$,$\tilde\rvec\gets\rvec$,$L_m\gets\max\lbrace 1,\mbox{round}(\iota/\epsilon_{(m-1)}\rbrace$.
\For{$j=1,\ldots,L_m$}\Comment{$L_m$ steps of the leapfrog integrator}
\State Set $\tilde\rvec=\rvec+(\epsilon_{m-1}/2)\nabla_{\alpha}l_{\alpha}|_{\alpha=\tilde\alphavec}$.
\State Set $\tilde\alphavec=\tilde\alphavec+\epsilon_{m-1}\tilde\rvec$.
\State Set $\tilde\rvec=\tilde\rvec+(\epsilon_{m-1}/2)\nabla_{\alpha}l_{\alpha}|_{\alphavec=\tilde\alphavec}$.
\EndFor
\State With probability $\bar\alpha=\min\lbrace1,\tfrac{\exp\lbrack l_{\alpha}(\tilde\alphavec)-0.5\tilde\rvec'\tilde\rvec\rbrack}{\exp\lbrack l_{\alpha}(\alphavec^{(m-1)})-0.5\rvec'\rvec\rbrack}\rbrace$, set $\alphavec^{(m)}\gets\tilde\alphavec$.
\If{$m\leq M_{\mbox{\scriptsize{adapt}}}}$ \Comment{dual averaging step}
\State Set $\bar H_m=(1-1/(m+t_0))\bar H_{m-1}+(1/(m+t_0))(\delta-\bar\alpha)$.
\State Set $\log(\epsilon_m)=\mu-\sqrt(m)\bar H_m/(\gamma),\log(\bar\epsilon_m)=m^{-\iota}\log(\epsilon_m)+(1-m^{-\iota})\log(\bar\epsilon_{m-1}).$
\Else
\State Set $\epsilon_{m}=\bar\epsilon_{M_{\mbox{\scriptsize adapt}}}$.
\EndIf
\end{algorithmic}
\end{algorithm}

In our empirical work, a
burn-in of 40,000 iterates was
employed, after which a Monte Carlo sample of size $J=50,000$ was collected. 
%These are very 
%conservative values, in that much smaller samples result in similar posterior estimates.

\vspace{-12pt}
\subsection{Approximate estimation using VB}\label{subsec:vb}
\vspace{-7pt}
The VB estimator approximates the augmented posterior 
$p(\varthetavec|\yvec)\propto p(\yvec|\varthetavec)p(\varthetavec)\equiv h(\varthetavec)$ with a 
tractable density $q_{\lambda}(\varthetavec)$. Here, $p(\yvec|\varthetavec)$ is the conditional
likelihood at Eq.~(\ref{eq:likelihood:y:cond:beta}), and $\lambdavec$ is a vector of 
`variational parameters' which are calibrated
by minimizing the Kullback-Leibler divergence between $q_{\lambda}(\varthetavec)$ and $p(\varthetavec|\yvec)$. It is straightforward to show \citep{OrmWan2010} that 
this is equivalent to maximizing the variational lower bound 
\begin{equation}\label{eq:Elb}
\mathcal{L}(\lambdavec)=\int q_{\lambda}(\varthetavec)\log\left\lbrace\frac{p(\yvec,\varthetavec)}{q_{\lambda}(\varthetavec)}\right\rbrace d\varthetavec=\dsE_q\left\lbrace\log(h(\varthetavec))-\log(q_{\lambda}(\varthetavec))\right\rbrace\,,
\end{equation}
with respect to $\lambdavec$. The expectation in Eq.~(\ref{eq:Elb}) is with respect to the 
variational approximation (VA) with density $q_\lambda$, and cannot be computed 
in closed form. Therefore, a stochastic gradient ascent (SGA) algorithm~\citep{HonRaiKuuTorKar2010,SalKno2013} 
is used to maximize ${\mathcal L}$. This employs an unbiased estimate $\reallywidehat{\nabla_{\lambda}\mathcal{L}(\lambdavec)}$
of the gradient of ${\mathcal L}$
to compute the update
\begin{equation*}%\label{eq:SDM}
\lambdavec^{(t+1)}=\lambdavec^{(t)} + \rhovec^{(t)}\circ\reallywidehat{\nabla_{\lambda}\mathcal{L}(\lambdavec^{(t)})}\,,
\end{equation*}
recursively.
If  $\lbrace \rhovec^{(t)} \rbrace_{t\geq 0}$ is a sequence of vector-valued learning rates that fulfil the
 Robbins-Monro conditions, then the sequence $\{\lambdavec^{(t)}\}_{t \geq 0}$ 
 converges to a local optimum~\citep{Bot2010}. The learning rates are set 
adaptively using the ADADELTA method as in~\cite{OngNotSmi2018}.
%of~\cite{Zei2012}.

For the SGA algorithm to be efficient, the estimate $\reallywidehat{\nabla_{\lambda}\mathcal{L}(\lambdavec)}$ 
should exhibit low variance. To achieve this we use the 
so-called `re-parameterization trick' \citep{KinWel2014,RezMohWie2014}. This 
expresses $\varthetavec$ as a function $\varthetavec=a(\zetavec,\lambdavec)$ of
another random variable $\zetavec$ that has a density $p_{\zeta}(\zetavec)$
that does not depend on $\lambdavec$. In this case, the lower bound is
\begin{equation}\label{eq:Elb3}
\begin{aligned}
\mathcal{L}(\lambdavec)&=\dsE_{p_{\zeta}}\left\lbrace\log h(a(\zetavec,\lambdavec))-\log q_{\lambda}(a(\zetavec,\lambdavec))\right\rbrace,
\end{aligned}
\end{equation}
where $\dsE_{p_{\zeta}}$ is an expectation with respect to $p_{\zeta}$.
Note that when differentiating
Eq.~\eqref{eq:Elb3} with respect to $\lambdavec$,
information from the
posterior density is used, whereas it is not when differentiating Eq.~(\ref{eq:Elb}).
Differentiating inside the expectation in Eq.~(\ref{eq:Elb3}) gives
\begin{eqnarray}
%\begin{equation*}%\label{eq:Elb4a}
\nabla_{\lambda}\mathcal{L}(\lambdavec) &= &\dsE_{p_{\zeta}}\left(\frac{\partial a(\zetavec,\lambdavec)'}{\partial\lambdavec}\nabla_{\varthetavec}\lbrace\log h(a(\zetavec,\lambdavec))-\log q_{\lambda}(a(\zetavec,\lambdavec))\rbrace-\nabla_{\lambda}\log q_{\lambda}(a(\zetavec,\lambdavec))\right)\nonumber\\
 &= &\dsE_{p_{\zeta}}\left(\frac{\partial a(\zetavec,\lambdavec)'}{\partial\lambdavec}\nabla_{\varthetavec}\lbrace\log h(a(\zetavec,\lambdavec))-\log q_{\lambda}(a(\zetavec,\lambdavec))\rbrace\right)\,,\label{eq:Elb4b}
%\end{equation*}
\end{eqnarray}
which follows from the `log-derivative trick' ($\dsE_q(\nabla_{\lambda}\log q_\lambda(\varthetavec))=0$).
An unbiased estimate of the expectation at Eq.~\eqref{eq:Elb4b}
can be computed by simulating from $p_{\zeta}$, and efficient implementations typically use just a single iterate of $\zetavec$.
%To further reduce the variance of the gradient estimate we follow~\cite{RoeWuDuv2017}
%and~\cite{OngNotSmi2018} who
%observe that if the variational density is exact (i.e.~$h(\varthetavec)\propto q_{\lambda}(\varthetavec)$), then
%$\nabla_{\lambda}\log(q_{\lambda}(\varthetavec))=0$.
%This implies that if $q_\lambda$ is a good approximation to $p(\varthetavec|\yvec)$, then
%\begin{equation}\label{eq:Elb4b}
%\nabla_{\lambda}\mathcal{L}(\lambdavec) \approx\dsE_{p_{\zeta}}\left(\frac{\partial a(\zetavec,\lambdavec)'}{\partial\lambdavec}\nabla_{\varthetavec}\lbrace\log h(a(\zetavec,\lambdavec))-\log q_{\lambda}(a(\zetavec,\lambdavec))\rbrace\right).
%\end{equation}
%However, even when $q_\lambda$ is a poor
%approximation, Eq.~\eqref{eq:Elb4b} is still likely to provide a gradient estimate with lower variance,
%so that we
%compute the gradient estimate using Eq.~\eqref{eq:Elb4b}. 

Successful application of variational methods requires $q_\lambda$ to be tractable
and a suitable transformation for the re-parameterization trick. 
We follow 
\cite{OngNotSmi2018}, who use the Gaussian 
VA $q_{\lambda}(\varthetavec)=\phi(\varthetavec;\muvec,\Upsilon)$ with a parsimonious 
factor covariance
structure, which meets both conditions. Here, 
$\Upsilon=\Psi\Psi'+\Delta^2$,
where $\Psi$ is a full rank $p_{\vartheta}\times K$ matrix with $K\ll p_{\vartheta}$,
$\dvec=(d_1,\ldots,d_{p_{\vartheta}})'$ and
$\Delta=\diag(\dvec)$. If $\Psi=\{\Psi_{i,j}\}$, then the elements $\Psi_{i,j}=0$ for $j>i$. 
For uniqueness, it is common to also assume $\Psi_{i,i}=1$, although
we do not because the lack of uniqueness does not
hinder the
optimization, and the unconstrained parametrization is more convenient.
To apply the re-parameterization trick, set $\varthetavec=\muvec+\Psi\xivec+\dvec\circ\deltavec$,
where 
 $\zetavec=(\xivec',\deltavec')'$, $\xivec\in\dsR^{K}$, $\deltavec\in\dsR^{p_{\vartheta}}$ and $p_{\zeta}(\zetavec)$ 
 is the density of a $N(\nullvec,I)$ distribution. 
 In this case, $\lambdavec=(\muvec',\mbox{vech}(\Psi)',\dvec')'$, which has gradient
 $\nabla_{\lambda}\mathcal{L}(\lambdavec)=(\nabla_{\mu}\mathcal{L}(\lambdavec)',
\nabla_{\mbox{\scriptsize{vech}}(\Psi)}\mathcal{L}(\lambdavec)',
\nabla_{d}\mathcal{L}(\lambdavec)')'$,
which can be computed analytically and efficiently; see Part~B.2 of the Web Appendix.
An unbiased estimate $\reallywidehat{\nabla_{\lambda}\mathcal{L}(\lambdavec)}$
is then computed using a sample from $p_{\zeta}$.
Algorithm~\ref{algo3} computes the VB estimates.

\begin{algorithm}
\caption{SGA for a Gaussian VA with a factor covariance structure.}\label{algo3}
Given $\lambdavec^{(0)}=\lbrace\muvec^{(0)},\Psi^{(0)},\dvec^{(0)}\rbrace$, $t=0$:
\begin{algorithmic}[1]
\While{Stopping rule is not satisfied}
\State Generate $(\xivec',\deltavec')'\sim\ND(\nullvec,I)$.
\State Construct the unbiased estimates $\reallywidehat{\nabla_{\mu}\mathcal{L}(\lambdavec^{(t)})}$, $\reallywidehat{\nabla_{\mbox{\scriptsize{vech}}(\Psi)}\mathcal{L}(\lambdavec^{(t)})}$ and $\reallywidehat{\nabla_{d}\mathcal{L}(\lambdavec^{(t)})}$ using the single sample $(\xivec',\deltavec')'$.
\State Compute the adaptive learning rate vector $\rhovec^{(t)}=\lbrace\rhovec_{\mu}^{(t)},\rhovec_{\mbox{\scriptsize{vech}}(\Psi)}^{(t)},\rhovec_{\delta}^{(t)}\rbrace$ using ADADELTA.
\State Set $\muvec^{(t+1)}=\muvec^{(t)}+\rhovec_{\mu}^{(t)}\circ\reallywidehat{\nabla_{\mu}\mathcal{L}(\lambdavec^{(t)})}$.
\State Set $\mbox{vech}(\Psi^{(t+1)})=\mbox{vech}(\Psi^{(t)})+\rhovec_{\mbox{\scriptsize{vech}}(\Psi)^{(t)}}\circ\reallywidehat{\nabla_{\mbox{\scriptsize{vech}}(\Psi)}\mathcal{L}(\lambdavec^{(t)})}$ and set $\Psi^{(t+1)}_{ij}=0$ for $i\geq j$.
\State Set $\dvec^{(t+1)}=\dvec^{(t)}+\rhovec_{d}^{(t)}\circ\reallywidehat{\nabla_{d}\mathcal{L}(\lambdavec^{(t)})}$.
\State Set $\lambdavec^{(t+1)} \gets \lbrace\muvec^{(t+1)},\Psi^{(t+1)},\dvec^{(t+1)}\rbrace$ and $t\gets t+1$.
\EndWhile
\end{algorithmic}
\end{algorithm}

In our empirical work, the calibrated value $\hat{\lambdavec}$ is set to the average value over the
last 10\% of steps.
A point estimate of the parameters is simply 
$\hat{\varthetavec}_{{\mbox{\tiny VB}}}=\dsE_{q_{\hat{\lambda}}}(\varthetavec)=\hat{\muvec}$.

\vspace{-12pt}
\subsection{Distributional and functional prediction}\label{subsec:pred:inf}
\vspace{-7pt}
For a new observation $Y_{n+1}$ of the response with covariate values
$\tilde{\xvec}_{n+1}=(\xvec_{n+1},\wvec_{n+1})$, the posterior predictive density
at Eq.~(\ref{eq:pred2}) is used as a distributional prediction. This can be
evaluated by considering a change of variables from $Y_{n+1}$ to $Z_{n+1}=\Phi_1^{-1}(F_Y(Y_{n+1}))$ as follows:
\begin{equation*}
p(y_{n+1}|\tilde{\xvec}_{(n+1)},\yvec) = \int p(y_{n+1}|\tilde{\xvec}_{(n+1)},\varthetavec)
p(\varthetavec|\yvec) \mbox{d}\varthetavec 
= \frac{p_Y(y_{n+1})}{\phi_1(z_{n+1})}\int p(z_{n+1}|\tilde{\xvec}_{(n+1)},\varthetavec)
p(\varthetavec|\yvec) \mbox{d}\varthetavec\,.
\end{equation*}
From Eq.~(\ref{eq:sreg}) the standardized pseudo-response has a conditional distribution
$Z_{n+1}|\tilde{\xvec}_{(n+1)},\varthetavec\sim 
N(m(\xvec_{n+1},\wvec_{n+1}),s_{n+1}^2\sigma^2_{n+1})$, which is independent
of the $(n+1)$ elements of $\tilde{\xvec}_{(n+1)}$, except for $\tilde{\xvec}_{n+1}$.
Thus, a Monte Carlo estimator of the posterior predictive density is
\begin{equation}
\hat{p}(y_{n+1}|\tilde{\xvec}_{(n+1)})\equiv
\frac{\hat{p}_Y(y_{n+1})}{\phi_1(\Phi_1^{-1}(\hat{F}_Y(y_{n+1})))}
\left\{
\frac{1}{J}\sum_{j=1}^J
\frac{1}{s_{n+1}^{[j]}\sigma_{n+1}^{[j]}}
\phi_1
\left( 
\frac{\Phi_1^{-1}(\hat{F}_Y(y_{n+1}))-m^{[j]}(\xvec_{n+1},\wvec_{n+1})}{s_{n+1}^{[j]}\sigma_{n+1}^{[j]}}
\right) 
\right\}\,.
\label{eq:phatyx}
\end{equation} 
Here, $m^{[j]}, s_{n+1}^{[j]}, \sigma_{n+1}^{[j]}$ are $m, s_{n+1},\sigma_{n+1}$ computed from draw $j$ of $\varthetavec$
from the posterior using the exact sampler
in Sec.~\ref{subsec:exactmcmc}, or a draw from $q_{\hat{\lambda}}(\varthetavec)$ for the VB estimator in Sec.~\ref{subsec:vb}. Because Eq.~(\ref{eq:phatyx}) is only a function of the element $\tilde{\xvec}_{n+1}$
of $\tilde{\xvec}_{(n+1)}$, we write
it henceforth as $\hat{p}(y_{n+1}|\tilde{\xvec}_{n+1})$.

A second estimate that is based on a point estimate $\hat{\varthetavec}$
is
\begin{equation*}
\hat{p}_{{\mbox{\tiny PE}}}(y_{n+1}|\tilde{\xvec}_{n+1})\equiv
\frac{\hat{p}_Y(y_{n+1})}{\phi_1(\Phi_1^{-1}(\hat{F}_Y(y_{n+1})))}
\left\{
\frac{1}{\hat s_{n+1}\hat \sigma_{n+1}}
\phi_1\left( 
\frac{\Phi_1^{-1}(\hat{F}_Y(y_{n+1}))-\hat m(\xvec_{n+1},\wvec_{n+1})}{\hat s_{n+1}\hat \sigma_{n+1}}
\right)
\right\}
\,,%\label{eq:phatyx2}
\end{equation*}  
with $\hat m,\hat s_{n+1},\hat \sigma_{n+1}$ computed from
$\hat{\varthetavec}$. This second estimate will typically be much faster to evaluate, and can be used with 
the VB estimate $\hat{\varthetavec}_{{\mbox{\tiny VB}}}$ or the
exact posterior mean.

We denote the
regression and variance functions as
$f(\xvec_{n+1},\wvec_{n+1})\equiv\dsE(Y_{n+1}|\xvec_{n+1},\wvec_{n+1})$
and $v(\xvec_{n+1},\wvec_{n+1})\equiv\mbox{Var}(Y_{n+1}|\xvec_{n+1},\wvec_{n+1})$, respectively. 
We stress that these are different than 
$\tilde m$ and $g$ in Eq.~(\ref{eq:latent:model:z}), which are the mean and variance functions for the 
pseudo-response. Estimates of $f$ and $v$ can be computed from the posterior predictive 
distribution at Eq.~(\ref{eq:pred2}) as follows.
Let $\bm{b}_{n+1}$ and $\vvec_{n+1}$ be the vectors of function basis terms evaluated at 
$\xvec_{n+1}$ and $\wvec_{n+1}$, respectively.
Then the Bayesian posterior predictive function estimators are:
\begin{equation}\begin{aligned}
\hat{f}(\xvec_{n+1},\wvec_{n+1}) &\equiv  \dsE(Y_{n+1}|\tilde{\xvec}_{n+1},\yvec)=\int\dsE\left( Y_{n+1}|\tilde{\xvec}_{n+1},\varthetavec\right)
p(\varthetavec|\yvec)\mbox{d}\varthetavec\,\\
\hat{v}(\xvec_{n+1},\wvec_{n+1}) &\equiv \mbox{Var}(Y_{n+1}|\tilde{\xvec}_{n+1},\yvec)=\int\mbox{Var}(Y_{n+1}|\tilde{\xvec}_{n+1},\varthetavec)p(\varthetavec|\yvec)\mbox{d}\varthetavec\,,
\end{aligned}
\label{eq:fns}
\end{equation}
where (by a change of variables from $Y_{n+1}$ to $Z_{n+1}$) the terms in the integrands are
\begin{eqnarray*}
\lefteqn{\dsE(Y_{n+1}|\tilde{\xvec}_{n+1},\varthetavec) 
	= \int F_Y^{-1}(\phi_1(z_{n+1}))p(z_{n+1}|\tilde{\xvec}_{n+1},\varthetavec) \mbox{d}z_{n+1} }\nonumber \\
	& &=\int F_Y^{-1}(\phi_1(z_{n+1}))\frac{1}{s_{n+1}\sigma_{n+1}}\phi_1\left(\frac{z_{n+1}-s_{n+1}\bm{b}_{n+1}'\bm{\beta}}{s_{n+1}\sigma_{n+1}}\right) \mbox{d}z_{n+1}\equiv 
	\hat{f}_\vartheta(\xvec_{n+1},\wvec_{n+1})\,,
\end{eqnarray*}
\begin{eqnarray*}
\lefteqn{\mbox{Var}(Y_{n+1}|\tilde{\xvec}_{n+)},\varthetavec)
	= \int \left(F_Y^{-1}(\phi_1(z_{n+1}))\right)^2 p(z_{n+1}|\tilde{\xvec}_{n+1},\varthetavec) \mbox{d}z_{n+1} -\hat{f}_\vartheta(\xvec_{n+1},\wvec_{n+1})^2} \nonumber \\
	& &= \int \left(F_Y^{-1}(\phi_1(z_{n+1}))\right)^2\frac{1}{s_{n+1}\sigma_{n+1}}\phi_1\left(\frac{z_{n+1}-s_{n+1}\bm{b}_{n+1}'\bm{\beta}}{s_{n+1}\sigma_{n+1}}\right) \mbox{d}z_{n+1}-\hat{f}_\vartheta(\xvec_{n+1},\wvec_{n+1})^2\,,
%	\label{eq:cVary0}
\end{eqnarray*}
$\sigma_{n+1}=\exp(\vvec_{n+1}'\alphavec)$ and
$s_{n+1}=[\exp(\vvec_{n+1}'\alphavec)+\bm{b}_{n+1}'P_\beta(\bm{\thetavec_{\beta}})^{-1}\bm{b}_{n+1}]^{-1/2}$.
The integrals with respect to $z_{n+1}$ above are computed using standard univariate numerical methods.
The integrals at Eq.~(\ref{eq:fns}) can be computed with draws from either the posterior using the exact estimator
or the calibrated VA when using the VB estimator. 
Estimators that are 
faster to compute can be obtained by simply conditioning on either
the posterior mean or 
$\hat{\varthetavec}_{{\mbox{\tiny VB}}}$, in a similar fashion as with the density estimator.

Last, other
distributional summaries---for example, quantiles, higher order moments or Gini 
coefficients---can be computed similarly.
\setlength{\abovedisplayskip}{0.1cm}
\setlength{\belowdisplayskip}{0.1cm}
\vspace{-15pt}
\section{P-Spline Copulas}\label{sec:application}
\vspace{-10pt}
In this section we construct regression copulas for a single covariate using cubic B-spline
bases for $\tilde m, g$. The knots are equally-spaced over the
range of the covariate, selected so that $\dim(\betavec)=22$, and $\dim(\alphavec)=12$.
The matrices $P_\beta,P_\alpha$ are
the precisions of stationary AR(2) models, each parameterized in terms of the disturbance
variance $\tau_j^2$ and two partial autocorrelations $-1<\psi_{j,1},\psi_{j,2}<1$, for $j=\alpha,\beta$; see~\cite{Barndorff1973}. 
Thus, $P_\beta,P_\alpha$ are 
of full rank, and $\thetavec_{\beta}=\{\tau^2_\beta,\psi_{\beta,1},\psi_{\beta,2}\}$, 
$\thetavec_\alpha=\{\tau^2_\alpha,\psi_{\alpha,1},\psi_{\alpha,2}\}$. This combination
of basis and type of prior for the basis coefficients
is widely called a `P-spline', although random walk priors are more popular.
However, 
the precision matrix of a random walk is of reduced rank, in which case
the distribution of $\tilde{\bm{Z}}$ with $\betavec,\alphavec$ integrated out is improper, and it does not have a proper copula density, so that such a prior cannot be used.  
%In our empirical work we found that a stationary AR(2) prior 
%provides more accurate prediction than an AR(1) prior.
An alternative is to employ the proper prior
suggested by~\cite{chib2006}, although we do not do so here.  
\vspace{-15pt}
\subsection{Real data examples}
\begin{figure}[htbp]
\caption{Marginal distributions of the response variables in the four test datasets.}
	\vspace{-10pt}
\begin{center}
\includegraphics[width=0.85\textwidth,angle=0]{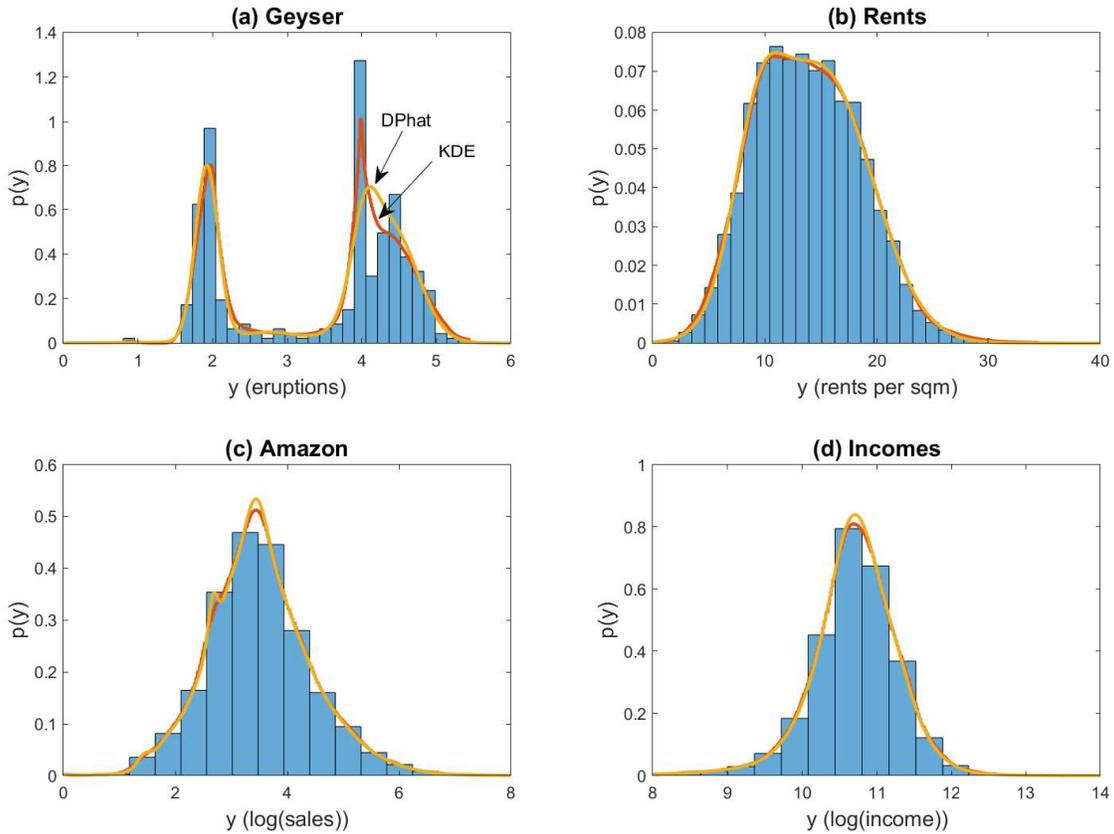}
\end{center}
	\vspace{-5pt}
Normalized histograms of the response ($Y$) of the four datasets in 
	Section~\ref{sec:application}, along with kernel (KDE) and Bayesian (DPhat) non-parametric density estimates of $p_Y$.
	The datasets are (a)~Geyser, 
	(b)~Rents, (c)~Amazon and (d)~Incomes.\label{fig:histograms:dat}
\end{figure}

\vspace{-10pt}
We illustrate our approach using 
the four real datasets listed in
Table~\ref{tab:summaries}. Each
has one covariate (although we consider multiple covariates in the next section), and we set $x_i=w_i$ throughout.  
Fig.~\ref{fig:histograms:dat}
plots histograms of the four response variables, along with KDE 
and Dirichlet process mixture (DPhat) non-parametric density estimates.
These are very similar, and we employ the KDE for $F_Y$, except where mentioned otherwise.
Note that the response in the
Geyser dataset in Fig.~\ref{fig:histograms:dat}(a) is bimodal, with which
most existing distributional regression methods would struggle. In contrast, it is straightforward
to account for this feature in our copula approach through the use of
a bimodal margin $F_Y$.
 With the Amazon and Incomes datasets, we consider the responses
on the logarithmic scale for clarity of illustration, although our regression copula is 
invariant to monotonic transformations of the response because 
the copula data $\bm{u}$ is unaffected.
%\footnote{We also employ the logarithmic
%	transformed response for the benchmark methods considered later, because these either were unstable or performed poorly without doing so.} 
%Scatterplots of the data 
%can also be found in Fig.~\ref{fig:functions2}(c,e,g,i).

We fit two variants
of the copula model. 
The first employs the copula function $C_H$, and is labelled `HPSC' for `heteroscedastic P-spline copula'. 
The second employs $C_1$ with the constraint 
$\alphavec=\bm{0}$ and is labelled `PSC' for (homoscedastic) `P-spline copula', and
is one of the regression copulas proposed by~\cite{KleSmi2017}.
Table~\ref{tab:models} lists key quantities of the two copulas.
%We employ the KDE for
%the margin $F_Y(y)$ of the copula models. 
Three benchmark models are also considered: the first is labeled PS and is
the P-spline smoother with Gaussian errors of~\cite{LanBre2004}, the 
second is labeled `HPS' and is the heteroscedastic P-spline smoother of~\cite{KleKneLanSoh2015}, while
the third is labeled `MLT' and is the `most likely transformation' model of~\citet{MoeHotBue2017}. 
For the latter we use Bernstein polynomials as suggested by the authors, 
and the method is known to be flexible and robust.

\vspace{-12pt}
\subsubsection{Exact versus approximate estimation}\label{subsec:modspecdat}
\begin{sidewaysfigure}[htbp]
\caption{Summaries of the variational lower bound ${\cal L}(\bm{\lambda})$ in the SGA.}
	\vspace{-15pt}
\begin{center}
	\includegraphics[width=0.9\textwidth,angle=0]{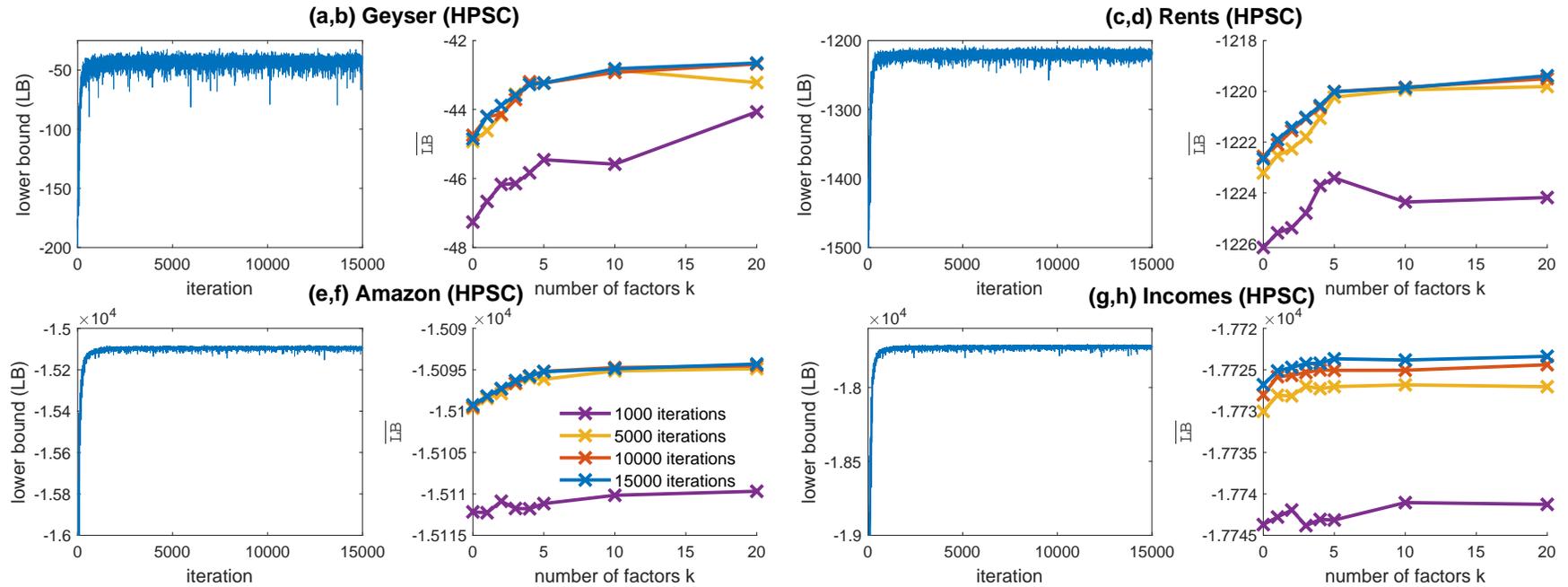}
\end{center}
	\vspace{-5pt}
The datasets
		are (a,b)~Geyser, (c,d)~Rents, (e,f)~Amazon, and (g,h)~Incomes. 
	Panels~(b,d,f,h) plot
	the average lower bound ($\overline {\mbox{LB}}$) over the last 10\% of steps, against the number of factors $K$ in the
	Gaussian factor variational approximation. Panels~(a,c,e,g) plot the variational lower bound against step 
	number for the approximation with $K=20$ factors.
	\label{fig:LBs:dat}
\end{sidewaysfigure}

\vspace{-7pt}
We first compare the VB approximate and the HMC exact posterior estimators
for the HPSC copula model. 
The VB estimator was fit using $K=0,1,2,3,4,5,10$ and $20$ factors, and for 1,5,10 and 15 thousand steps.
Fig.~\ref{fig:LBs:dat}(b,d,f,h) plots the
mean lower bound value over the last 10\% of steps ($\overline{\mbox{LB}}$) against $K$ for each of the four step sizes
and each of the datasets. 
Increasing $K$ up to 5 improves the accuracy of the VA, but further increases
have little impact.
Fig.~\ref{fig:LBs:dat}(a,c,e,g) plots ${\cal L}(\lambdavec)$ against the step number
for a VA with $K=20$ factors, 
and in each case the SGA algorithm converges rapidly. 
Figs.~A and~B in the Web Appendix plot the mean and standard deviation of the 
coefficients $(\betavec,\alphavec)$ from the VA, against their exact posterior means and standard 
deviations. These show that the variational estimates of the posterior means are highly accurate, 
but---as is usual in VB inference---the posterior standard deviations are underestimated.
Computation times are reported in Table~A in the Web Appendix and show
that the VB estimator is much faster than the exact method 
and practical to implement, even for a copula of 
dimension $n=40,981$.
 
%\textbf{The anova type measures $\frac{1}{n}\frac{\Var(Y_i|x_i)}{\sigma_Y^2}$,
%where $\sigma_Y^2$ is an estimator of the marginal variance of $F_Y$ are given in Table~\ref{tab:anova}.}
 
\vspace{-12pt}
\subsubsection{Predictive accuracy}\label{subsubsec:pred}
\vspace{-7pt}
To compare the accuracy of the five models (PSC, HPSC, PS, HPS and MLT) 
we compute the predictive logarithmic score
by ten-fold cross-validation. For a given 
dataset, we partition the data into 10 (approximately) equally-sized sub-samples,
denoted as
$\{(y_{i,k},x_{i,k},w_{i,k});i=1,\ldots,n_k\}$ for $k=1,\ldots,10$. For sub-sample $k$, we compute the 
density estimator using the remaining 9 sub-samples as the training data, and denote these
as $\hat p_k(y|x,w)$. For our copula model we use the density estimator $\hat p_{\tiny{PE}}$ in Sec.~\ref{subsec:pred:inf}.
The ten-fold mean logarithmic score is then
$
\overline{\mbox{LS}}_{CV}=\frac{1}{10}\sum_{k=1}^{10}\frac{1}{n_k}\sum_{i=1}^{n_k}\log \hat{p}_k(y_{i,k}|x_{i,k},w_{i,k})\,.
$ 

Table~\ref{tab:lsc} reports
the $\overline{\mbox{LS}}_{CV}$ values, where the posterior of the copulas is computed either
exactly using MCMC or HMC, or approximately using VB, with scores given for both cases; we make four
observations.
First, in all examples both copula models---which account fully for the non-Gaussian distribution of the 
responses---outperform the two benchmark PS and HPS models. Second, the performance
of the copula models estimated using VB is very similar to that of the copula models estimated by exact methods.
Third, in every case the HPSC outperforms the 
PSC, showing that the added flexibility of the heteroscedastic copula translates into 
improved distributional predictions -- something
that we demonstrate further below.
Last, in all examples HPSC is competitive with the benchmark MLT model, which 
also allows the entire predictive distribution to vary with the covariates. 

\vspace{-12pt}
\subsubsection{Mean and variance function estimates}
\vspace{-7pt}
To compare the distributional regression estimates, 
Fig.~\ref{fig:functions} plots the posteriors of 
$f,v$, computed as in Sec.~~\ref{subsec:pred:inf} 
for the Rents dataset. Posterior mean and 95\% intervals are given for $f$
in the left-hand panels, and for $v$ in the right-hand panels. 
Panels~(a,b) compare the posteriors from the HPSC model computed exactly using HMC, and
approximately using VB, and they are very similar, further illustrating the high accuracy 
of the VB estimator. 
Panels~(c,d) compare the posteriors 
from the HPSC model using the three different approaches to estimating $F_Y$.
These are the kernel estimator (KDE), the Dirichlet process mixture (DPhat),
and integrating out $F_Y$ using its draws (DP) as in~\cite{GraLis2017}. The posteriors 
are similar, and the approach used to estimate the margin has little effect
on the distributional regression estimates for this example. 
Panels~(e,f) compare the function estimates from the HPSC model against those of the benchmark 
MLT model, and they differ substantially -- particularly for the variance function $v$. 

Finally, panels~(g,h) compare the function estimates 
from the two different regression copulas HPSC and PSC. 
While the estimates of 
$f$ are similar, those for $v$ are very different. 
This is because the regression model for the pseudo-response of the PSC is homoscedastic with respect
to the covariate, whereas that for the HPSC is heteroscedastic. Fig.C in the Web Appendix 
plots the posterior estimates of $f,v$ for the other three datasets. 
Similar results to the Rents dataset are found, where estimates of $f$ from the two regression copula
models are similar, but those of $v$
differ substantially. 
\vspace{-12pt}
\subsubsection{Dependence metrics and prediction}\label{subsubsec:dep}
\vspace{-7pt}
The improved fit of the HPSC over PSC is because the dependence structure of $C_H$ is a much more flexible function of the covariates
than $C_1$ is with $\alphavec=\bm{0}$. To illustrate this we construct
pairwise dependence metrics as follows. Set $\xvec^+=\wvec^+=(\xvec',x_{n+1},x_{n+2})'$,
and $\tilde{\xvec}^+=\{\xvec^+,\xvec^+\}$, 
then compute Spearman's rho for
the bivariate sub-copula $C_H^{n+1,n+2}$ with $\thetavec$ integrated out with
respect to its posterior; ie: 
\[
\hat{\rho}^S(x_{n+1},x_{n+2}) \equiv \int \rho^S_{n+1,n+2}(\tilde{\xvec}^+,\thetavec)
p(\thetavec|\yvec)\mbox{d}\thetavec\,,
\]
where $\rho^S_{n+1,n+2}$ is given in Sec.~\ref{sec:propertiesC2} part~(iii).  
The integration is computed %approximately using draws from the VA, or exactly 
using 
draws from the posterior. For the PSC, the coefficients
$\bm{\alpha}=\bm{0}$, and integration is only with respect to $\thetavec_\beta$.
The metric $\hat \rho^S$ is evaluated on a bivariate grid 
for $(x_{n+1},x_{n+2})$ over the range of the covariate, and its values plotted as a surface.
The process can be replicated for
the other dependence metrics.

Fig.~\ref{fig:dep:spear} plots the surfaces of $\hat{\rho}^S$ for the Rents dataset and
both regression copulas.  For both copulas, $\hat \rho^S(x_{n+1},x_{n+2})$ declines
as $|x_{n+1}-x_{n+2}|$ increases, which is to be expected from any effective regression smoothing method.
However, correlation is locally varying for the HPSC only. For example, correlation is 
higher for values of the covariate (area) around 20 and 80. 
Equivalent surfaces for the other three datesets, along with the upper quantile dependence
 and Kendall's tau, are
given in Part~C of the Web Appendix,
and the same features can be seen. This local variation in the dependence structure of the HPSC
ensures the level of smoothing in the regression and variance functions in Fig.~\ref{fig:functions}
(and Fig.C of the Web Appendix)
 are `locally adaptive' with
respect to the covariate.
%-- a feature of accurate regression smoothers in general.

\begin{figure}[htbp]
\caption{Predictive densities for the Incomes dataset for four different values of age.}
	\vspace{-15pt}
\begin{center}
	\includegraphics[width=0.99\textwidth,angle=0]{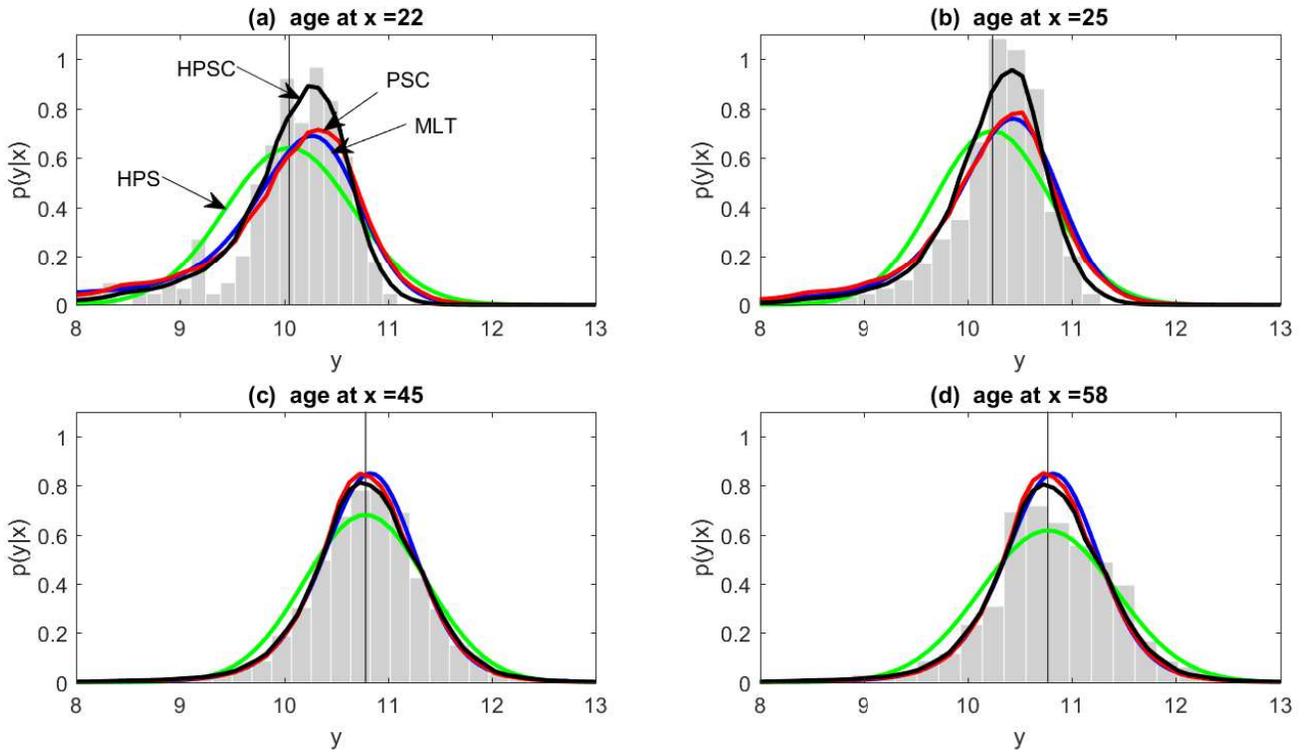}
\end{center}
	\vspace{-5pt}
The different ages are (a) 22 years old, (b) 24 years old,
	(c) 45 years old, and (d) 58 years old. Densities are from the PSC (red), HPSC (black), HPS (green) and MLT (blue) regression models. Also plotted are histograms
	of the sub-samples of individuals with these four ages in the Incomes dataset.\label{fig:incomes:pred:dens}
\end{figure}

In fact, the entire distributional regression fit is locally adaptive to the value of the covariate.
To illustrate this, we 
compute predictive densities for the Incomes dataset from both copula models. 
Fig.~\ref{fig:incomes:pred:dens} plots these for four values of the covariate (age), 
along with those from the benchmark HPS and MLT models.
Because age is measured discretely, we also provide histograms
of the salaries of all individuals of these ages. First, 
because the HPS model is conditionally Gaussian,
the predictive distributions are also, and are inconsistent with the histograms.
Second,
even though the two 
copula models share the same
margin $F_Y$, their predictive densities differ. Those
from the HPSC copula model are more consistent with the histograms, which accords with
the increased accuracy measured by
the scores in Table~\ref{tab:lsc}. The MLT densities are similar to 
those of the PSC copula model, and are also dominated by those from the HPSC copula model. 

\vspace{-12pt}
\subsection{Simulation study}
\vspace{-7pt}
We undertake a simulation study to
illustrate the efficacy of our copula-based approach to 
semi-parametric distributional regression.
Constructing a simulation design is challenging because
all aspects of the distribution are unknown functions of the covariates.
Therefore, we base our designs on the five distributional regression methods
fitted to the four real datasets in the previous subsection, giving 20 data generating processes (DGPs).
From each DGP we simulate 100 datasets (called `replicates' here),
and then refit all five methods to every replicate.
Accuracy of a method for each fitted replicate is assessed by using it to predict
the densities of the observations in an additional 101$^{\mbox{\small st}}$ replicate. Thus, we are assessing the accuracy of out-of-sample density forecasting.
Full details
on the simulation study are given in Part~D of the Web Appendix. 

\begin{sidewaysfigure}[htbp]
	\caption{Simulation results for the DGPs constructed from fits to the Incomes dataset.}
		\vspace{-10pt}
	\begin{center}
		\includegraphics[width=0.9\textwidth,angle=0]{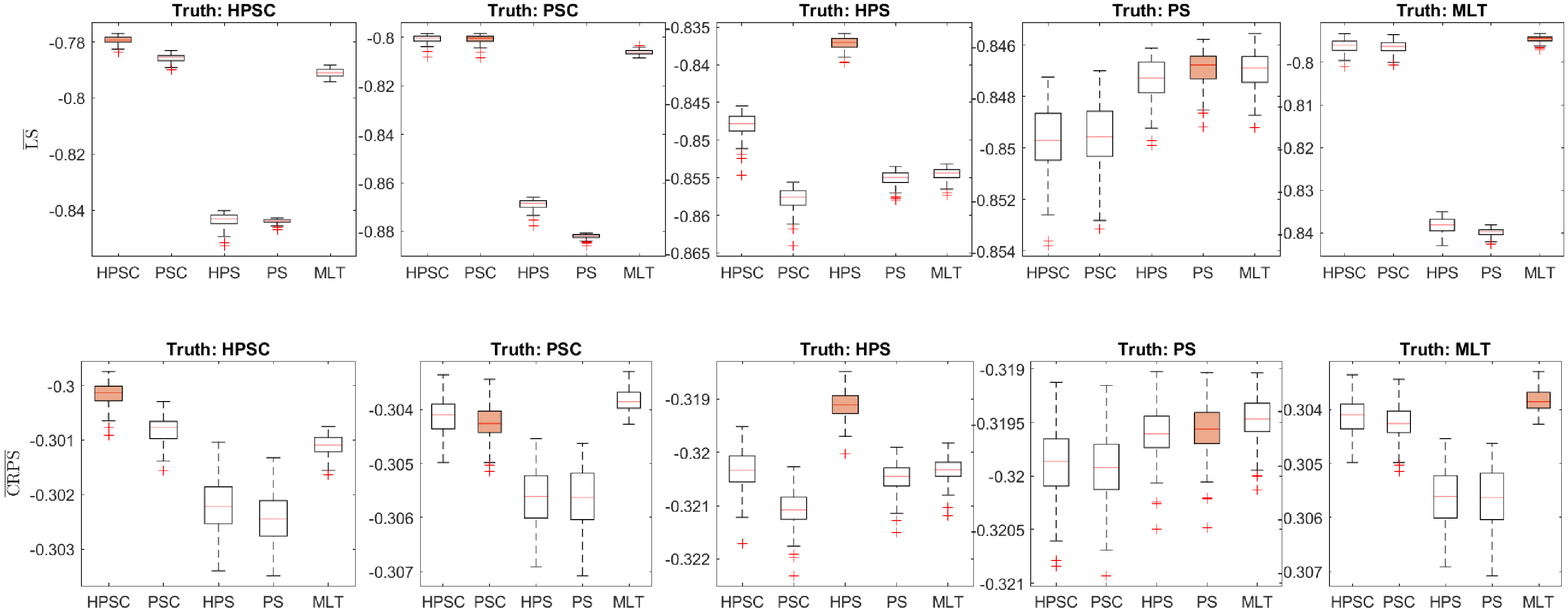}
	\end{center}
	\vspace{-5pt}
		The top panels report the mean logarithmic score ($\overline{\mbox{LS}}$), and the bottom panels the mean CRPS ($\overline{\mbox{CRPS}}$). These are out-of-sample density forecasting metrics averaged over observations in a 101st replicate. Results are orientated so that higher values correspond to greater accuracy.
	The columns give results for replicates simulated from each of the DGPs obtained from fitting the five distributional regression models to the original data. 
	In each panel boxplots of the metrics for each the 100 replicates are given, with one boxplot for each of the five methods.
	The shaded boxplots are for the cases where the method matches the DGP used to generate the data, which will typically be most accurate.
\label{fig:simincomes}
\end{sidewaysfigure}

Fig.~\ref{fig:simincomes} gives boxplots of the
 mean logarithmic score ($\overline{\mbox{LS}}$) and mean continuous ranked probability score ($\overline{\mbox{CRPS}}$)
 \citep{GneRaf2007} of predictions for the DGPs based on the Incomes dataset (the other 15 DGPs
are given in Part~D of the Web Appendix). The shaded boxplots are for the method that 
 matches the DGP in each panel, and these are (unsurprisingly) either the best, or equal best, at recapturing the DGP. 
 Our focus is therefore on the next best performers, and  we make three 
 observations. First, when the DGP is either copula model, the HPSC either equals or out-performs
 the PSC, highlighting its superiority as a regression copula.
 Second, when the DGP is the HPS, then the regression copula HPSC is best at recapturing this DGP. 
 Last, the HPSC either equals or out-performs the MLT benchmark method for all DGPs and metrics,  
 except for the PS DGP. Results for the other 15 DGPs are similar.
\setlength{\abovedisplayskip}{0.1cm}
\setlength{\belowdisplayskip}{0.1cm}
\vspace{-15pt}
\section{Radial Basis Copula for Electricity Prices}\label{sec:electricity}
\vspace{-10pt}
The relationship between intra-day electricity spot price and demand  is
used by participants in 
wholesale markets to formulate
optimal
bidding strategies~\citep[pp.53--72]{krischen2004}. However, its estimation using regression 
methods is difficult
because prices have a
very heavy right tail, and all aspects of their distribution  
vary extensively with demand, day and time of day~\citep{bunn2016}.
To account for this,
we construct a regression copula from trivariate radial bases for $\tilde m, g$, 
combined with horseshoe priors for regularization. We apply
it to high-frequency Australian electricity price and demand 
data, and compare our 
approach to other distributional regression methods.
\vspace{-12pt}
\subsection{Electricity data and regression copula model}
\vspace{-7pt}
The Australian national electricity market (NEM) is a wholesale market 
where generators, distributors and third party participants bid for the sale and 
purchase of electricity one day ahead of transmission;
see~\cite{ignatieva2016} and~\cite{smith2018} for current
descriptions of the market. We consider half-hourly market-wide price $P_i$ 
and total market demand $D_i$
from 1 Jan 2014 to 31 Dec 2018, so that $n=87,648$.
Total market demand is the sum of demand across the five regions in the NEM, while the market-wide price is the demand-weighted average price across the five regions, constructed from data available at {\tt www.aemo.com.au}.
The three covariates are demand, time of day ($\mbox{TOD}_i$) and day number ($\mbox{Day}_i$), and set
$\xvec_i=\wvec_i=(D_i,\mbox{TOD}_i,\mbox{Day}_i)'$, with each covariate scaled
to the unit interval.
For $\tilde m$ and $g$ we employ thin plate spline radial basis functions (RBF) of the form
$b_j(\xvec)=\delta(\xvec-\kvec_j)^2\log(\delta(\xvec-\kvec_j))$ for knot
$\kvec_j$. The distance function 
$\delta(x_1,x_2,x_3)=||(x_1,\sin(\pi x_2),x_3)||$ is the Euclidean distance with a
sine transformation on the second element to ensure the basis is periodic on $[0,1)$ 
for $\mbox{TOD}_i$. The knots are set equal to
a random sample (stratified by time of day) of 240 and 96 covariate values
for $\tilde m$ and $g$, respectively.
We follow~\cite{KleSmi2017} and use the horseshoe prior
for the regularization at Eq.~(\ref{eq:prior:beta}), and provide details in Part~E of the Web Appendix.

\begin{figure}[htbp]
\caption{Marginal distribution summaries of the logarithm of half-hourly electricity prices.}
	\vspace{-15pt}
\begin{center}
\includegraphics[width=0.875\textwidth,angle=0]{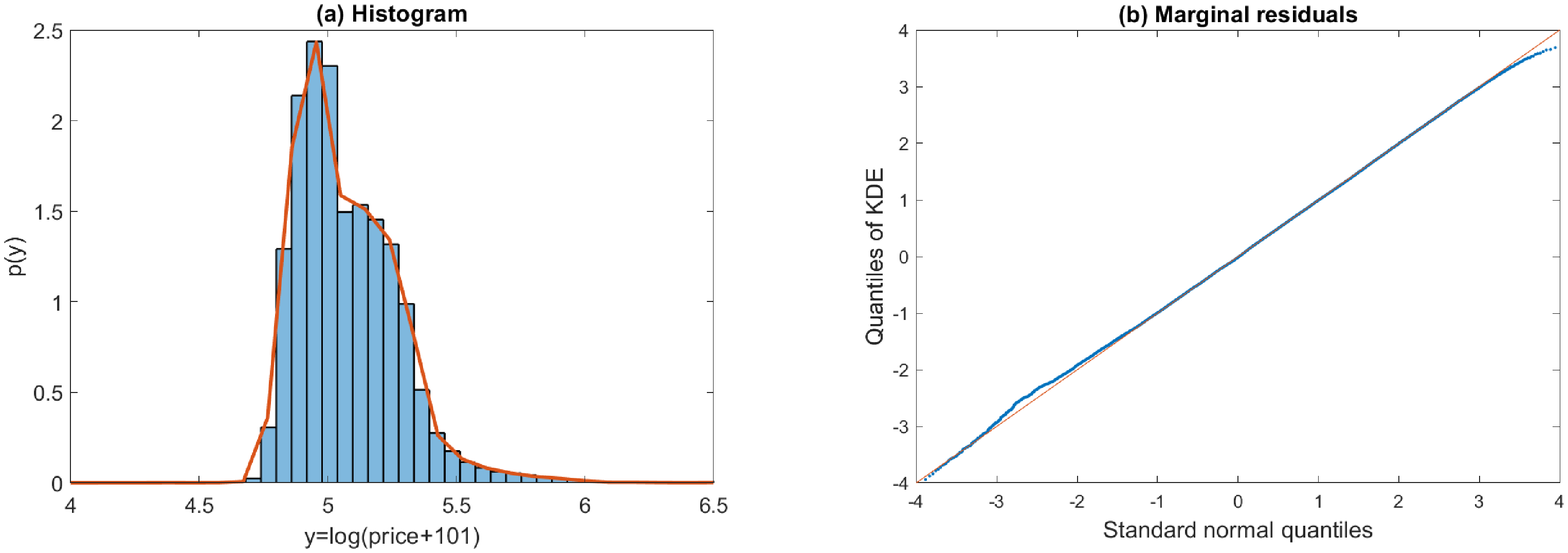}
\includegraphics[width=0.875\textwidth,angle=0]{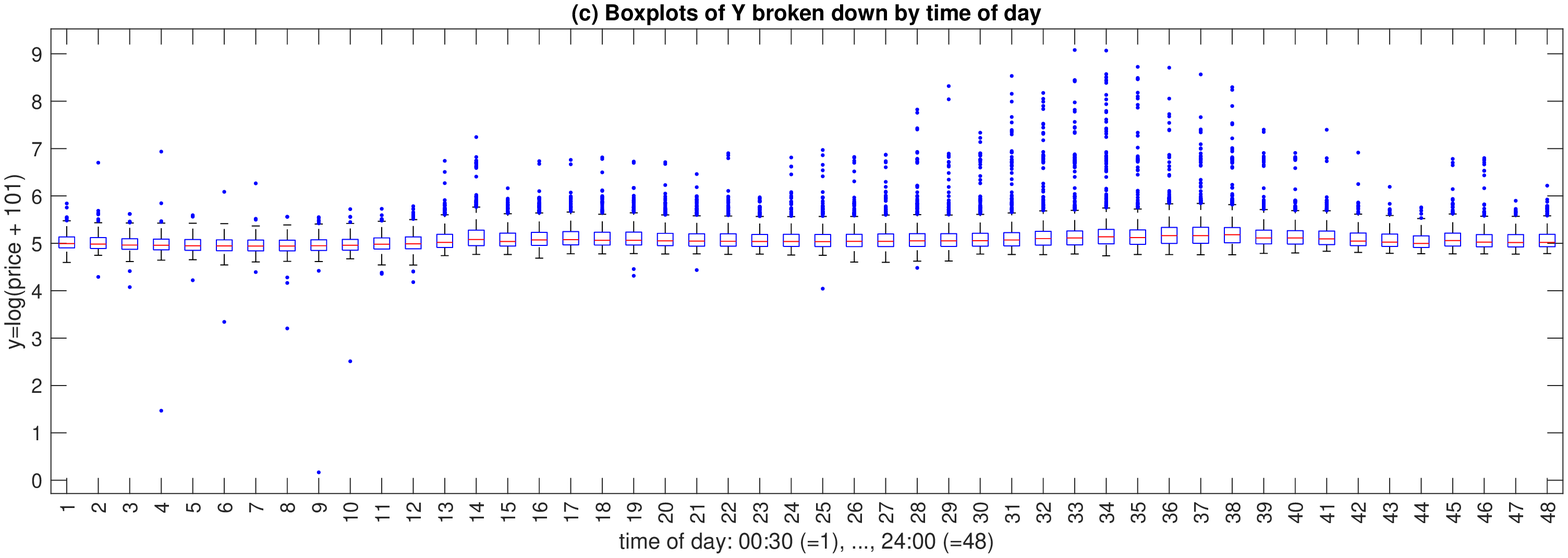}
\end{center}
	\vspace{-5pt}
Summaries are for $Y=\log(\mbox{Price}+101)$. Panel~(a)
plots a histogram and the KDE $\hat F_Y$ over the range $4<Y<6.5$, although the tails of the marginal distribution extend further. 
Panel~(b) provides a quantile-quantile plot of the residuals from the marginal fit $\hat F_Y$. Panel~(c) provides
boxplots of $Y$ broken down by time of day, revealing the diurnal variation in the distribution.
\label{fig:histograms:electricity}
\end{figure}

Due to the extreme skew in electricity prices we  set $Y_i=\log(P_i+101)$, where
we add 101 before taking the logarithm because the minimum observed
price in our data is $-\$99.82$ (prices can be negative in the NEM).
Fig.~\ref{fig:histograms:electricity}(a) plots a histogram of the response and KDE $\hat F_Y$, 
showing that even on the logarithmic scale the
distribution of prices is positively skewed and heavy-tailed. Panel~(b) gives a quantile-quantile
plot highlighting the accuracy of the KDE estimator for $F_Y$. 
Panel~(c) contains boxplots of the response broken down by the time of day, and
reveals the strong diurnal variation in the entire distribution of prices. 
\vspace{-12pt}
\subsection{Empirical results}
\begin{figure}[htbp]
	\caption{The mean and variance of the logarithm of price as a function of demand from the regression copula model.}
		\vspace{-15pt}
	\begin{center}
		\includegraphics[width=0.95\textwidth,angle=0]{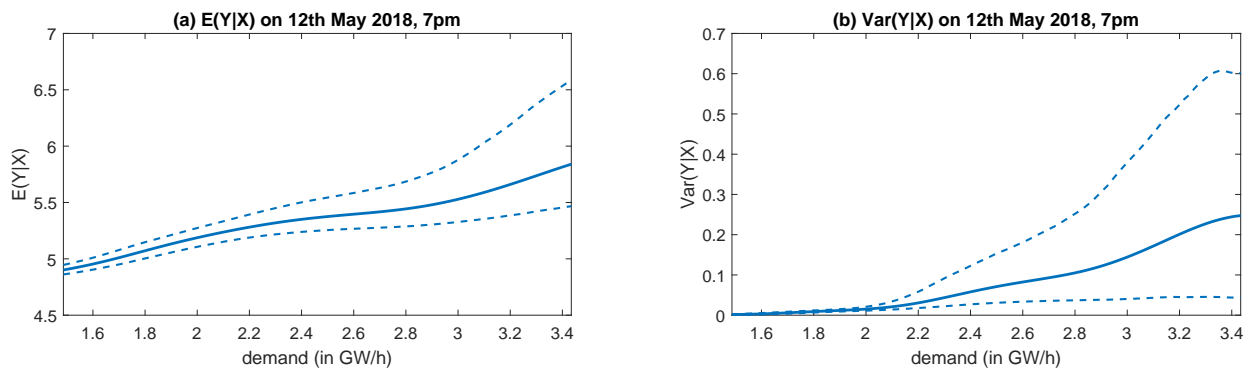}
	\end{center}
	\vspace{-5pt}
	Panel~(a) plots the mean function $f$ and panel~(b) the variance function $v$ against demand (in GW/h) for 12 May 2018 at 19:00 (the peak demand time of day). The (variational) posterior mean
	estimates of the functions are given in bold, while the approximate 95\% posterior intervals are
	given in dashed lines. 
	\label{fig:moments:electricity}
\end{figure}

\vspace{-7pt}
We estimate our regression copula (labeled `HRBFC') using VB with $K=20$ factors for the approximation. A plot of ${\cal L}(\lambdavec)$ against
step number (see Web Appendix) indicates reliable 
convergence of the SGA algorithm. There is a strong (non-additive) effect
of the three covariates on price. For example, Fig.~\ref{fig:moments:electricity}
plots a `slice' of the trivariate mean $f$ and variance $v$
functions against demand $D_i$ for 
12 May 2018 at 19:00, which is the time of day with the highest mean price. 
They are the variational posteriors, computed as at Eq.~(\ref{eq:fns}), and  
show the positive relationship between the first two moments of price
and demand. 

\begin{figure}[htbp]
\caption{Predictive distributions of the logarithm of electricity prices from the regression copula model (HRBFC).}
	\vspace{-10pt}
	\begin{center}
\includegraphics[width=0.95\textwidth,angle=0]{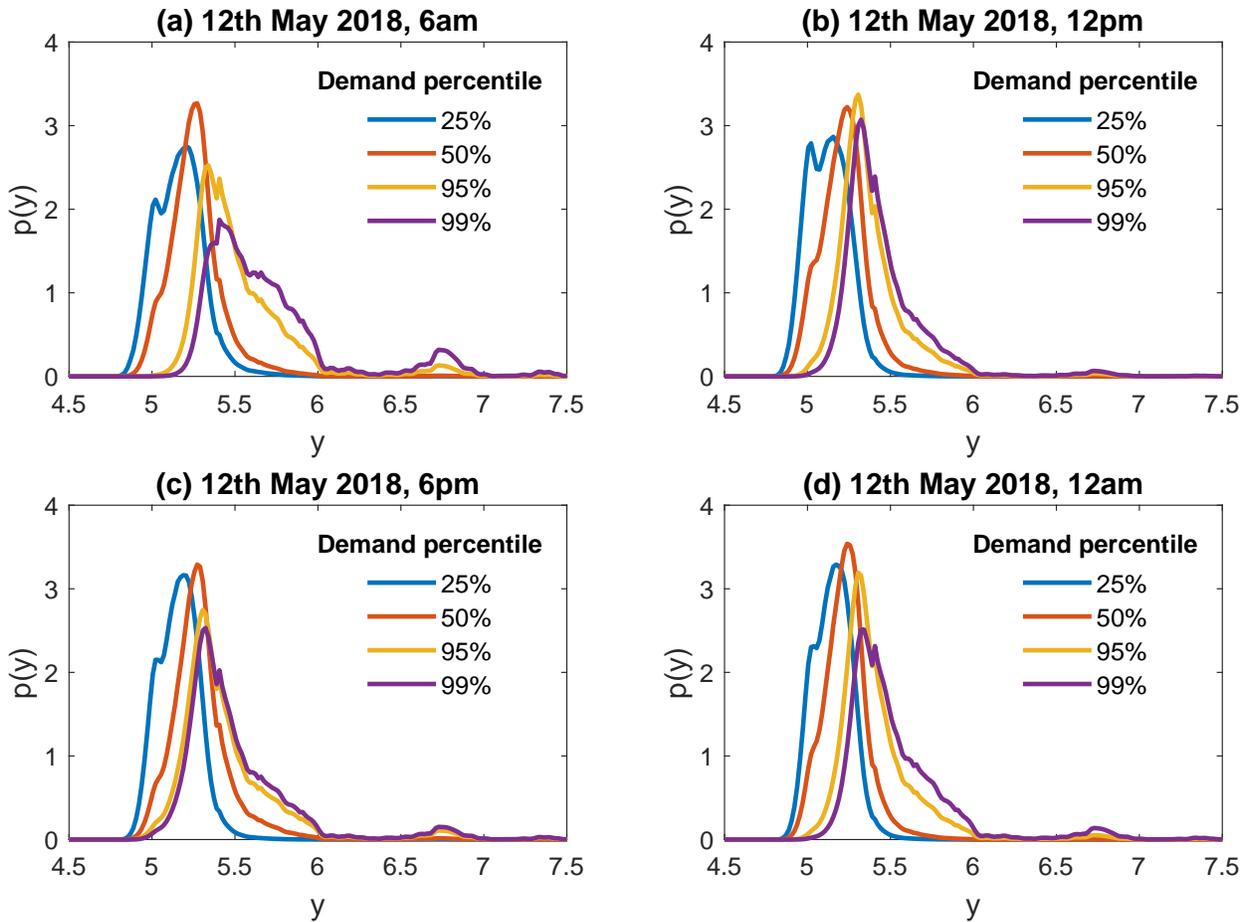}
\end{center}
	\vspace{-5pt}
	The four panels provide predictions for 12 May 2018 at~(a) 06:00, (b)~12:00, (c)~18:00 and (d)~24:00.
In each panel, the predictive densities are constructed at four levels of demand corresponding to 
the 0.25, 0.5, 0.95 and 0.99 percentiles of demand at each time of day. 
\label{fig:dens:tod:electricity}
\end{figure}

To illustrate the impact of demand on the entire distribution,
Fig.~\ref{fig:dens:tod:electricity} plots the predictive densities of $Y$
on 12 May 2018 at~(a) 06:00, (b)~12:00, (c)~18:00 and (d)~24:00.
In each panel, densities are constructed at four levels of demand that correspond to 
the 0.25, 0.5, 0.95 and 0.99 percentiles of demand at each time of day. Increases
in demand accentuate the upper tail, consistent with the nonlinear 
impact of demand shocks on price spikes documented previously~\citep{higgs2008,smith2018}. 
Further plots of predictive densities over the four years (see Part~F of the 
Web Appendix) show the upper tail is increasingly sensitive  to demand, 
matching the increasing frequency of price spikes during the period.

\begin{figure}[htbp]
\caption{Mean quantile score for the electricity price example.}
	\vspace{-15pt}
\begin{center}
\includegraphics[width=0.5\textwidth,angle=0]{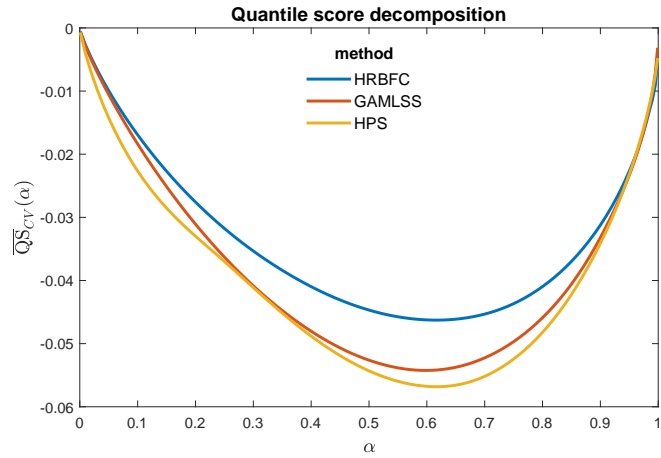}
\end{center}
	\vspace{-5pt}
The 10-fold cross-validated mean quantile score function ($\overline{\mbox{QS}}_{CV}(\alpha)$) orientated so that higher values 
correspond to greater accuracy, and plotted against quantile $0<\alpha<1$ for the regression copula model (HRBFC), and GAMLSS, HPS benchmarks.\label{fig:QS:electricity}
\end{figure}

Last, we compare our regression copula to two benchmarks models. 
The first is the approach
of~\cite{RigSta2005} (labeled `GAMLSS') where we tried several distributions
and found the \texttt{ST2} to give the best fit. We found convergence problems when specifying
all parameters as additive splines of the three covariates, and were restricted to only
allow the mean and variance to do so.
The second benchmark is a heteroscedastic regression model with additive P-spline terms for the three
covariates (labeled `HPS'). To measure the accuracy 
of the distributional forecasts for the three models, Table~\ref{tab:lsc:electricty} 
reports the cross-validated mean score metric $\overline{\mbox{LS}}_{CV}$ defined in Sec.~\ref{subsubsec:pred},
plus a 10-fold cross-validated mean CRPS metric ($\overline{\mbox{CRPS}}_{CV}$). 
The radial basis regression copula model clearly dominates the GAMLSS and HPS benchmarks. Fig.~\ref{fig:QS:electricity} plots the 
(cross-validated) mean quantile score $\overline{QS}_{CV}(\alpha)=\frac{1}{10}\sum_{k=1}^{10}\frac{1}{n_k}\sum_{i=1}^{n_k} \widehat{\mbox{QS}}_k(y_{i,k};\alpha|x_{i,k},w_{i,k})$ for each method and $\alpha\in(0,1)$, where $\widehat{\mbox{QS}}_k(y;\alpha|\xvec,\wvec)\equiv -\mbox{QS}_{\alpha}(\hat{F}_k^{-1}(\alpha),y)$ is defined in
\cite{GneRan2011}. All scores are orientated so that higher values indicate greater accuracy. The 
figure reveals the greater accuracy of 
the HRBFC model at all quantiles, except for the extreme tails where the three
models are similar. Predictive density forecasts from GAMLSS and HPS are
provided in Part~F of the Web Appendix, and show they (unlike those from the HRBFC model) are only weakly effected by
increases in demand, which  is inconsistent with previous analyses~\citep{higgs2008,ignatieva2016,smith2018}.
\setlength{\abovedisplayskip}{0.1cm}
\setlength{\belowdisplayskip}{0.1cm}

\vspace{-15pt}
\section{Discussion}\label{sec:discussion}
\vspace{-12pt}
This paper proposes modeling the entire 
distribution of a vector of regression response values, conditional on covariates, using
a copula decomposition. To do so, a new copula $C_H$ is
constructed from a heteroscedastic semi-parametric regression for a pseudo-response. 
When combined with non-parametric or other margins, the resulting regression model
is flexible in both the distributional shape and the functional
relationship between the covariates and response. Our approach is very general, scalable and 
numerically stable. We show in our empirical work that it 
improves predictive accuracy for non-Gaussian data,
relative to a number of leading benchmark regression 
approaches.

\begin{figure}[htbp]
	\caption{Comparison of different posterior estimates of the regression function $f$ 
		(left-hand side) and  variance function $v$ (right-hand side) for the Rents data.} 
	{
	\centering{\includegraphics[width=0.73\textwidth,angle=0]{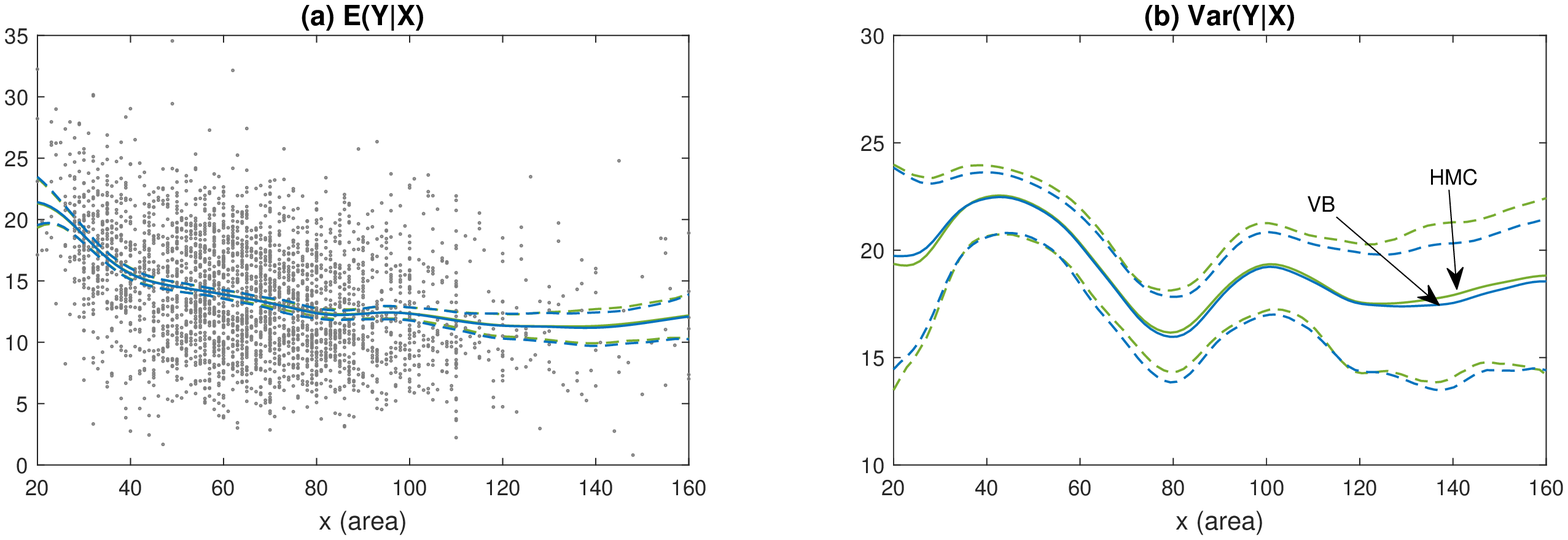}}\\%\vspace{0.1cm}
	\centering{\includegraphics[width=0.73\textwidth,angle=0]{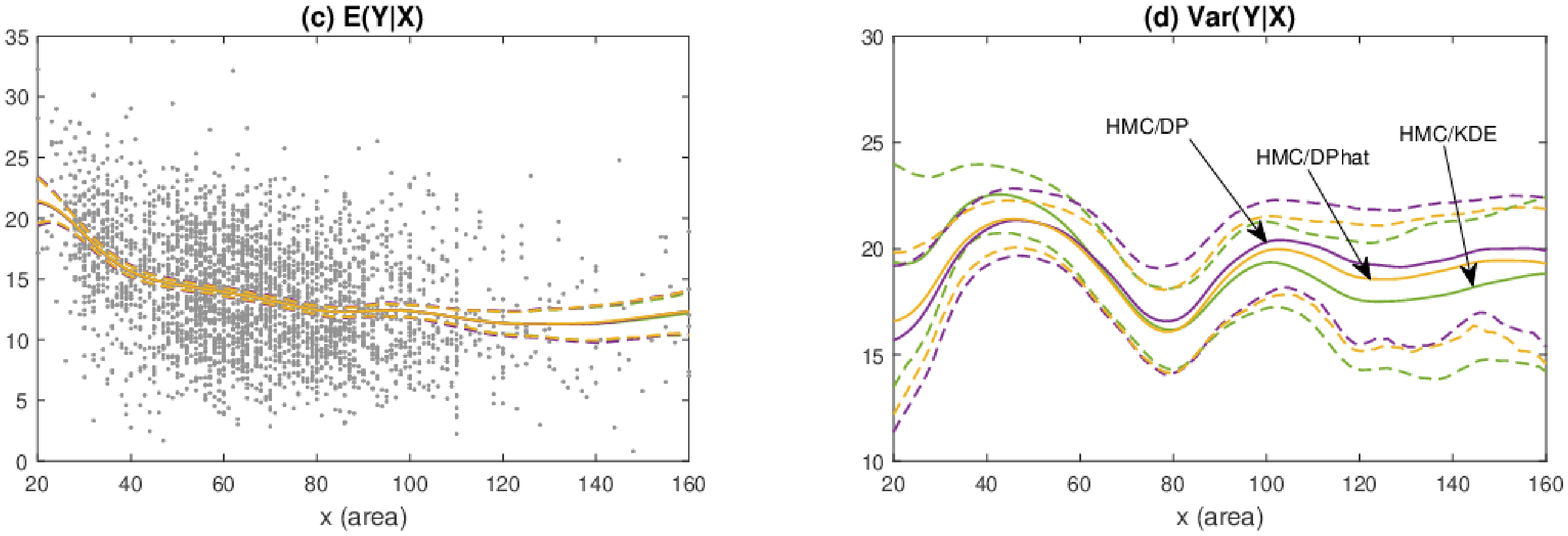}}
	\centering{\includegraphics[width=0.73\textwidth,angle=0]{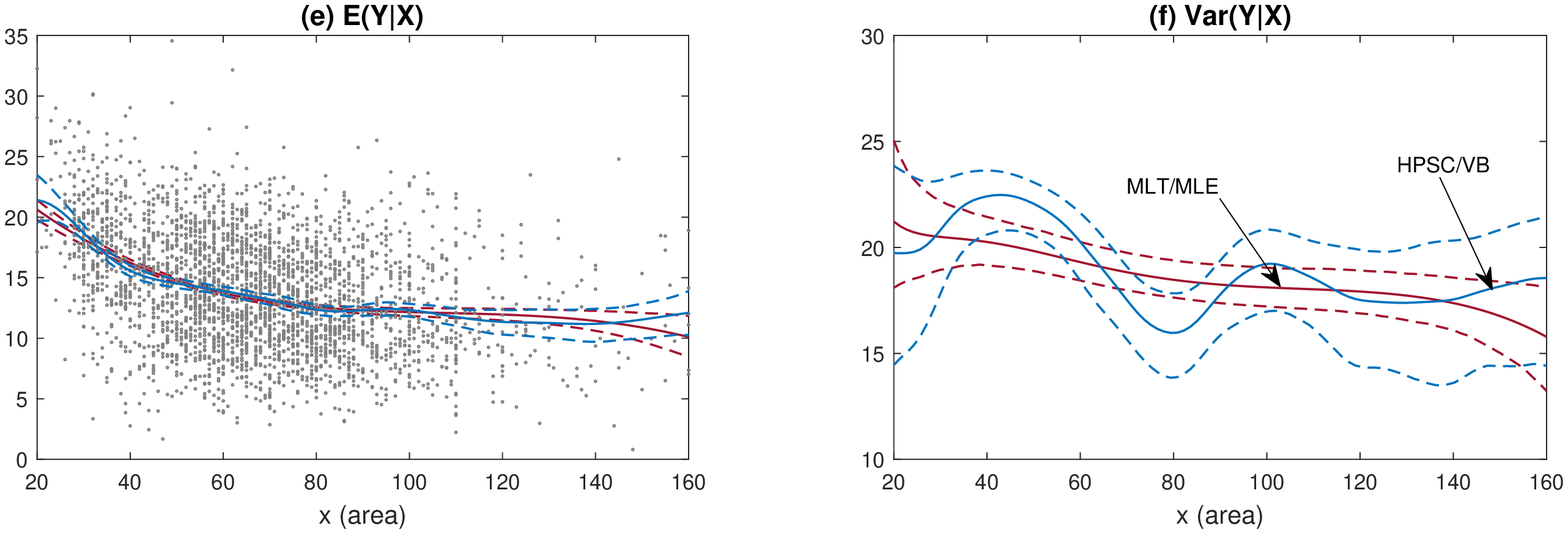}}
		\centering{\includegraphics[width=0.73\textwidth,angle=0]{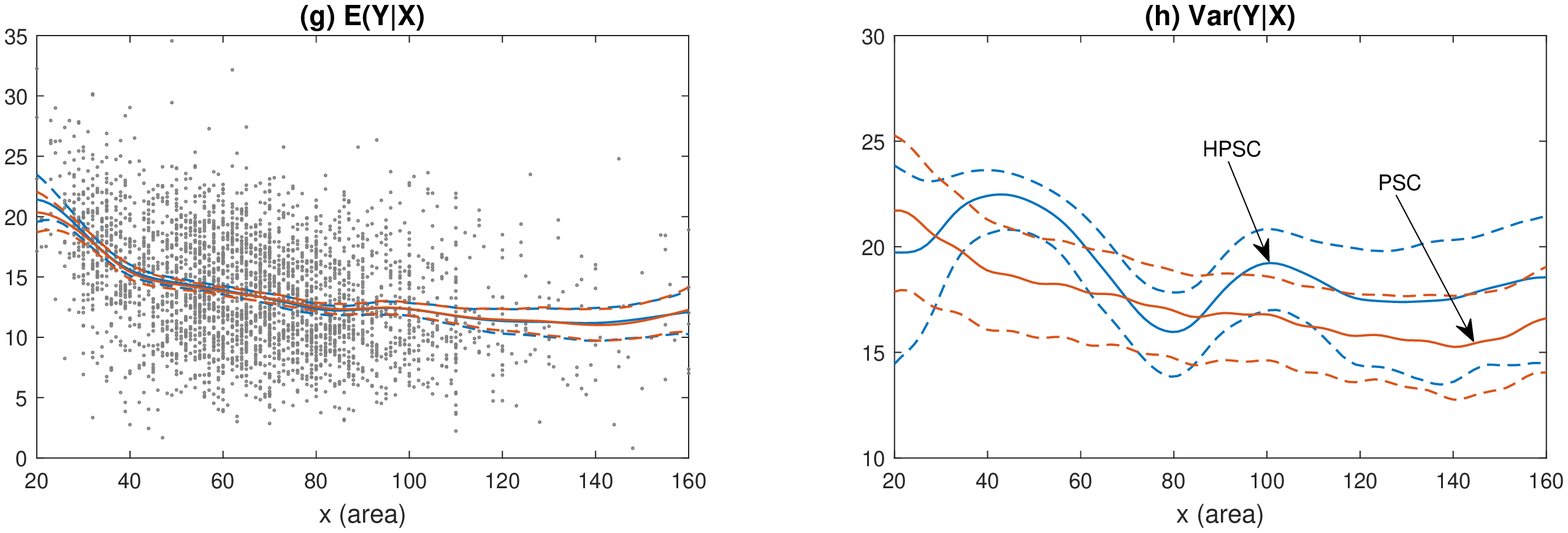}}\\%\vspace{0.1cm}
	}
		The posterior means of $f$ and $v$ are given as solid 
		lines, and 95\% posterior intervals by dashed lines. 
		Scatterplots of the data are included on the left-hand panels.
		Panels~(a,b) compare function estimates from the HPSC model
		computed using HMC and VB. 
		Panels~(c,d) compare function estimates computed using 
		HMC from the HPSC model using three different marginal estimators discussed in
		the text: KDE, DPhat and DP. 
		Panels~(e,f) compare function estimates from the HPSC model (computed using HMC) with those from the benchmark MLT model. Panels~(g,h) show estimates of the regression function $f$ 
		(left-hand side) and variance function $v$ (right-hand side) from the HPSC model,
		compared to those from the PSC model.\label{fig:functions}
\end{figure}

%Low-dimensional copulas have been used previously to capture dependence between multiple response variables in multivariate regression models~\citep{PitChaKoh2006,OakRit2000,OhPat2017},
%sometimes with dependence parameters that are unknown functions of covariates
%\citep{AcaCraYao2011,CraSab2012}. However, these all use copulas in a very different
%way than the $n$-dimensional implicit copulas
%considered here. 
A number of authors construct the $n$-dimensional implicit Gaussian copulas of Gaussian
processes \citep{wauthier2010,Wilson2010}. However, these are very different copulas than 
those constructed here. \citet{KleSmi2017} propose constructing the copula of Bayesian regularized smoothers, which 
is equivalent to our implicit copula when $\alphavec=\bm{0}$. This paper extends their
work by allowing for heteroscedasticity in the pseudo-response, which yields a copula
with a much richer dependence structure as shown in Fig.~\ref{fig:dep:spear}.
This makes the distributional regression
locally adaptive, as can be seen in the mean and variance function estimates in Fig.~\ref{fig:functions}, and increases predictive accuracy. However, our proposed copula is more difficult to estimate, and the MCMC schemes discussed by~\citet{KleSmi2017}---who do not consider alternatives---are infeasible.  To address this, we develop efficient exact estimation with a HMC step
for generating $\alphavec$, and approximate estimation using VB. The empirical work demonstrates the 
efficacy these methods using five diverse real datasets. In every case, our fitted
copula model is more accurate than
both the simpler regression copula $C_1$ with $\alphavec=\bm{0}$ and the benchmark models. Moreover,
estimation and prediction is fast, allowing the application of the distributional
regression methodology to large datasets.

%\newpage
\singlespacing %EDIT
%%EDIT \renewcommand{\baselinestretch}{-2.0}
%\footnotesize
%\renewcommand{\baselinestretch}{1.0}
\bibliography{litliste}

\begin{table}[htbp]
	\caption{Details and source of the four univariate datasets.}
	\vspace{-20pt}
\begin{center}
	{\small
		\centering\renewcommand\arraystretch{1.25}
		\begin{tabular}{ccccc}
			\hline\hline
			\multirow{2}{*}{}{Dataset} & \multirow{2}{*}{}{{$n$}}& \multirow{2}{*}{}{{Covariate}} & \multirow{2}{*}{}{{Response}}  &\multirow{2}{*}{}{{Source}}
			\\\hline
			Geyser & 299 & waiting time (min) & eruption time (min) & \cite{MASS}\\
			Rents & 3,082 &  apartment area (m$^2$) & residential rent  (EUR/m$^2$)  &\cite{fahkne13} \\
			Amazon  &   31,925  & website visit duration (min) & $\log$(sales) ($\log(\mbox{USD})$) &\cite{PanSmiDan2014} \\     
			Incomes & 40,981 & worker age (years) & $\log$(income)  (EUR) &\cite{KleKneLanSoh2015} \\\hline\hline
		\end{tabular}
	}
\end{center}
	\vspace{-5pt}
The columns give (in order) the dataset name, number of observations, covariate,
response variable, and published source of the data.\label{tab:summaries}
\end{table}

\begin{table}[h]
\caption{Key quantities of the two P-spline regression copulas PSC and HPSC.}
	\vspace{-20pt}
\begin{center}
	{\small
\renewcommand\arraystretch{1.25}
\begin{tabular}{ccc}
  \hline\hline
 Quantity & PSC & HPSC
    \\\hline
  $\varthetavec$ &  $\varthetavec=\lbrace\betavec,\thetavec_{\beta}\rbrace$ & $\varthetavec=\lbrace\betavec,\alphavec,\thetavec_{\beta},\thetavec_{\alpha}\rbrace$ \\
 $s_i$ &  $s_i=(1+\bvec_i'P_\beta(\thetavec_{\beta}^{-1})\bvec_i)^{-1/2}$ & $s_i=(\exp(\vvec_i'\alphavec)+\bvec_i'P_\beta(\thetavec_{\beta}^{-1})\bvec_i)^{-1/2}$ \\
 $S$  &  $S(\xvec,\thetavec_{\beta})$ & $S(\xvec,\wvec,\alphavec,\thetavec_{\beta})$\\
$R$ & $R(\xvec,\thetavec_{\beta})=S(I+BP_\beta(\thetavec_{\beta}^{-1})B')S'$ & $R(\tilde{\xvec},\alphavec,\thetavec_{\beta})=S(\exp(V\alphavec)+BP_\beta(\thetavec_{\beta}^{-1})B')S'$\\ 
    \hline\hline
\end{tabular}
}
\end{center}
	\vspace{-5pt}
 Reported from the top to bottom rows are: augmented parameters, normalizing factor, normalizing matrix, and parameter matrix of the Gaussian copula $C_1$.\label{tab:models}
\end{table}

\begin{table}[htbp]
\caption{Predictive accuracy of distributional regression methods for four test datasets.}
\begin{center}
		\vspace{-20pt}
{\small
\renewcommand\arraystretch{1.25}
\begin{tabular}{cccccccc}
  \hline\hline
   &\multicolumn{7}{c}{Model / Estimation Method} \\ \cline{2-8}
 & \multicolumn{2}{c}{PSC/} & \multicolumn{2}{c}{HPSC/} & PS/ & HPS/ & MLT/  \\ 
Dataset &VB &MCMC &VB &HMC &MCMC &MCMC & MLE \\ \hline
Geyser &  -189.80 & -190.08 & \textbf{-187.52} & -188.19  &  -409.86 & -349.56 & -259.38 \\
    Rents    & -89,105 & -89,015 &  \textbf{-88,949} & -88,959 & -89,285 & -89,277 & -88,973 \\
 %   Nigeria   & -46,084  & -46,059 &  \textbf{-45,982} & -46,054 & -46,369 & -46,307 & -46,144 & \textbf{-45,845}\\
    Amazon   & -42,328 & -42,253 &  \textbf{-42,207} & -42,213 & -43,064 & -42,898 & -42,409 \\     
    Incomes   & -33,339 & -32,722  & \textbf{-32,396}  & -32.530 & -35,259 & -35,102 & -32,840 \\\hline\hline
\end{tabular}
}
\end{center}
	\vspace{-5pt}
	{\small
 The 10-fold cross-validated mean predictive logarithmic scores ($\overline{\mbox{LS}}_{CV}$) multiplied by $n$ for presentation. Higher values indicate greater accuracy.
	 The models are the two regression copula models (PSC and HPSC),
 	and the benchmark Gaussian P-spline (PS), its heteroscedastic version (HPS) and 
 	the `most likely transformation' method (MLT).
 	The Bayesian posterior of the copulas are computed either exactly using MCMC or HMC,
 	or approximately using VB, with scores given for both cases.}
 	\label{tab:lsc}
\end{table}

\begin{table}[htbp]
	\caption{Two measures of predictive accuracy for the Australian electricty spot price data.}
	\begin{center}
		\vspace{-15pt}
		{\small
			\renewcommand\arraystretch{1.25}
			\begin{tabular}{cccc}
				\hline\hline
				Metric & HRBFC& HPS &GAMLSS \\ \hline
				$\overline{\mbox{LS}}_{CV}$ & \textbf{0.8416}  &0.4351  &0.6926 \\
				$\overline{\mbox{CRPS}}_{CV}$ &\textbf{-0.0677} &-0.0809 &-0.0772 \\\hline\hline
			\end{tabular}
		}
	\end{center}
	\vspace{-5pt}
	The 10-fold cross-validated mean predictive logarithmic score  ($\overline{\mbox{LS}}_{CV}$) and continuous ranked probability score ($\overline{\mbox{CRPS}}_{CV}$), for density forecasts from the three distributional regressions.
	Higher values indicate greater predictive accuracy.
	The models are the radial basis regression copula (HRBFC),
	heteroscedastic P-spline  (HPS) and 
	GAMLSS with family~\texttt{ST2}.
	%The Bayesian posterior of the copula is computed using VB.
	\label{tab:lsc:electricty}
\end{table}

\begin{figure}[htbp]
	\caption{Spearman's rho from the regression copulas for the Rents dataset.}
		\vspace{-15pt}
	\begin{center}
	\includegraphics[width=0.825\textwidth,angle=0]{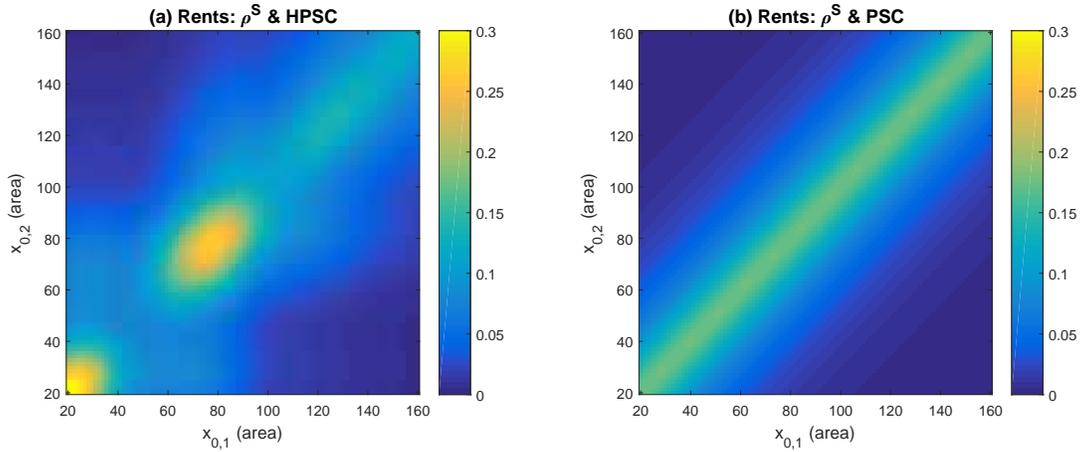}
	\end{center}
	\vspace{-5pt}
	Each panel plots estimates of Spearman's rho $\hat\rho^S(x_{n+1},x_{n+2})$ as bivariate functions
	of $(x_{n+1},x_{n+2})$ over the range of the covariate (area). Panel~(a) gives 
		values for the HPSC, and panel~(b) for the PSC. The localized variation in $\hat\rho^S$
		in panel~(a)
		corresponds to local adaptivity in the distributional regression from the copula model. Analogous
		plots for the other three datasets, and other dependence metrics (Kendall's tau, upper and lower quantile dependence) are given in the Web Appendix.\label{fig:dep:spear}
\end{figure}

%\begin{figure}[t]
%\caption{Electricity IV.}
%\begin{center}
%\includegraphics[width=0.5\textwidth,angle=0]{Figures/HRBFC_LB.eps}
%\end{center}
%s.\label{fig:LB:electricity}
%\end{figure}

\newpage

\doublespacing
\noindent
\begin{center}{\bf \Large{Web Appendix for\\\vspace{1em}}}
{\bf \Large{\lq \centering Bayesian Inference for Regression Copulas\rq}}\end{center}
\begin{center}
	{\bf Contents}
\end{center}
\begin{itemize}
	\item[] {\bf Part~A}: Proofs and derivations.
	\item[] {\bf Part~B:} Derivation of the derivatives required to implement the exact and approximate inferential schemes for the HPSC and PSC in Section~4, and pseudo code for the exact sampler of Section~3.
	\item[] {\bf Part~C}: Additional figures and tables referred to in Section~4.1 of the manuscript.
	\item[] {\bf Part~D}: Details and additional figures on the simulation in Section~4.2 of the manuscript and additional results.
	\item[] {\bf Part~E}:  Details and derivatives required to implement the approximate inferential scheme for the HRBFC in Section~5.
	\item[] {\bf Part~F}: Additional figures referred to in Section~5 of the manuscript.
\end{itemize}

\newpage

\appendix
\renewcommand{\thesection}{Part~\Alph{section}}
\renewcommand\thefigure{\Alph{figure}}   
\renewcommand\thetable{\Alph{table}} 
\newpage
\vspace{-15pt}
\section{Proofs and Derivations}\label{sec:proof}
\vspace{-15pt}
\numberwithin{equation}{section}

\setcounter{equation}{0}
\subsection{Margin of $Y_i$}
\vspace{-10pt}
Regression models are usually specified conditional on parameters for the mean, variance and possibly
other moments. In contrast, the definition of the regression copula model at Eq.~(1) is unconditional 
on such parameters, and the margin $F_Y$ of $Y_i$ is assumed to be invariant with respect to the covariates
$\tilde{\xvec}_i$. 
However, $Y_i$ is dependent on $\tilde{\xvec}_i$ when also conditioning on the mean and variance functions $\tilde m, g$ of the pseudo-response, as we now show. First, from Eq.~(8) the normalized response has 
distribution $Z_i|\tilde{\xvec}_i,\tilde{m},g \sim N(m(\xvec_i,\wvec_i),s_i^2\sigma_i^2)$.  From Theorem~1,
$U_i=\Phi_1(Z_i)$, so that $Y_i=F_Y^{-1}(\Phi_1(Z_i))$, and the Jacobean of the transformation
is $p_Y(y_i)/\phi_1(z_i)$. Then the density of the conditional distribution is:
\[
p(y_i|\tilde{\xvec}_i,\tilde m,g)=p(z_i|\tilde{\xvec}_i,\tilde m,g)\frac{p_Y(y_i)}{\phi_1(z_i)} =\frac{1}{s_i\sigma_i}
\phi_1\left(\frac{z_i-m(\xvec_i,\wvec_i)}{s_i\sigma_i}\right)\frac{p_Y(y_i)}{\phi_1(z_i)}\,.
\]
Thus, the distribution of $Y_i$ is a function of $\tilde{\xvec}_i=\{\xvec_i,\wvec_i\}$ when also conditioning
on $\tilde m,g$.

\subsection{Proof of Theorem~1}
\vspace{-10pt}
Recall that $\tilde{\xvec}=\{\xvec,\wvec\}$ and $\thetavec=\{\thetavec_\alpha,\thetavec_\beta\}$, and
note that the distribution 
$Z_i|\tilde{\xvec},\thetavec$ is standard normal because
\begin{equation*}\begin{aligned}
p(z_i|\tilde{\xvec},\thetavec)&=\int p_{Z_i}(z_i|\tilde{\xvec},\alphavec,\thetavec_\beta)\,p(\alphavec|\thetavec_{\alpha})\mathrm{d}\alphavec=\int \phi_1(z_i)\,p(\alphavec|\thetavec_{\alpha})\mathrm{d}\alphavec=\phi_1(z_i)\,,
\end{aligned}\end{equation*}
while 
%from Eq.~(\ref{eq:c1}),
$p_Z(\zvec|\tilde{\xvec},\alphavec,\thetavec_\beta)=c_1(\uvec|\bm{x},\bm{w},\alphavec,\thetavec_{\beta})\prod_{i=1}^n \phi_1(z_i)$.
Then, the implicit copula density~\cite[Sec 3.1]{Nel2006} of $\bm{Z}|\tilde{\xvec},\thetavec$ is given by
\begin{equation*}\begin{aligned}
c_H(\uvec|\tilde{\xvec},\thetavec)&=\frac{p(\zvec|\tilde{\xvec},\thetavec)}{\prod_{i=1}^n \phi_1(z_i)}=\frac{\int p_Z(\zvec|\tilde{\xvec},\alphavec,\thetavec_\beta)\,p(\alphavec|\thetavec_{\alpha})\mathrm{d}\alphavec}{\prod_{i=1}^n \phi_1(z_i)}\\
&=\frac{\int c_1(\uvec|\bm{x},\bm{w},\alphavec,\thetavec_{\beta})\prod_{i=1}^n \phi_1(z_i) p(\alphavec|\thetavec_{\alpha})\mathrm{d}\alphavec}{\prod_{i=1}^n \phi_1(z_i)}=\int c_1(\uvec|\bm{x},\bm{w},\alphavec,\thetavec_{\beta}) 
p(\alphavec|\thetavec_{\alpha})\mathrm{d}\alphavec\,,
\end{aligned}\end{equation*}
which is the required expression for $c_H$ in Theorem~1. Similarly, if $F_Z(\uvec|\tilde{\xvec},\thetavec)$ denotes
the joint distribution function of $\bm{Z}|\tilde{\xvec},\thetavec$, then its implicit copula function~\citep[Sec.3.1]{Nel2006} is
\begin{eqnarray*}
	C_H(\uvec|\tilde{\xvec},\thetavec) &= &F(\zvec|\tilde{\xvec},\thetavec)=\int\int p_Z(\zvec|\tilde{\xvec},\alphavec,\thetavec_\beta)p(\alphavec|\thetavec_\alpha)
	\mbox{d}\alphavec \mbox{d}\zvec \\
	&= &\int\int \phi(\zvec;\bm{0},R)\mbox{d}\zvec\, p(\alphavec|\thetavec_\alpha) \mbox{d}\alphavec 
	= \int \Phi(\zvec;\bm{0},R) p(\alphavec|\thetavec_\alpha) \mbox{d}\alphavec \\
	& = &\int C_1(\uvec|\xvec,\wvec,\alphavec,\thetavec_\beta)p(\alphavec|\thetavec_\alpha) \mbox{d}\alphavec\,.
\end{eqnarray*}

\subsection{Proof of Theorem~2}
First, note that 
$|C_1(\uvec|\xvec,\wvec,\alphavec,\thetavec_\beta)\,p(\alphavec|\thetavec_\alpha)|
\leq M(\uvec)\,p(\alphavec|\thetavec_\alpha)$, where $M(\uvec)$ is the 
upper Fr\'{e}chet-Hoeffding bound. Also note that from 
the definition of $R$,
% at Eq.~(\ref{eq:correlationmatrix}),
\[
\lim\limits_{ \gamma_\beta \rightarrow 0} R = S^\star \Sigma S^\star= I_n\,,
\]
where $S^\star=\mbox{diag}(\exp(-\vvec_1'\alphavec/2),\ldots,\exp(-\vvec_n'\alphavec/2))$\,.
Then, by Theorem~1 and Lebesgue's dominated convergence 
theorem,
\begin{eqnarray*}
	\lim\limits_{ \gamma_\beta \rightarrow 0} C_H(\uvec|\tilde{\xvec},\thetavec) &= &
	\int \lim\limits_{ \gamma_\beta \rightarrow 0} C_1(\uvec|\xvec,\wvec,\thetavec_\beta,\alphavec)\,p(\alphavec|\thetavec_\alpha)\mbox{d}\alphavec\\
	&= &
	\int \lim\limits_{ \gamma_\beta \rightarrow 0} \Phi(\zvec;\bm{0},R)\,p(\alphavec|\thetavec_\alpha)\mbox{d}\alphavec =\int \Phi(\zvec; \bm{0},I)\,p(\alphavec|\thetavec_\alpha)\mbox{d}\alphavec\\
	&= &   \Phi(\zvec; \bm{0},I) =\Phi((\Phi_1^{-1}(u_1),...,\Phi_1^{-1}(u_n))'; 0,I) \\
	&= &\prod_{i=1}^n \Phi_1(\Phi_1^{-1}(u_i))= \Pi(\uvec)\,.
\end{eqnarray*}

\subsection{Derivation of Dependence Metrics for $C_H^{ij}$}
To derive the lower quantile dependence metric at part~(i), 
\begin{eqnarray*}
	\lambda^L_{ij}(q|\tilde{\xvec},\thetavec)
	&= &\frac{C_H^{ij}(q,q)}{q}= \int \frac{C_1^{ij}(q,q)\,p(\alphavec|\thetavec_\alpha)}{q} \mbox{d}\alphavec=\int \lambda^L_1(q|\xvec,\wvec,\alphavec,\thetavec_\beta)\,p(\alphavec|\thetavec_\alpha) \mbox{d}\alphavec\,.
\end{eqnarray*}
The derivation of the upper quantile dependence is similar.

To derive the metrics at part~(ii), first note that for any bivariate copula function $C$, if $M(u,v)=\min(u,v)$ is the 
upper Fr\'{e}chet-Hoeffding bound, then
$|C(q,q)/q| \leq  M(q,q)/q = 1 $. Denote the $(i,j)$th element of $R$ 
%at Eqn.~(\ref{eq:correlationmatrix})
as $r_{ij}$, and the sub-copulas
of $C_1$ and $C_H$ in these elements as $C_1^{ij}$ and $C_H^{ij}$. 
Then, by Theorem~1 and Lebesgue's dominated convergence 
theorem,
\[
\lambda^L_{ij}  = \lim\limits_{q \downarrow 0} \frac{C_H^{ij}(q,q)}{q}
=  \int\lim\limits_{q \downarrow 0}\frac{C_1^{ij}(q,q)}{q} p(\alphavec|\thetavec_\alpha) \mbox{d}\alphavec = 0
\]
because $C_1^{ij}$ is a Gaussian copula with zero tail dependence, so that
$\lim\limits_{q \downarrow 0}\frac{C_1^{ij}(q,q)}{q}=0$. 
The derivation of $\lambda^U$ is similar.

The derivation of $\rho^S_{ij}$ in part~(iii) follows from the definition
of Spearman's correlation, and its expression for a Gaussian copula, as follows:
\begin{eqnarray*}
	\rho^S_{ij}(\tilde{\xvec},\thetavec)
	&= &12 \int C_H^{ij}(u,v)\mbox{d}(u,v)-3 = 12 \int \int C_1^{ij}(u,v)\,p(\alphavec|\thetavec_\alpha)\mbox{d}(u,v)\mbox{d}\alphavec -3\\
	&= & \int \left( 12\int C_1^{ij}(u,v)\mbox{d}(u,v)-3+3\right)\,p(\alphavec|\thetavec_\alpha) \mbox{d}\alphavec -3\\
	&= &\frac{6}{\pi} \int \arcsin(r_{ij}/2)\,p(\alphavec|\thetavec_{\alpha})\mbox{d}\alphavec\,.
\end{eqnarray*}
The derivation of $\tau^K_{ij}$ is similar. 
\newpage

\vspace{-15pt}
\setlength{\abovedisplayskip}{0.1cm}
\setlength{\belowdisplayskip}{0.1cm}
\section{Details on Estimation}
\vspace{-10pt}

\vspace{-12pt}
\subsection{Exact Estimation}
\vspace{-7pt}

We now review the implemented steps for exact inference for the HPSC. 
Recall that for this copula $\thetavec_\beta=\{\tau^2_\beta,\psi_{\beta,1},\psi_{\beta,2}\}$ and
$\thetavec_\alpha=\{\tau^2_\alpha,\psi_{\alpha,1},\psi_{\alpha,2}\}$, and the precision matrices
can be factorized as $P_\beta(\thetavec_\beta)=\frac{1}{\tau^2_\beta} P(\psivec_\beta)$ and 
$P_\alpha(\thetavec_\alpha)=\frac{1}{\tau^2_\alpha} P(\psivec_\alpha)$, where $P(\psivec)$ is the usual band
two scaled precision matrix of an AR(2) process with partial autoregressive coefficients $\psivec$.
The complete algorithm is provided at the end of this Section in Algorithm~\ref{exactsampler}. The sampler for the PSC is obtained by simply skipping the generation steps of $\alphavec$ and $\thetavec_{\alpha}.$
\vspace{-10pt}
\subsubsection{Gibbs update for  $\betavec$}
\vspace{-5pt}
We generate from   
the conditional posterior 
$p(\betavec|\xvec,\wvec,\yvec,\lbrace\thetavec\setminus\betavec\rbrace)=p(\betavec|\xvec,\zvec,\lbrace\thetavec\setminus\betavec\rbrace)$ which is Gaussian with mean
$\muvec_{\beta}=\Sigma_{\beta}B'\Sigma^{-1}S^{-1}\zvec$ 
and covariance matrix $\Sigma_{\beta}=(B'\Sigma^{-1}B+P_{\beta}(\thetavec_{\beta}))^{-1}$.  %\textbf{note that this is not $\Omega$ any more as in the homosecdastic case. Mike, can you please check and compute as well that I am really correct?}

\vspace{-10pt}
\subsubsection{MH step for  $\thetavec_{\beta}$}\label{blubb}
\vspace{-5pt}
We generate each element of $\thetavec_{\beta}$ separately, relying on analytical derivatives for the proposal densities and by making the algorithm extremely efficient as we only store the unique rows of the design matrices $B$.

A Metropolis-Hastings step is used
to generate $\tilde\tau_{\beta}^2=\log(\tau_{\beta}^2)$, where a normal distribution matching the mode and curvature is used to 
approximate its conditional. Note
that
\begin{equation*}\label{eq:logFC:upsilon}
  \begin{aligned}
	  &l_{{\tilde\tau_{\beta}^2}}\equiv \log(p({\tilde\tau_{\beta}^2}|\xvec,\wvec,\yvec,\lbrace\thetavec\setminus\tau_{\beta}^2\rbrace))\propto -\frac{{\tilde\tau_{\beta}^2}}{2}\left(p_1-1\right)-\frac{1}{2\exp({\tilde\tau_{\beta}^2})}\betavec'P(\psivec_{\beta})\betavec-\left(\frac{\exp({\tilde\tau_{\beta}^2})}{b_{\tau_{\beta}^2}}\right)^{\tfrac{1}{2}}\\&
\quad\quad\quad\qquad\qquad\qquad\quad\;\,-\frac{1}{2}\sum_{i=1}^{n}\log(s_i^2)
-\frac{1}{2}\left(\zvec'(S\Sigma S')^{-1}\zvec-2\betavec'B'\Sigma^{-1}S^{-1}\zvec\right).
  \end{aligned}
\end{equation*}
Approximating $l_{{\tilde\tau_{\beta}^2}}$ by a second order Taylor expansion around the current state ${{\tilde{\tau_{\beta}^2}}}^{(m)}$,
and taking the exponent yields a Gaussian proposal density involving the score function and the second derivative of the conditional posterior above. 
Analytical expressions for these are given in Appendix~B.1 of~\citet{KleSmi2017} after replacing $P((\psi_{\beta,1},\psi_{\beta,2})')$ by $P(\psi_{\beta})$, $S^2$ by $S\Sigma S'$, and $\betavec'B'\Sigma^{-1}S^{-1}\zvec$ by $\betavec'B'S^{-1}\zvec$.

To improve sampling behaviour, we transform $\psi_{\beta,j}\in[-1+\epsilon,1-\epsilon]$ component-wise for $j=1,2$ onto the real line via 
$\tilde{g}:[-1+\epsilon,1-\epsilon]\to\dsR$, $\tilde\psi_{\beta,j}\equiv \tilde{g}(\psi_{\beta,j}) = \log\left((\psi_{\beta,j}+(1-\epsilon))/(1-\epsilon-\psi_{\beta,j})\right)$,
and set $\epsilon=0.05$. The conditional log-posterior is
\begin{equation*}\label{eq:logFC:xi}
  \begin{aligned}
	  l_{\tilde\psi_{\beta,j}}\equiv \log(p(\tilde\psi_{\beta,j}|\xvec,\wvec,\yvec,\lbrace\thetavec\setminus\psi_{\beta,j}\rbrace))&\propto \log\left(\frac{\partial\psi_{\beta,j}}{\partial\tilde\psi_{\beta,j}}\right)
-\frac{1}{2}\left(\zvec'(S\Sigma S')^{-1}\zvec-2\betavec'B'\Sigma^{-1}S^{-1}\zvec\right)\\&\quad-\frac{1}{2}\sum_{i=1}^{n}\log(s_i^2)-\frac{\betavec'P(\psivec_\beta)\betavec}{2\tau_{\beta}^2}.
  \end{aligned}
\end{equation*}
For $j=1,2$ separately we then perform MH-steps with Gaussian proposal computed from the gradient and Hessian of $l_{\tilde\psi_{\beta,j}}.$
First and second derivatives with respect to $\tilde\psi_{\beta,j}$ are
\begin{equation}\label{eq:tau2:beta}
\begin{aligned}
\frac{\partial l_{\tilde\psi_{\beta,j}^2}}{\partial\tilde\psi_{\beta,j} }&=   \frac{\partial}{\partial{\tilde\psi_{\beta,j}}} \log\left(\frac{\partial{\psi_{\beta,j}}}{\partial{\tilde\psi_{\beta,j}}}\right) - \frac{1}{2} \sum_{i=1}^{n} \left(s_i^{-2}\frac{\partial}{\partial{\tilde\psi_{\beta,j}}}s_{i}^{2} -\frac{z_i^2\sigma_i^2}{(s_i^{2})^2} \frac{\partial}{\partial{\tilde\psi_{\beta,j}}}s_i^2 \right) +\betavec'B'\Sigma^{-1}\frac{\partial}{\partial{\tilde\psi_{\beta,j}}}S^{-1}\zvec \\
&\quad - \frac{1}{2} \betavec'\frac{\partial}{\partial{\tilde\psi_{\beta,j}}}P_{\beta}(\thetavec_{\beta})\betavec + \frac{1}{2} \frac{\partial}{\partial{\tilde\psi_{\beta,j}}} \log(\det(\Delta_{\beta}))\\
\frac{\partial^2 l_{\tilde\psi_{\beta,j}^2}}{\partial\tilde\psi_{\beta,j}^2 }&=  \frac{\partial^2}{\partial{\tilde\psi_{\beta,j}^2}} \log\left(\frac{\partial{\psi_{\beta,j}}}{\partial{\tilde\psi_{\beta,j}}}\right) - \frac{1}{2} \sum_{i=1}^{n} \frac{\partial}{\partial{\tilde\psi_{\beta,j}}}\left(s_i^{-2}\frac{\partial}{\partial{\tilde\psi_{\beta,j}}}s_{i}^{2} -\frac{z_i^2\sigma_i^2}{(s_i^{2})^2} \frac{\partial}{\partial{\tilde\psi_{\beta,j}}}s_i^2 \right) + \betavec'B'\Sigma^{-1}\frac{\partial^2}{\partial{\tilde\psi_{\beta,j}^2}}S^{-1}\zvec \\
&\quad - \frac{1}{2} \betavec'\frac{\partial^2}{\partial{\tilde\psi_{\beta,j}^2}}P_{\beta}(\thetavec_{\beta})\betavec + \frac{1}{2} \frac{\partial^2}{\partial{\tilde\psi_{\beta,j}^2}} \log(\det(\Delta_{\beta})), \\
\end{aligned}
\end{equation}
for which we computed the following derivatives:
% Let $\gamma_{\beta,0}=\frac{\tau_{\beta}^2}{(1-\tilde\psi_{\beta,1}^{2})(1-\tilde\psi_{\beta,2}^{2})}$ and $\tilde P_{\beta}$ such that $\gamma_{\beta,0} (1-\tilde\psi_{\beta,1}^{2})(1-\tilde\psi_{\beta,2}^{2})\tilde P_{\beta}^{-1}=P(\thetavec_{\beta})^{-1}$. Then,
\begin{equation*}
\begin{aligned}
%s_{i}^{2} &= \left(1 + \gamma_{\beta,0} (1-\tilde\psi_{\beta,1}^{2})(1-\tilde\psi_{\beta,2}^{2})\bvec_{i}'\tilde P_{\beta}^{-1}\bvec_{i}\right)^{-1} \mbox{ and hence }\\
\frac{\partial{s_{i}^{2}}}{\partial{\tilde\psi_{\beta,j}}} &= - (s_{i}^{2})^{2} \frac{\partial}{\partial{\tilde\psi_{\beta,j}}} \bvec_{i}'P_{\beta}(\thetavec_{\beta})^{-1}\bvec_{i} \\
%&= - \frac{\gamma_{0} (1-\tilde\psi_{\beta,2}^{2})}{(s_{i}^{2})^{2}} \left[-2\tilde\psi_{\beta,1} \bvec_{i}'\tilde P_{\beta}^{-1}\bvec_{i} + (1-\tilde\psi_{\beta,1}^{2}) \frac{\partial}{\partial{\tilde\psi_{\beta,1}}} \bvec_{i}'\tilde P_{\beta}^{-1}\bvec_{i}\right] \\
\frac{\partial^2{s_{i}^{2}}}{\partial{\tilde\psi_{\beta,j}^{2}}} &= 2(s_{i}^{2})^{3}\left( \frac{\partial}{\partial{\tilde\psi_{\beta,1}}}\bvec_{i}'P_{\beta}(\thetavec_{\beta})^{-1}\bvec_{i} \right)^2 - (s_{i}^{2})^{2} \frac{\partial^2}{\partial{\tilde\psi_{\beta,j}}^2}\bvec_{i}'P_{\beta}(\thetavec_{\beta})^{-1}\bvec_{i},
\end{aligned} 
\end{equation*}
and where
\begin{equation*}
\begin{aligned}
(\Delta_{\beta})_{ij}&=\begin{cases}1 & \mbox{if } i=j\mbox{ and } i\neq 1,2\\
-\psi_{\beta,2} & \mbox{if } i=j+2 \\
-\psi_{\beta,1}(1-\psi_{\beta,2}) & \mbox{if } i=j+1\mbox{ and } i\neq 1,2\\
\sqrt{1-\psi_{\beta,1}^2}\sqrt{1-\psi_{\beta,2}^2} & \mbox{if } i=j=1\\
(1-\psi_{\beta,2}^2) & \mbox{if } i=j=2\\
-\psi_{\beta,1}\sqrt{1-\psi_{\beta,2}^2} & \mbox{if } i=2\mbox{ and } j=1.\end{cases}
\end{aligned}
\end{equation*}
Then, we use that $\frac{\partial}{\partial{\tilde\psi_{\beta,j}}}(\Delta_{\beta})_{ij}=\frac{\partial\psi_{\beta,j}}{\partial{\tilde\psi_{\beta,j}}}\frac{\partial}{\partial{\psi_{\beta,j}}}(\Delta_{\beta})_{ij}$, with
\begin{equation*}
\begin{aligned}
\frac{\partial\psi_{\beta,j}}{\partial{\tilde\psi_{\beta,j}}}&=\frac{2\exp(\tilde\psi_{\beta,j})(1-\epsilon)}{(1+\exp(\tilde\psi_{\beta,j}))^2}\\
\frac{\partial}{\partial{\tilde\psi_{\beta,j}}}\log\left(\frac{\partial\psi_{\beta,j}}{\partial{\tilde\psi_{\beta,j}}}\right)&=1-\frac{2\exp(\tilde\psi_{\beta,j})}{1+\exp(\tilde\psi_{\beta,j})}.
\end{aligned}
\end{equation*}
Then, 
\begin{equation*}
\begin{aligned}
\frac{\partial}{\partial{\psi_{\beta,j}}}P(\psivec_{\beta}) &= \left(\frac{\partial}{\partial{\psi_{\beta,j}}}\Delta_{\beta}'\right)\Delta_{\beta}  + \Delta_{\beta}' \frac{\partial}{\partial{\psi_{\beta,j}}}\Delta_{\beta} \\
\frac{\partial}{\partial{\psi_{\beta,1}}}\delta_{11} 
&= - \frac{\psi_{\beta,1}^{2} \sqrt{(1-\psi_{\beta,2}^{2})}}{\sqrt{1-\psi_{\beta,1}^{2}}} \\
\frac{\partial}{\partial{\psi_{\beta,2}}}\delta_{11} &=  - \frac{\psi_{\beta,2}^{2} \sqrt{1-\psi_{\beta,1}^{2}}}{\sqrt{1-\psi_{\beta,2}^{2}}} \\
\frac{\partial}{\partial{\psi_{\beta,1}}}\delta_{21} &=- \sqrt{1-\psi_{\beta,2}^{2}} \hspace{1cm} \frac{\partial}{\partial{\psi_{\beta,2}}}\delta_{21} = \frac{\psi_{\beta,1}\psi_{\beta,2}}{\sqrt{1-\psi_{\beta,2}^{2}}}\\
\frac{\partial}{\partial{\psi_{\beta,1}}}\delta_{22} &= 0 \hspace{2.95cm} \frac{\partial}{\partial{\psi_{\beta,2}}}\delta_{22} =  - \frac{\psi_{\beta,2}}{\sqrt{1-\psi_{\beta,2}^{2}}} \\
\frac{\partial}{\partial{\psi_{\beta,1}}}\delta_{31} &= 0 \hspace{2.95cm} \frac{\partial}{\partial{\psi_{\beta,2}}}\delta_{31} = -1 \\
\frac{\partial}{\partial{\psi_{\beta,1}}}\delta_{32} &= -(1-\psi_{\beta,2}) \hspace{1.1cm} \frac{\partial}{\partial{\psi_{\beta,2}}}\delta_{32} = \psi_{\beta,1}, 
\end{aligned}
\end{equation*}
\begin{equation*}
\begin{aligned}
\frac{\partial}{\partial{\psi_{\beta,1}}}\log(\det(P(\psivec_\beta))^{\frac{1}{2}}) &= -\frac{1}{1-\psi_{\beta,1}^{2}}  \\\frac{\partial}{\partial{\psi_{\beta,2}}}\log(\det(P(\psivec_\beta))^{\frac{1}{2}})&=- \frac{2\psi_{\beta,2}^{2}}{(1-\psi_{\beta,2}^{2})^{2}}.
\end{aligned}
\end{equation*}

\vspace{-10pt}
\subsubsection{HMC for $\alphavec$}\label{subsec:hmc}
\vspace{-5pt}

See the main paper for further details. Here we just derive the analytical expression of derivatives of $s_i^2$, $\kappa_{1,i}$ and $\kappa_{2,i}$ with respect to $\etavec_{\alpha}$. These are
\begin{equation*}
\begin{aligned}
\frac{\partial{s_{i}^{2}}}{\partial{\etavec_{\alpha}}} &= \frac{\partial{}}{\partial{\etavec_{\alpha}}} (\exp(\etavec_{i, \alpha}) + b_{i}'P_{\beta}(\thetavec_{\beta})^{-1}b_{i})^{-1} =- \exp(\etavec_{i, \alpha})(s_{i}^{2})^{2}\\
\frac{\partial{\kappa_{1,i}}}{\partial{\etavec_{\alpha}}} &= -\kappa_{1,i}+1,\qquad\quad\frac{\partial{\kappa_{2,i}}}{\partial{\etavec_{\alpha}}} = -\kappa_{2,i}+\frac{1}{2} (s_i^2)^{1/2}.
\end{aligned}
\end{equation*}

\vspace{-10pt}
\subsubsection{MH step for  $\thetavec_{\alpha}$}\label{blubb2}
\vspace{-5pt}
Generation of the components of $\thetavec_{\alphavec}$ are done with the equivalent approach as we described already for $\thetavec_{\beta}$. The conditional log posterior 
\begin{equation}\label{eq:logFC:tau2:alpha}
  \begin{aligned}
	  &l_{{\tilde\tau_{\alpha}^2}}\equiv \log(p({\tilde\tau_{\alpha}^2}|\xvec,\wvec,\yvec,\lbrace\thetavec\setminus\tau_{\alpha}^2\rbrace))\propto -\frac{{\tilde\tau_{\alpha}^2}}{2}\left(p_2-1\right)-\frac{1}{2\exp({\tilde\tau_{\alpha}^2})}\alphavec'P(\psivec_{\alpha})\alphavec-\left(\frac{\exp({\tilde\tau_{\alpha}^2})}{b_{\tau_{\alpha}^2}}\right)^{\tfrac{1}{2}}.
  \end{aligned}
\end{equation}
Then, we have that
\begin{equation}\label{eq:logFC:psi:alpha}
  \begin{aligned}
	  l_{\tilde\psi_{\alpha,j}}\equiv \log(p(\tilde\psi_{\alpha,j}|\xvec,\wvec,\yvec,\lbrace\thetavec\setminus\psi_{\alpha,j}\rbrace))&\propto \log\left(\frac{\partial\psi_{\alpha,j}}{\partial\tilde\psi_{\alpha,j}}\right)
-\frac{\alphavec'P(\psivec_{\alpha})\alphavec}{2\tau_{\alpha}^2}.
  \end{aligned}
\end{equation}
First and second derivatives of~\eqref{eq:logFC:tau2:alpha} with respect to $\tilde\tau^2_{\alpha}$ are
\begin{equation*}\label{eq:tau2:alpha}
\begin{aligned}
\frac{\partial l_{\tilde\tau_{\alpha}^2}}{\partial\tilde\tau_{\alpha}^2 }&=    -\frac{1}{2}(p_{\alpha}-1) - \frac{1}{2} \left(\frac{\exp(\tilde\tau_{\alpha}^2)}{b_{0}}\right)^{\frac{1}{2}} + \frac{1}{2 \exp(\nu)} \alphavec'P_{\alpha}(\thetavec_{\alpha})\alphavec \\
\frac{\partial^2 l_{\tilde\tau_{\alpha}^2}}{\partial(\tilde\tau_{\alpha}^2)^2 }&=    -\frac{1}{4}\left(\frac{\exp(\tilde\tau_{\alpha}^2)}{b_{0}}\right)^{\frac{1}{2}} - \frac{1}{2\exp(\tilde\tau_{\alpha}^2)} \alphavec'P_{\alpha}(\thetavec_{\alpha})\alphavec\\
\end{aligned}
\end{equation*}
Furthermore, derivatives of~\eqref{eq:logFC:psi:alpha} with respect to $\tilde\psi_{\alpha,1}$ are
\begin{equation*}
\begin{aligned}
\frac{\partial{l_{\tilde\psi_{\alpha,j}}}}{\partial{\tilde\psi_{\alpha,j}}} &\propto \frac{\partial}{\partial{\tilde\psi_{\alpha,j}}} \log(\det(\Delta_{\alpha})) - \alphavec' \frac{1}{2\tau_{\alpha}^{2}} \frac{\partial}{\partial{\tilde\psi_{\alpha,j}}}P(\tilde\psi_{\alpha,j})\alphavec + \frac{\partial}{\partial{\tilde\psi_{\alpha,j}}}\left(\log\left(\frac{\partial{\psi_{\alpha,j}}}{\partial{\tilde\psi_{\alpha,j}}}\right)\right) \\
\frac{\partial^2{l_{\tilde\psi_{\alpha,j}}}}{\partial{\tilde\psi_{\alpha,j}^2}} &\propto \frac{\partial^2}{\partial{\tilde\psi_{\alpha,j}^2}} \log(\det(\Delta_{\alpha})) - \alphavec' \frac{1}{2\tau_{\alpha}^{2}} \frac{\partial^2}{\partial{\tilde\psi_{\alpha,j}^2}}P(\psivec_{\alpha})\alphavec + \frac{\partial^2}{\partial{\tilde\psi_{\alpha,j}^2}}\left(\log\left(\frac{\partial{\psi_{\alpha,j}}}{\partial{\psivec_{\alpha}}}\right)\right), \\
\end{aligned}
\end{equation*}
and where the derivatives of $\log(\det(\Delta_{\alpha}))$ and $P_{\alpha}(\thetavec_{\alpha})$ with respect to $\tilde\psi_{\alpha,j}$ follow the formulas given in Subsection~\ref{blubb}.

\vspace{-10pt}
\subsubsection{Exact Sampler}
\vspace{-5pt}

\begin{algorithm}
\caption{Exact Sampler for the HPSC}\label{exactsampler}
Given $\thetavec^{(0)},\delta,\iota,M,M_{\mbox{\scriptsize adapt}}$:
\begin{algorithmic}[1]
\State Set $\gamma=0.05$, $t_0=10$, $\kappa=0.75$ as in~\citet{HofGel2014}.
\State Set $\epsilon_0$ using Algorithm~4 of~\citet{HofGel2014} and $\bar\epsilon_0=1$.
\For{$j=1,\ldots,M$}
\State Sample $\betavec^{(m)}\sim\ND(\muvec,\Sigma_{\beta})$.
\State Perform MH step for $\tau_{\beta}^2$ to get $(\tau_{\beta}^2)^{(m)}$.
\State Perform MH step for $\psi_{\beta,1}$ to get $\psi_{\beta,1}^{(m)}$.
\State Perform MH step for $\psi_{\beta,2}$ to get $\psi_{\beta,2}^{(m)}$.
\State Perform one step of Algorithm~1 of the main paper to get $\alphavec^{(m)}$.
\State Perform MH step for $\tau_{\alpha}^2$ to get $(\tau_{\alpha}^2)^{(m)}$.
\State Perform MH step for $\psi_{\alpha,1}$ to get $\psi_{\alpha,1}^{(m)}$.
\State Perform MH step for $\psi_{\alpha,2}$ to get $\psi_{\alpha,2}^{(m)}$.
\EndFor
\end{algorithmic}
\end{algorithm}

\vspace{-12pt}
\subsection{Variational Inference}\label{app:subsec:vb}
\vspace{-7pt}
To implement Algorithm~2 for variational inference, the following derivatives need 
to be evaluated, which we do analytically.
\begin{equation*}\begin{aligned}
\nabla_{\mu}\mathcal{L}(\lambdavec)&=\dsE_{p_{\zeta}}\left(\nabla_{\varthetavec}\log(h(\muvec+\Psi\xivec+\dvec\circ\deltavec))-(\Psi\Psi'+\Delta^2)^{-1}(\Psi\xivec+\dvec\circ\deltavec)\right),\\
\nabla_{\mbox{vech}(\Psi)}\mathcal{L}(\lambdavec)&=\dsE_{p_{\zeta}}\left(\nabla_{\varthetavec}\log(h(\muvec+\Psi\xivec+\dvec\circ\deltavec))-(\Psi\Psi'+\Delta^2)^{-1}(\Psi\xivec+\dvec\circ\deltavec)\xivec'\right),\\
\nabla_{d}\mathcal{L}(\lambdavec)&=\dsE_{p_{\zeta}}\left(\nabla_{\varthetavec}\log(h(\muvec+\Psi\xivec+\dvec\circ\deltavec))-(\Psi\Psi'+\Delta^2)^{-1}(\Psi\xivec+\dvec\circ\deltavec)\deltavec'\right).
\end{aligned}\end{equation*}
The inverse of $(\Psi\Psi'+\Delta^2)$ is computed efficiently using the
Woodbury formula. The
derivatives of $\log(h(\varthetavec))=l_{\varthetavec}$ with respect to the transformations of $\lbrace \thetavec_{\beta},\thetavec_{\alpha}\rbrace$
are given above in Sections~\ref{blubb} and~\ref{blubb2}, respectively, the one for $\alphavec$ is given in the main paper, Section 3.2.  Finally, the gradient of $\betavec$ is
\[
\nabla_{\beta}l_{\varthetavec}=B'\Sigma^{-1}S^{-1}\zvec-(B'\Sigma^{-1}B+P_{\beta}(\thetavec_{\beta}))\betavec.
\]

%\begin{algorithm}
%\caption{Stochastic gradient ascent for variational approximation with factor covariance structure.}\label{algo3}
%Given $\lambdavec^{(0)}=\lbrace\muvec^{(0)},\Psi^{(0)},\dvec^{(0)}\rbrace$, $t=0$:
%\begin{algorithmic}[1]
%\While{Stopping rule is not satisfied}
%\State Generate $(\xivec',\deltavec')'\sim\ND(\nullvec,I)$.
%\State Construct unbiased estimates of $\reallywidehat{\nabla_{\mu}\mathcal{L}(\lambdavec^{(t)})}$, $\reallywidehat{\nabla_{\Psi}\mathcal{L}(\lambdavec^{(t)})}$ and $\reallywidehat{\nabla_{d}\mathcal{L}(\lambdavec^{(t)})}$ using the single sample $(\xivec',\deltavec')'$.
%\State Compute the adaptive learning rate $\rhovec^{(t)}=\lbrace\rhovec_{\mu}^{(t)},\rhovec_{\Psi}^{(t)},\rhovec_{\delta}^{(t)}\rbrace$ using ADADELTA.
%\State Set $\muvec^{(t+1)}=\muvec^{(t)}+\rhovec_{\mu}^{(t)}\circ\reallywidehat{\nabla_{\delta}\mathcal{L}(\lambdavec^{(t)})}$.
%\State Set $\Psi^{(t+1)}=\Psi^{(t)}+\rhovec_{\mu}^{(t)}\circ\reallywidehat{\nabla_{\delta}\mathcal{L}(\lambdavec^{(t)})}$ and set $\Psi^{(t+1)}_{ij}=0$ for $i\geq j$.
%\State Set $\deltavec^{(t+1)}=\deltavec^{(t)}+\rhovec_{\delta}^{(t)}\circ\reallywidehat{\nabla_{\delta}\mathcal{L}(\lambdavec^{(t)})}$.
%\State Set Set $\lambdavec^{(t+1)} \gets \lbrace\muvec^{(t+1)},\Psi^{(t+1)},\deltavec^{(t+1)}\rbrace$ and $t\gets t+1$.
%\EndWhile
%\end{algorithmic}
%\end{algorithm}

\newpage
\vspace{-15pt}
\section{Additional Figures and Tables in Section 4.1}
\vspace{-10pt}

\begin{table}[htbp]
	\centering\renewcommand\arraystretch{1.5}
	\begin{tabular}{ccccc}
		\hline\hline
		&\multicolumn{4}{c}{Regression Copula  / Estimation Method} \\
		Dataset & PSC/VB & PSC/MCMC & HPSC/VB & HPSC/HMC   \\\hline
		Geyser & 3.8 & 12.1 & 3.9 & 16.87 \\
		Rents &   11.1  & 37.3   & 10.1  & 43.5  \\
		Amazon   &   86.5 &  171.8 & 81.2   &  167.3 \\     
		Incomes  &   118.7  & 188.0  &  101.7   & 208.9 \\\hline\hline
	\end{tabular}
	\caption{Computing times (in seconds) to undertake 1,000 sweeps or steps
		for all methods, three regression copulas and the four datasets. 
		The regression copulas are PSC and HPSC, plus the additive HPSC. 
		The Bayesian posteriors are computed either exactly using MCMC or HMC,
		or approximately using VB. All computations were done in Matlab and exploited the many replicated value of the covariates where possible.}\label{tab:times_fast}
\end{table}

\begin{table}[h!]
	\centering\renewcommand\arraystretch{1.5}
	\begin{tabular}{c|cccc}
		\hline\hline
		& Geysesr & Rents   & Amazon &  Incomes \\\hline
		predIF$_{\mbox{\scriptsize MCMC}}$ & [0.84 1.06 1.10] &   [0.92 1.03 1.21]   & [1.04 1.05 1.05] & [0.92 1.02 1.20]\\
		predIF$_{\mbox{\scriptsize HMC}}$  &  [0.82 1.13 2.71] &  [0.68 1.21 2.43]  & [0.90 1.19 3.45] & [0.91 1.02 1.32]\\\hline\hline
	\end{tabular}
	\caption{Inefficiency factors predIF $1+2\sum_{m=1}^{\infty}\rho_{m}$ of the exact methods MCMC and HMC, where $\rho_m$ are the autocorrelations of the predictions $\dsE(Y|x)$. Computations have been conducted in the R-package \texttt{coda}. Reported are min/mean/max for each data set (columnwise) and for MCMC (first row) and HMC (second row) of 1,000 samples obtained from 50,000 samples and a thinning parameter of 50.}\label{tab:predIF}
\end{table}

%\begin{figure}[t]
%{\includegraphics[width=0.99\textwidth,angle=0]{../Figures/params_mean.eps}}%\hspace{1cm}
%\caption{Parameter estimates of $\betavec$ (blue) and $\alphavec$ (if $\neq\nullvec$) of VB against HMC/MCMC. The datasets are (a,b) \texttt{Rents}, (c,d) \texttt{Nigeria}, (e,f) \texttt{Amazon},
%and (g,h) \texttt{Incomes}. HPSC is shown in panels (a,c,e,g) and PSC in (b,d,f,h).}\label{fig:params}
%\end{figure}

\begin{figure}[h!]
	\includegraphics[width=0.99\textwidth,angle=0]{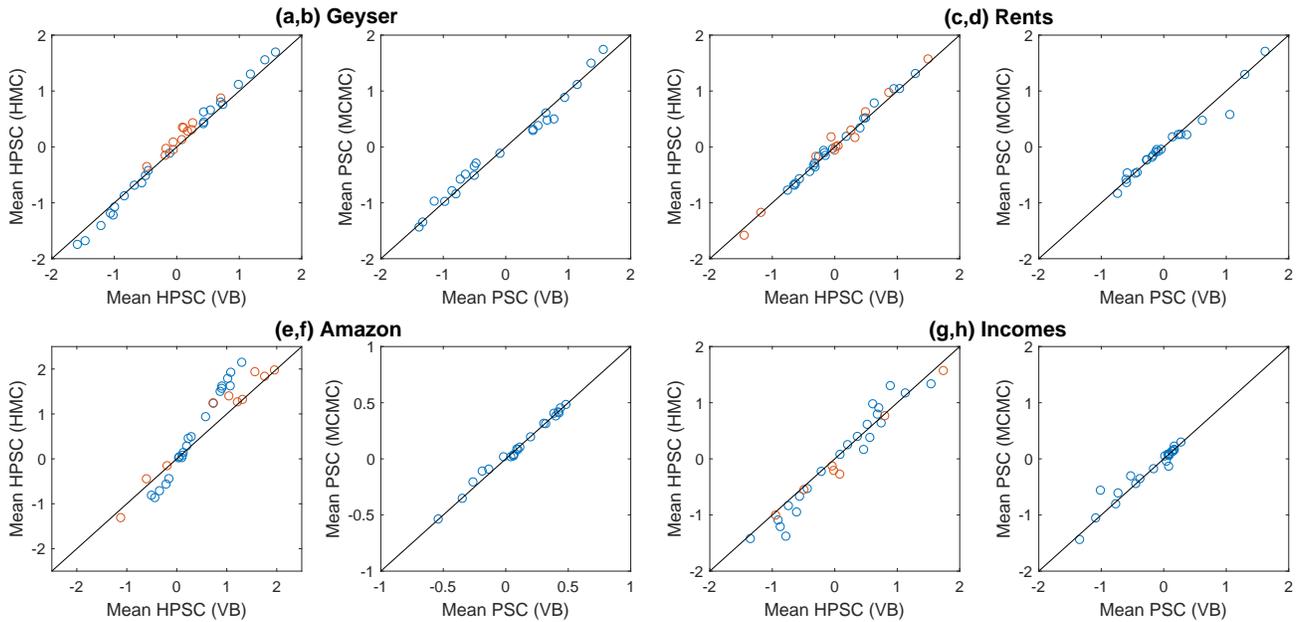}
	\caption{Parameter estimates of $\betavec$ (blue) and $\alphavec$ (if $\neq\nullvec$, red) of VB against HMC/MCMC. The datasets are (a,b)~Geyser, (c,d)~Rents, (e,f)~Amazon, and (g,h)~Incomes. HPSC is shown in panels~(a,c,e,g) and PSC in panels~(b,d,f,h). The close alignment shows the 
		very high degree of accuracy of the location of the VA $q_\lambda$.}
	\label{fig:params}
\end{figure}

\begin{figure}[h!]
{\includegraphics[width=0.99\textwidth,angle=0]{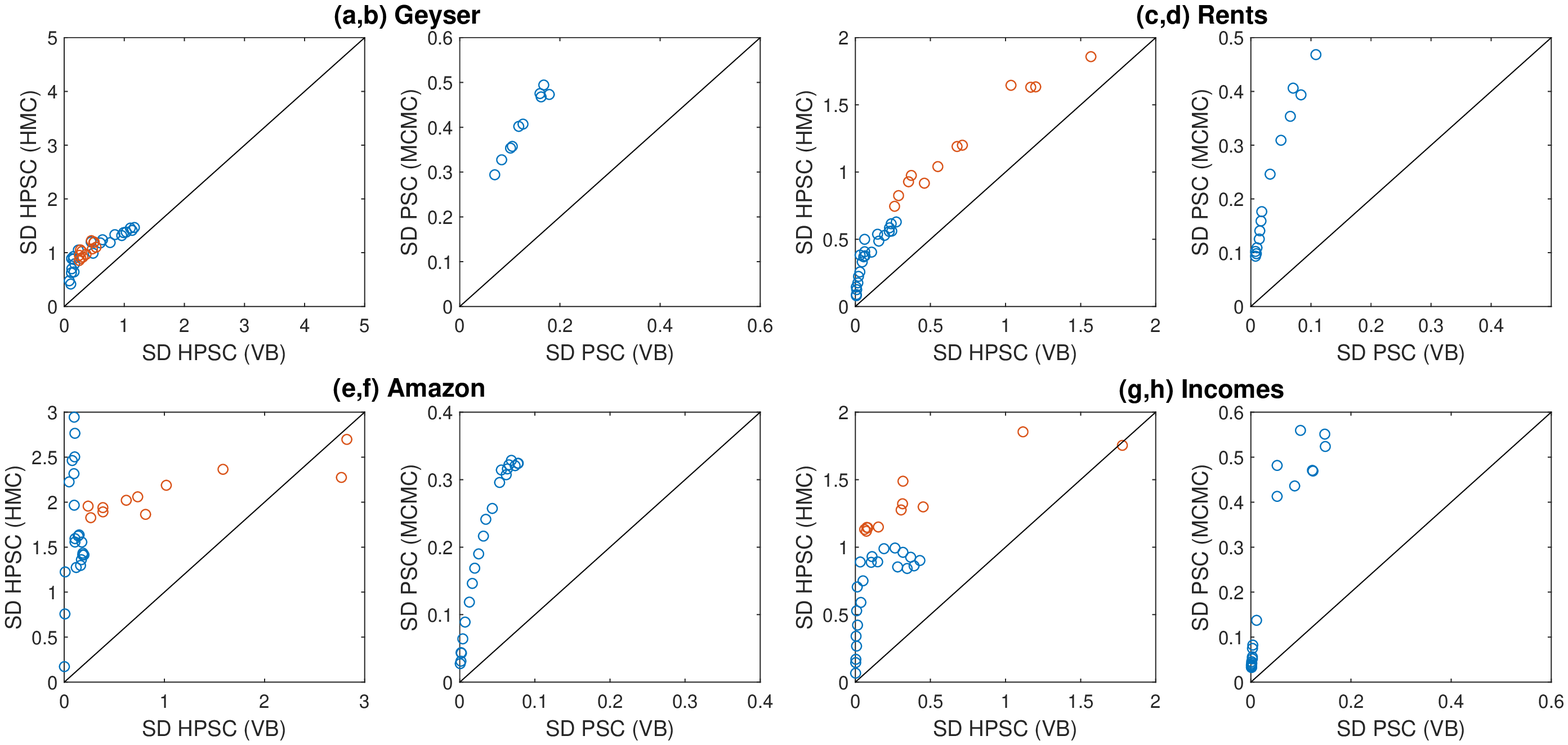}}%\hspace{1cm}
\caption{Standard deviations  of $\betavec$ (blue) and $\alphavec$ (if $\neq\nullvec$, red) of VB against HMC/MCMC. The datasets are (a,b) Geyser, (c,d) Rents, (e,f) Amazon,
and (g,h) Incomes. HPSC is shown in panels (a,c,e,g) and PSC in (b,d,f,h).
Under-estimation of the posterior standard deviation of parameter values is common when 
using Gaussian variational approximations, although the posterior standard deviation 
of functionals can still be reasonably accurate; see the examples in~\cite{OngNotSmi2018}.}\label{fig:params:sd}
\end{figure}

\begin{figure}[t]
	\centering{\includegraphics[width=0.93\textwidth,angle=0]{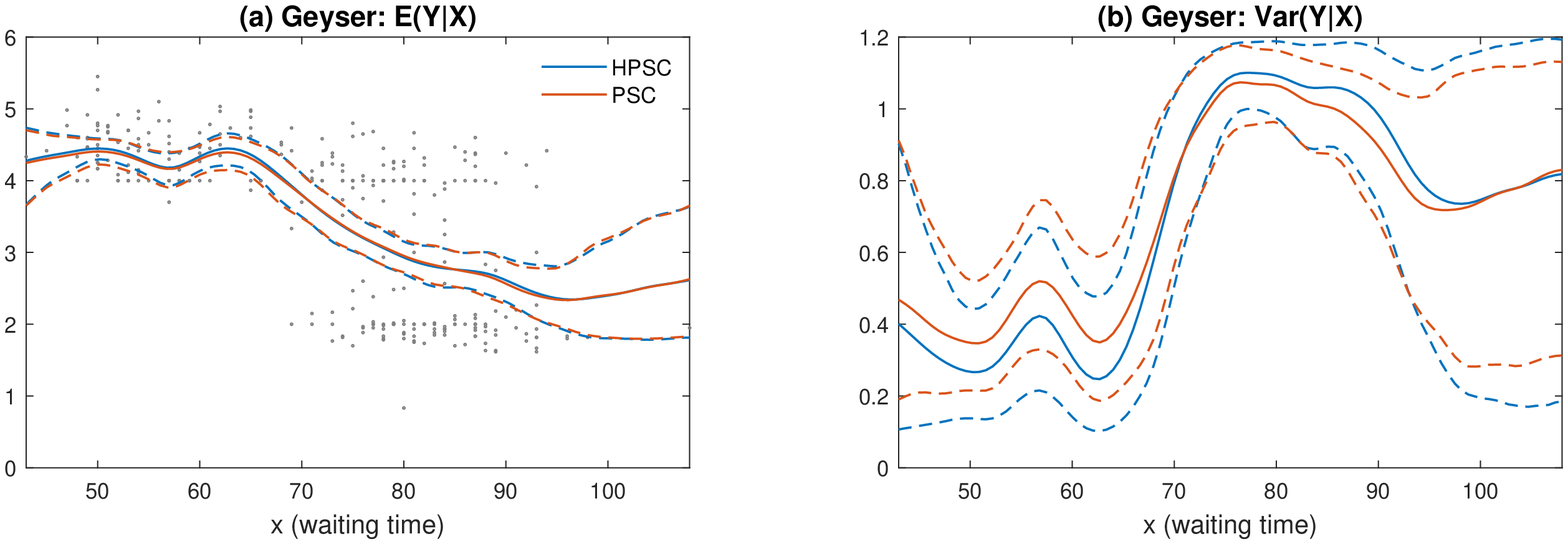}}\\\vspace{0.075cm}
	
	\centering{\includegraphics[width=0.93\textwidth,angle=0]{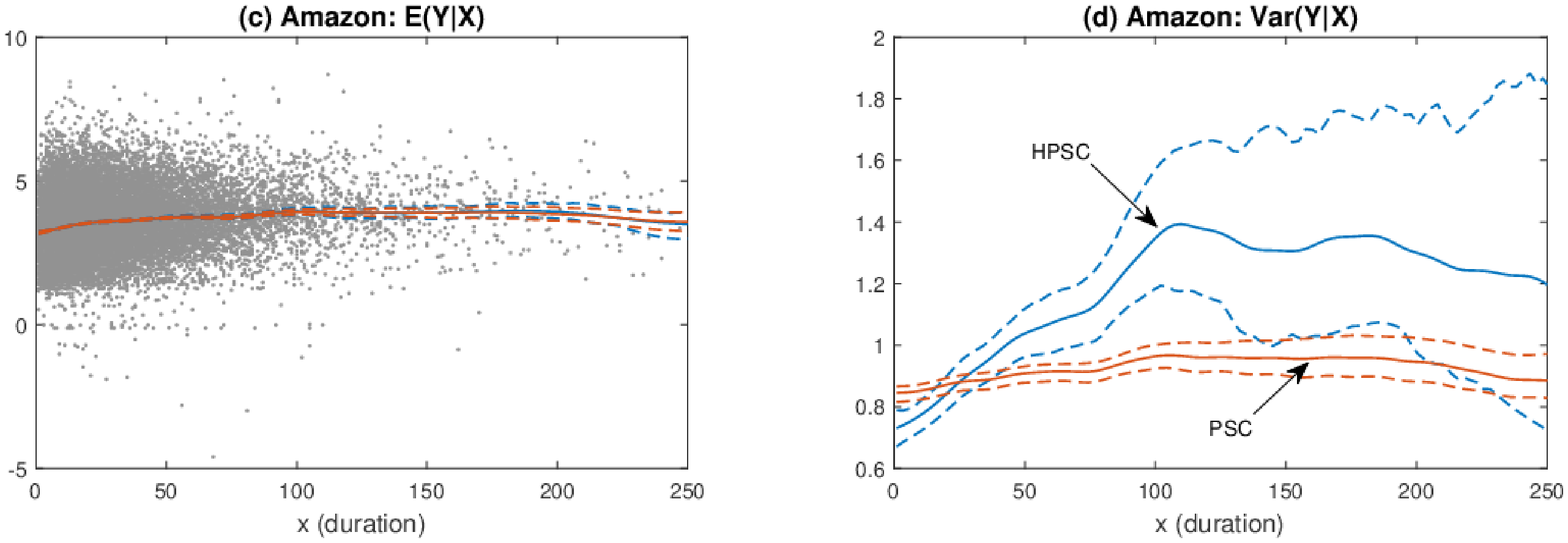}}\\%\vspace{0.1cm}
	\centering{\includegraphics[width=0.93\textwidth,angle=0]{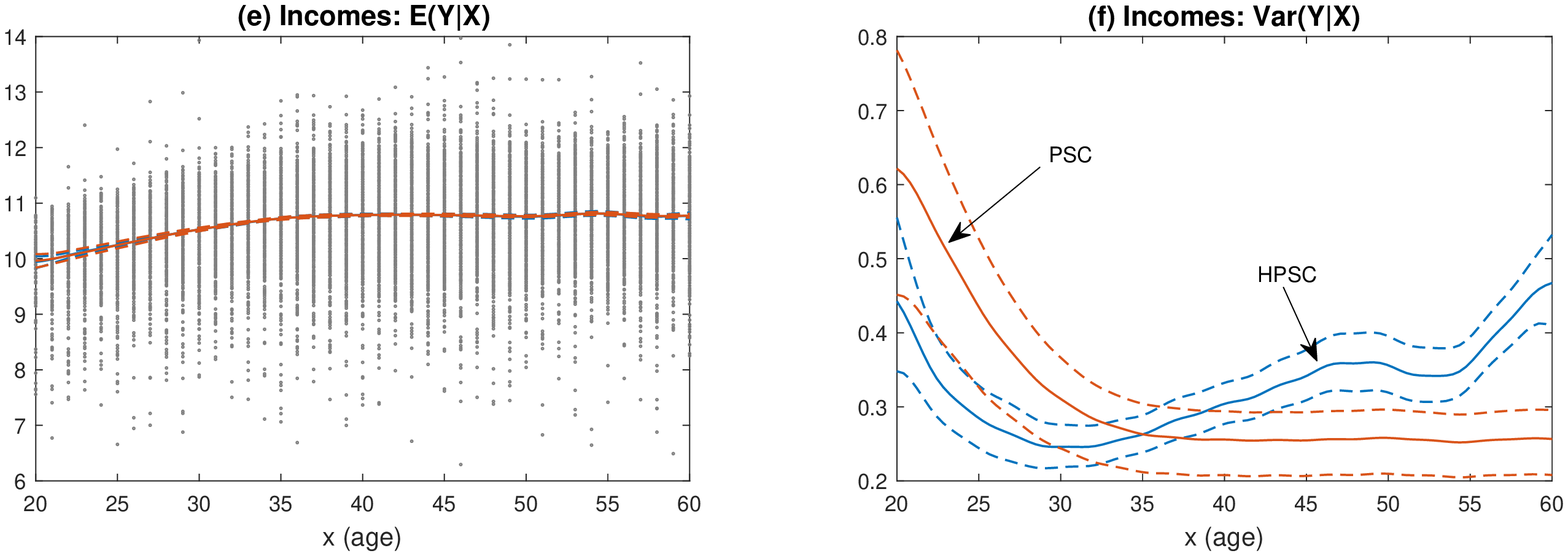}}
	\caption{Estimates of the regression function $f$ 
		(left-hand side) and variance function $v$ (right-hand side) from the HPSC model,
		compared to those from the PSC model.
		The posterior means of each function are given as a solid 
		lines, and 95\% posterior intervals by dashed lines. Scatterplots of the data are included in the left-hand panels.
		All function estimates employ the same KDE estimator for the margins $F_Y$, and the datasets
		are (a--b) Geyser, (c--d) Amazon, (e--f) Incomes.}\label{fig:functions2}
\end{figure}

\begin{figure}[t]
	\centering{\includegraphics[width=0.815\textwidth,angle=0]{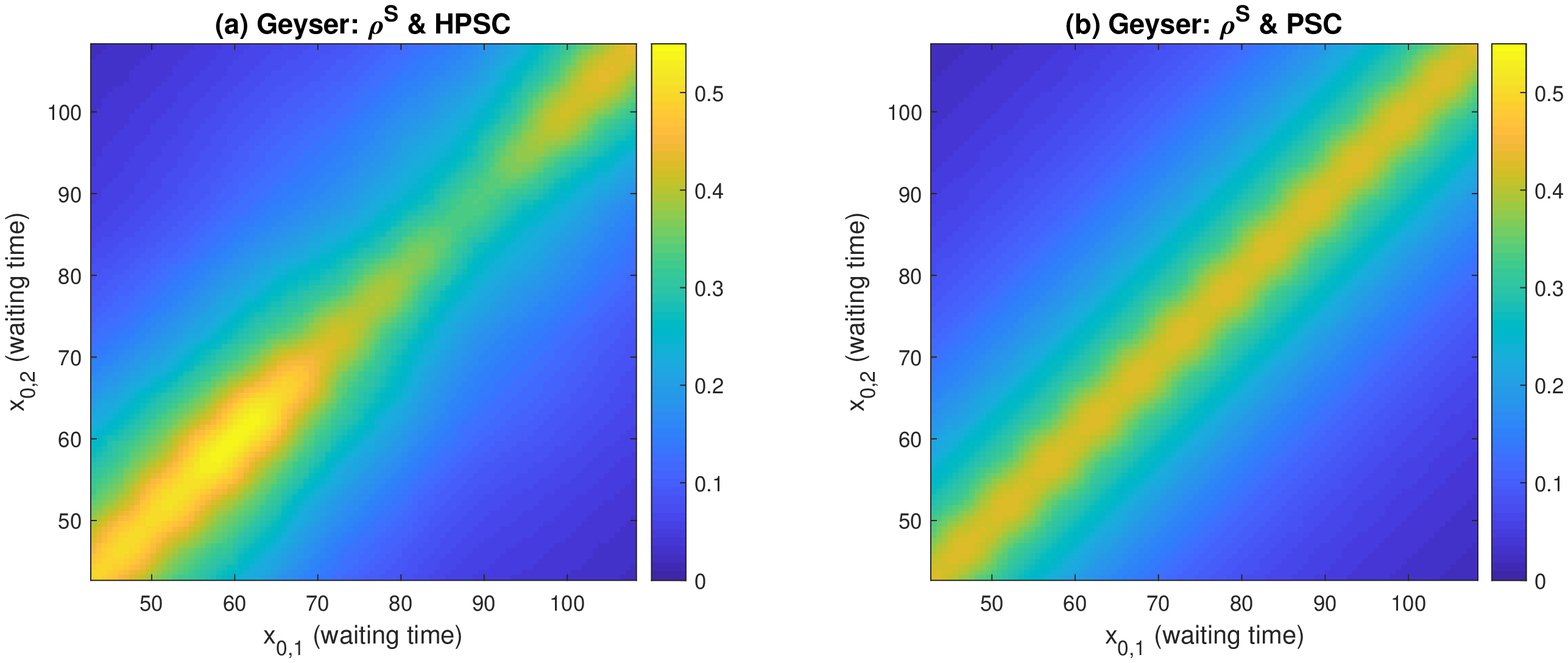}}\\
\centering{\includegraphics[width=0.825\textwidth,angle=0]{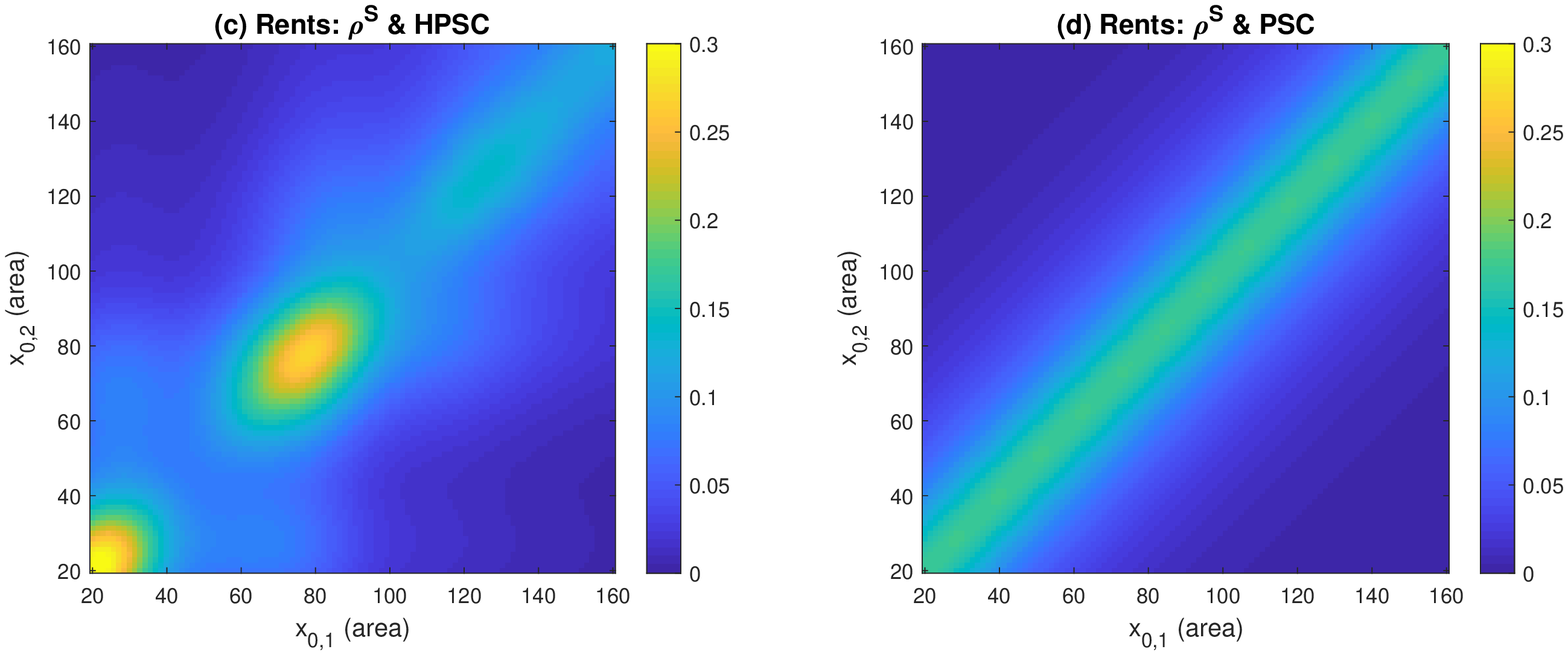}}\\
	%\hspace{0.6cm}
	\centering{\includegraphics[width=0.825\textwidth,angle=0]{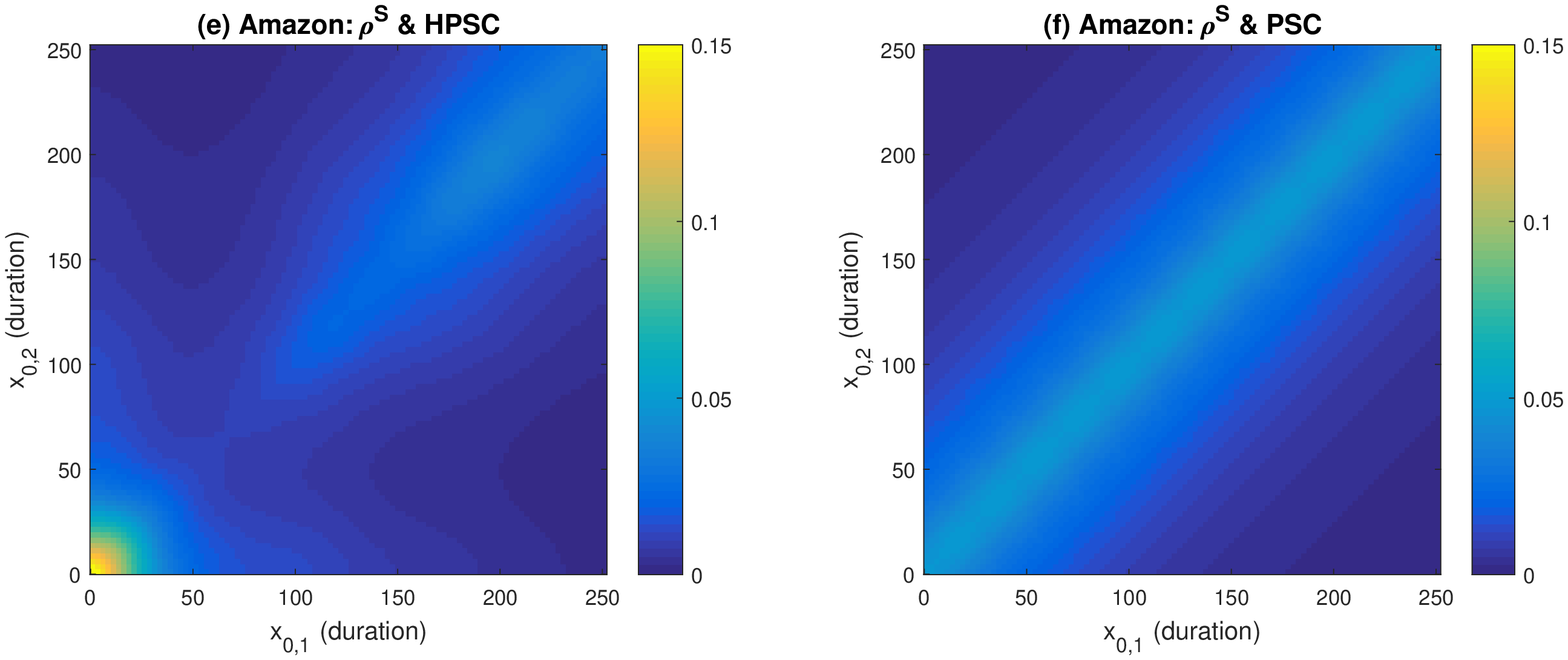}}\\%\vspace{0.2cm}
	\centering{\includegraphics[width=0.825\textwidth,angle=0]{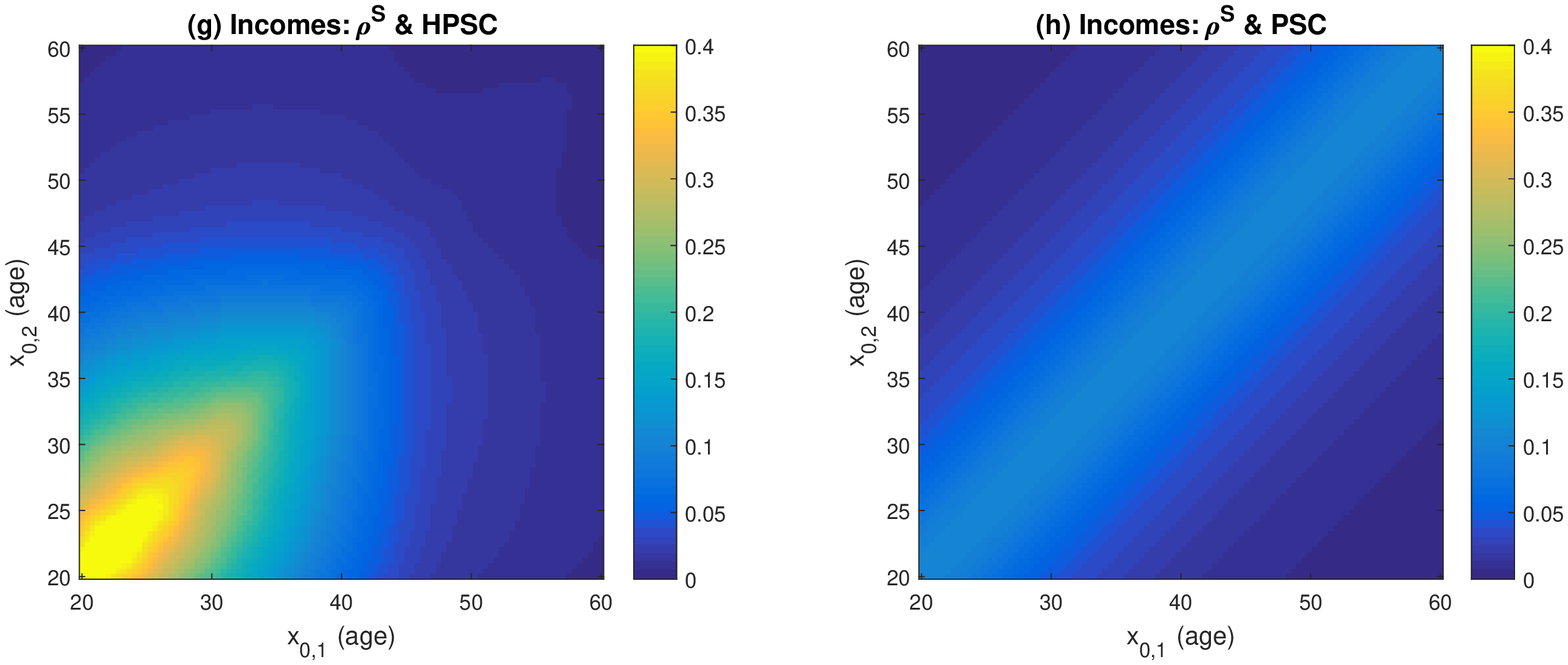}}
	%\centering{\includegraphics[width=0.975\textwidth,angle=0]{Figures/dep_amazon.eps}}\hspace{0.3cm}
	%\centering{\includegraphics[width=0.975\textwidth,angle=0]{Figures/dep_incomes.eps}}
	\caption{Estimates of Spearman's rho $\hat\rho^S(x_{0,1},x_{0,2})$. The lefthand panels give 
		values for the HPSC, and the righthand panels for the PSC.
		The datasets are (a,b)~Rents, (c,d)~Nigeria, (e,f)~Amazon, and (g,h)~Incomes.}\label{fig:dep:spear}
\end{figure}

\begin{figure}[t]
\centering{\includegraphics[width=0.815\textwidth,angle=0]{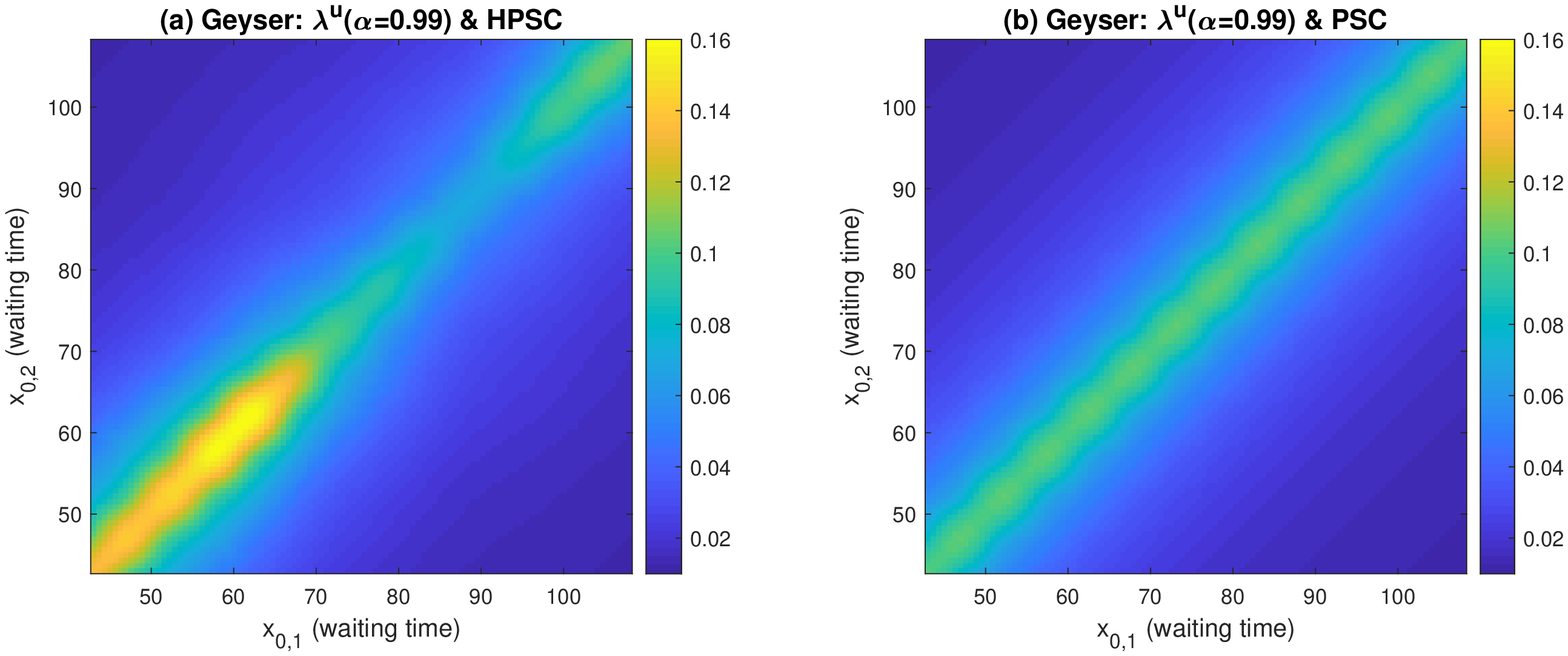}}\\
\centering{\includegraphics[width=0.825\textwidth,angle=0]{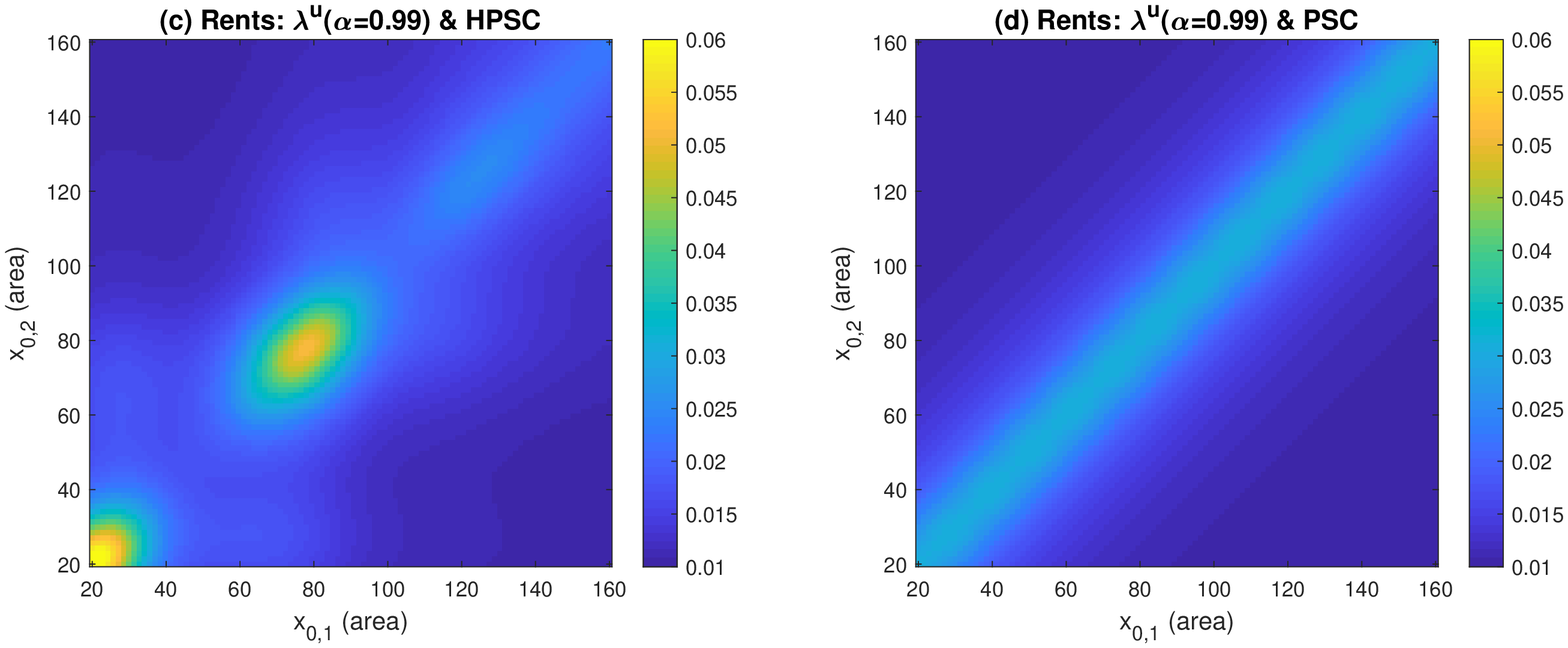}}\\%\hspace{0.3cm}
%\centering{\includegraphics[width=0.825\textwidth,angle=0]{../Figures/dep_nigeria_R1.eps}}\\
%\hspace{0.6cm}
\centering{\includegraphics[width=0.825\textwidth,angle=0]{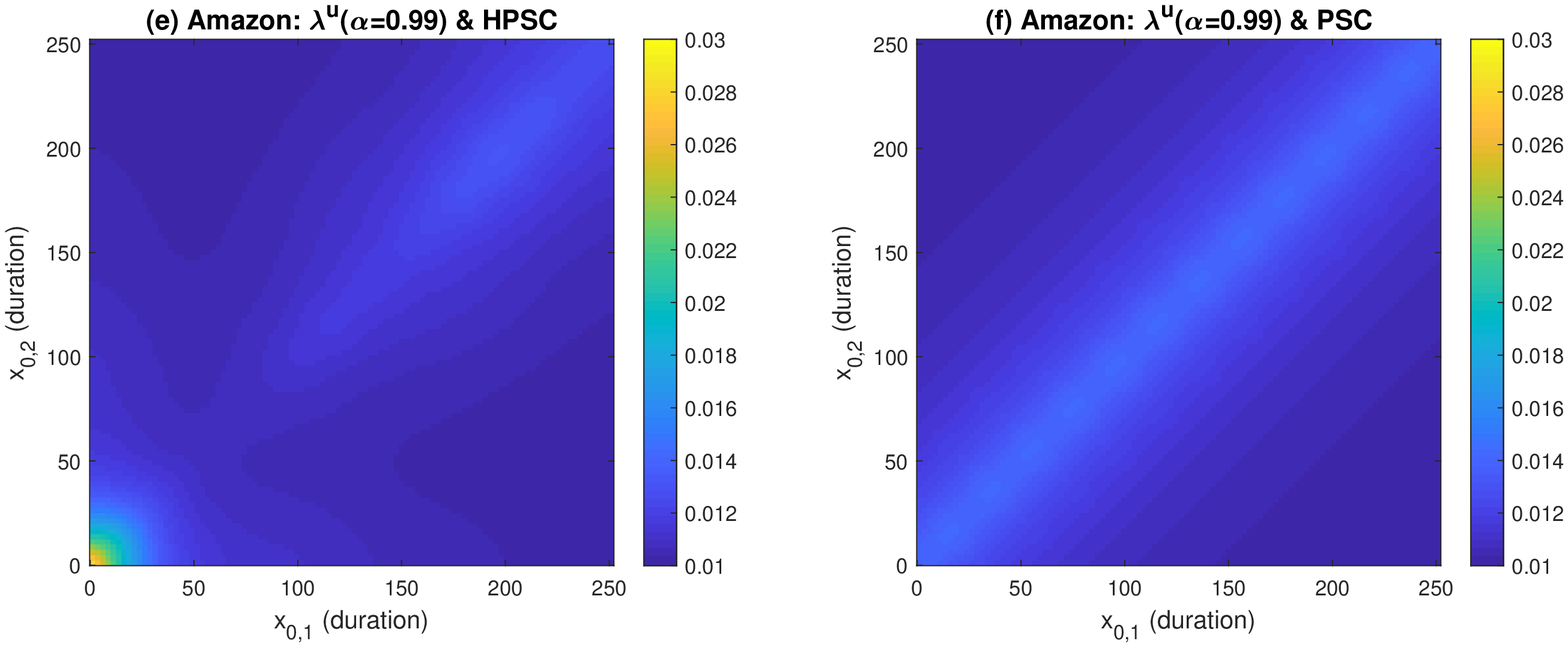}}\\%\vspace{0.2cm}
\centering{\includegraphics[width=0.825\textwidth,angle=0]{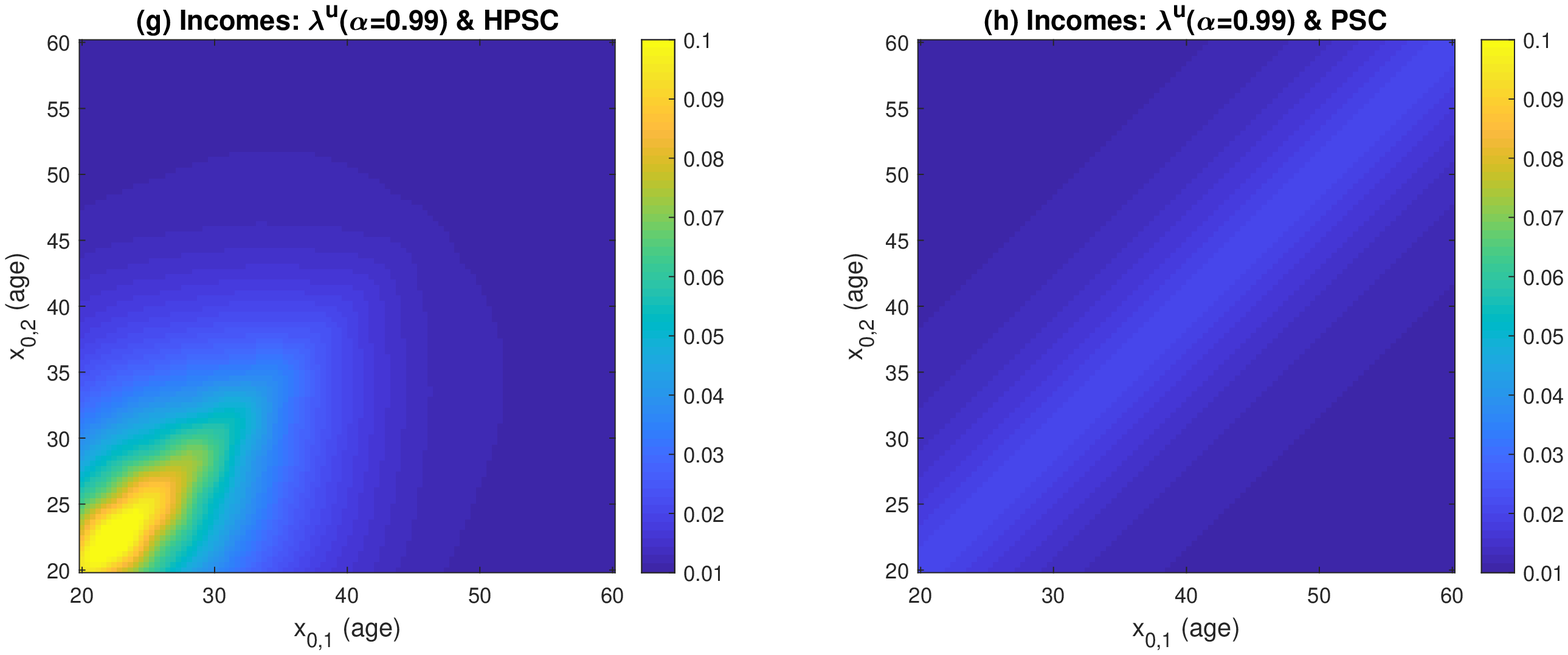}}\\
%\centering{\includegraphics[width=0.975\textwidth,angle=0]{../Figures/dep_amazon.eps}}\hspace{0.3cm}
%\centering{\includegraphics[width=0.975\textwidth,angle=0]{../Figures/dep_incomes.eps}}
\caption{Estimates of the upper quantile dependence $\lambda^U(0.99|x_{0,1},x_{0,2})$. The lefthand panels give values for the
HPSC, and the righthand panels for the PSC. The datasets are (a,b) Geyser, (c,d) Rents, (e,f) Amazon,
and (g,h) Incomes.}\label{fig:dep:qu}
\end{figure}

\begin{figure}[t]
\centering{\includegraphics[width=0.815\textwidth,angle=0]{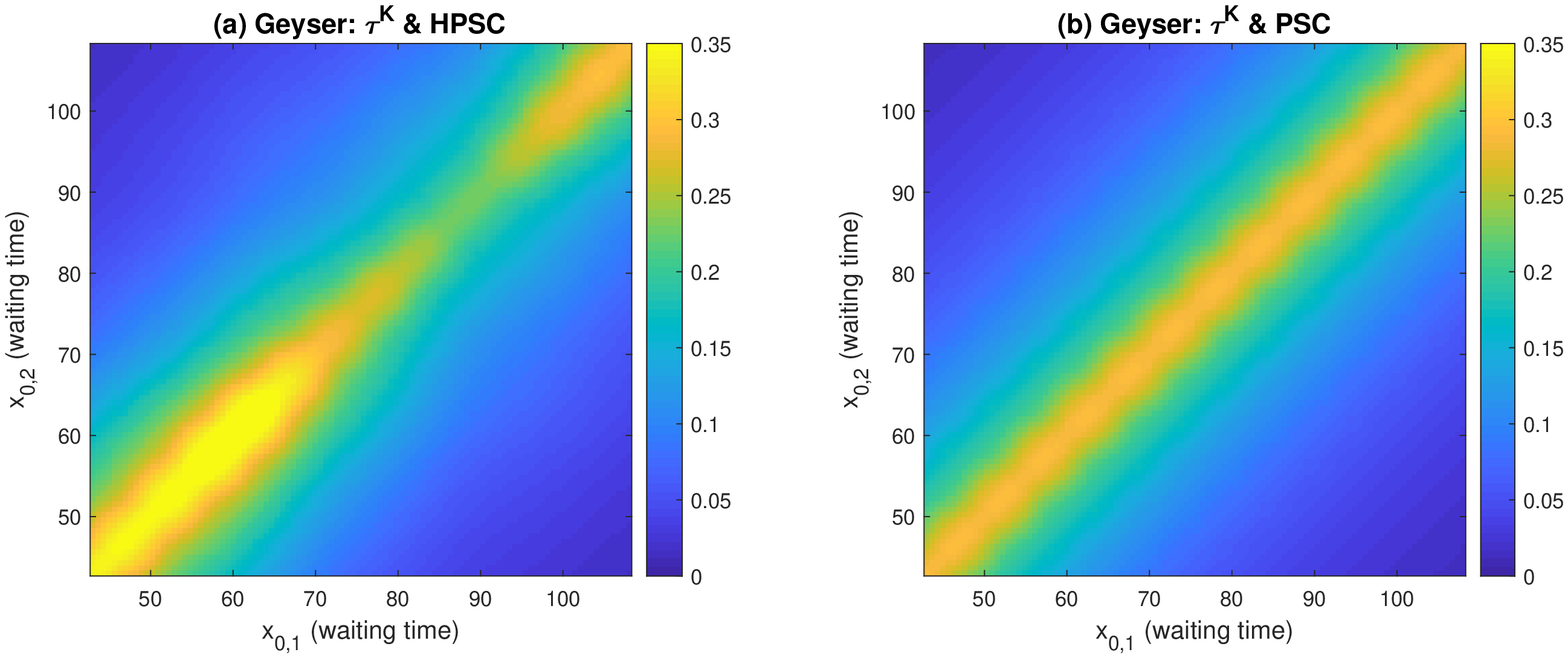}}\\
\centering{\includegraphics[width=0.825\textwidth,angle=0]{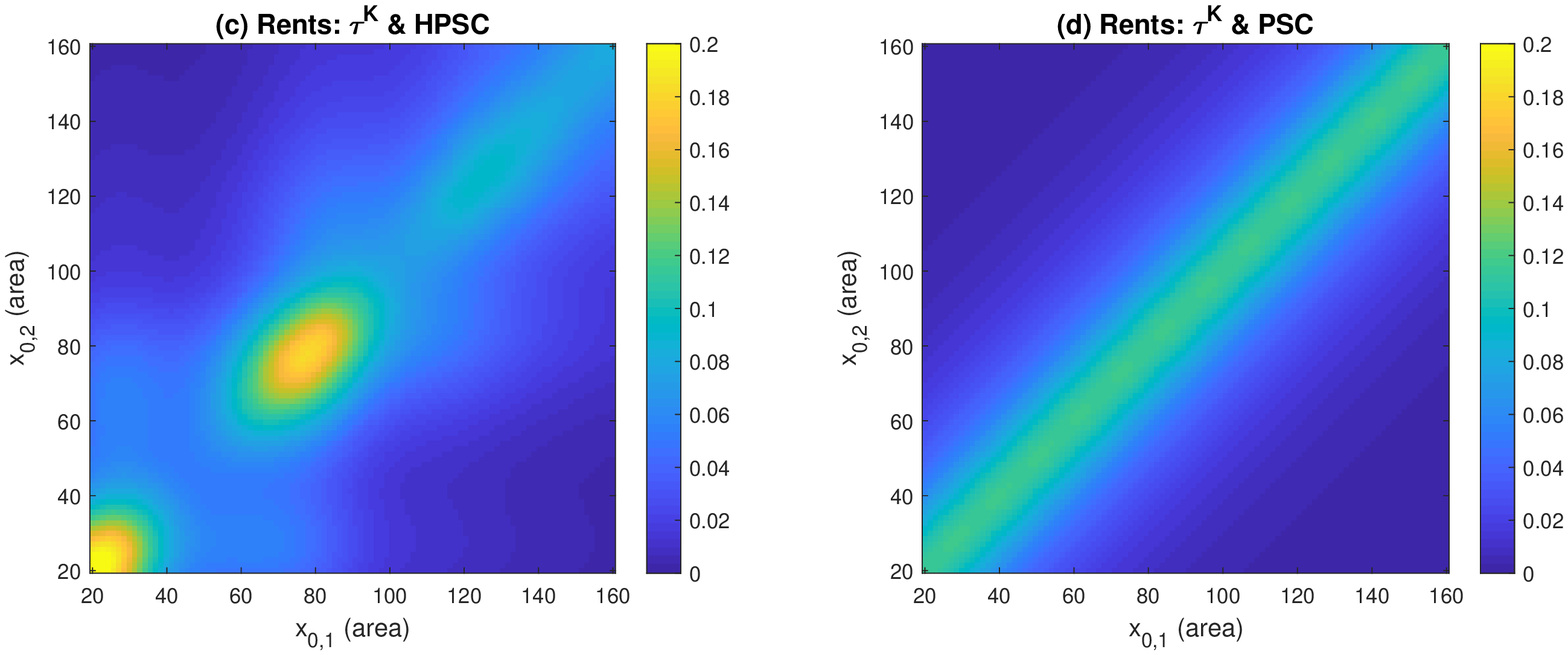}}\\%\hspace{0.3cm}
%\hspace{0.6cm}
\centering{\includegraphics[width=0.825\textwidth,angle=0]{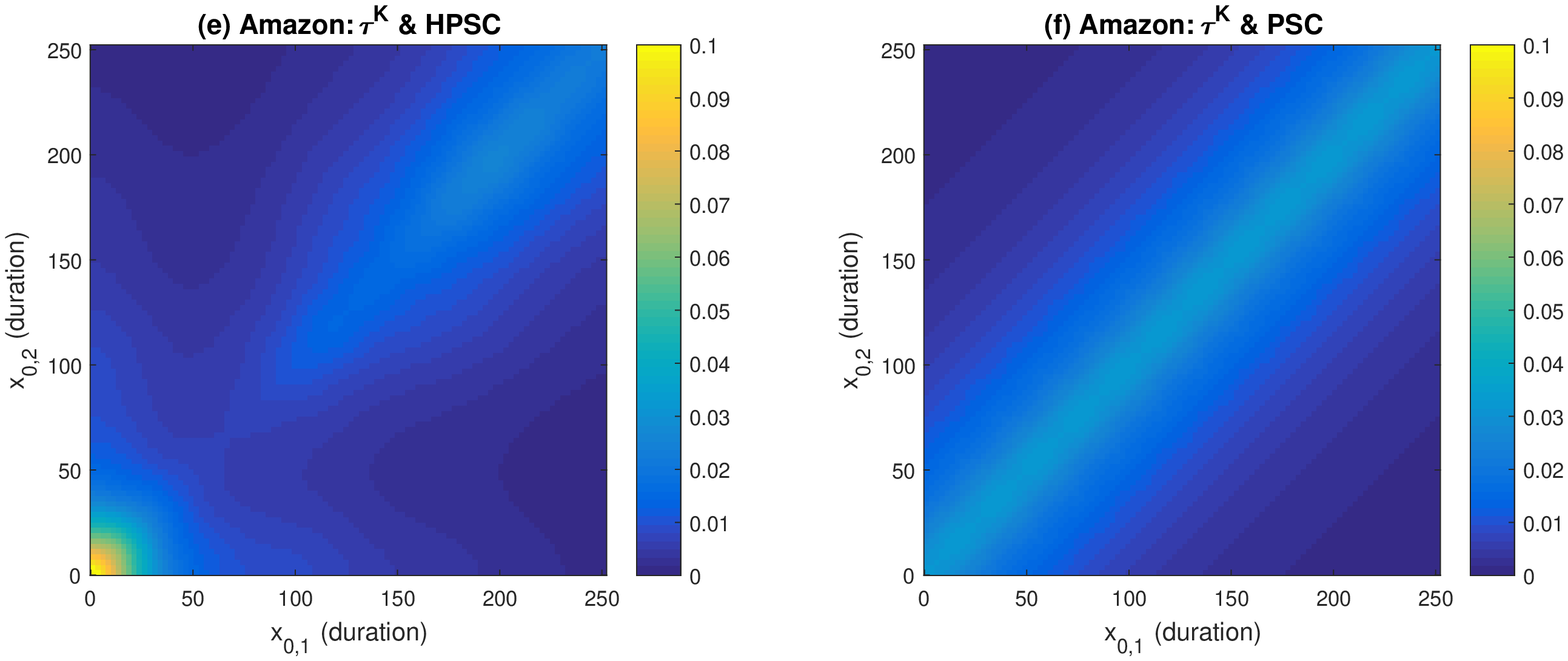}}\\%\vspace{0.2cm}
\centering{\includegraphics[width=0.825\textwidth,angle=0]{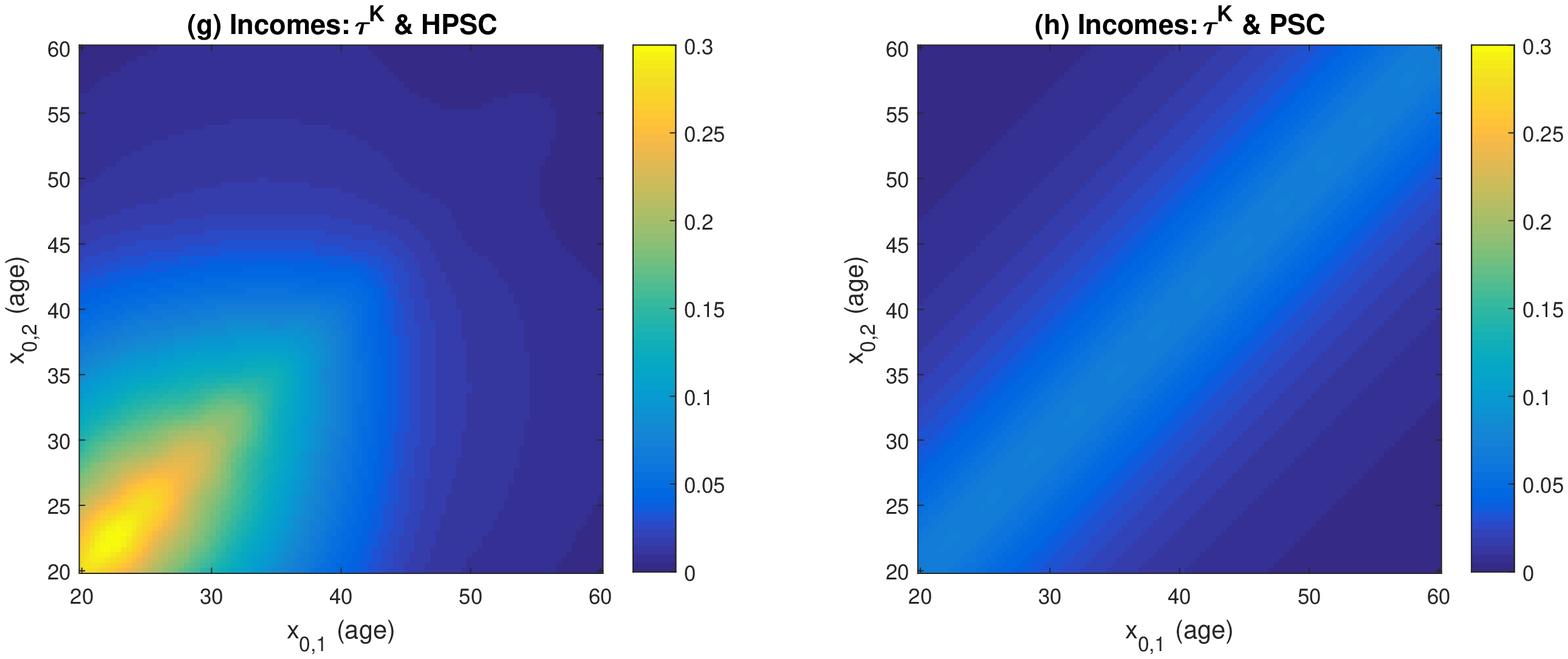}}\\
%\centering{\includegraphics[width=0.975\textwidth,angle=0]{../Figures/dep_amazon.eps}}\hspace{0.3cm}
%\centering{\includegraphics[width=0.975\textwidth,angle=0]{../Figures/dep_incomes.eps}}
\caption{Estimates of Kendall's tau $\tau^K(x_{0,1},x_{0,2})$. The lefthand panels give values for the
HPSC, and the righthand panels for the PSC. The datasets are (a,b) Geyser, (c,d) Rents, (e,f) Amazon,
and (g,h) Incomes.}\label{fig:dep:kendall}
\end{figure}

\clearpage

\newpage
\vspace{-15pt}
\section{Simulation Study Details}
\vspace{-10pt}
\numberwithin{equation}{section}
In this part of the appendix we provide further details for the simulation study in 
Section~4.2, along with additional results. 
The simulation designs are based on the five distributional regression methods each 
fitted to the four real datasets, giving  a total of 20 data generating processes (DGPs) in the
simulation study. Each DGP has covariate values given by those in the original dataset.
For the DGPs based on the Amazon and Incomes datasets, to speed up
the computations the replicates were based on sub-samples
of $n=3,000$ and $n=5,000$ randomly selected covariate values from the original datasets.
Similarly, we also use the faster distributional regression 
prediction $\hat{p}_{{\mbox{\tiny PE}}}$ defined in Section~3.4 for the regression copula models, 
where the point estimate used is  $\varthetahatvec_{{\mbox{\tiny VB}}}$ obtained by VB.

From each DGP we simulate 100 datasets (called `replicates' here), where for each we generate  $n$ observations of the response, but use the same covariate values as the original dataset.
We then refit all five methods to every replicate.
Accuracy of a method for each fitted replicate is assessed by using the fitted model
to evaluate the predictive distributions of
the observations in an additional 101$^{\mbox{\small st}}$ replicate simulated from the DGP. 
Thus, we are assessing
the accuracy of out-of-sample density forecasting from the same DGP.
From these predictions we can construct two density forecasting measures of accuracy: the mean logarithmic score ($\overline{\mbox{LS}}$), and the mean continuous rank probability score ($\overline{\mbox{CRPS}}$)
\citep{GneRaf2007}, when the mean is over the density forecasts
for the observations in the 101st replicate.

\begin{sidewaysfigure}[t]
	\caption{Simulation results for the DGPs constructed from fits to the {\bf Geyser} dataset.}
	\begin{center}
		\includegraphics[width=0.9\textwidth,angle=0]{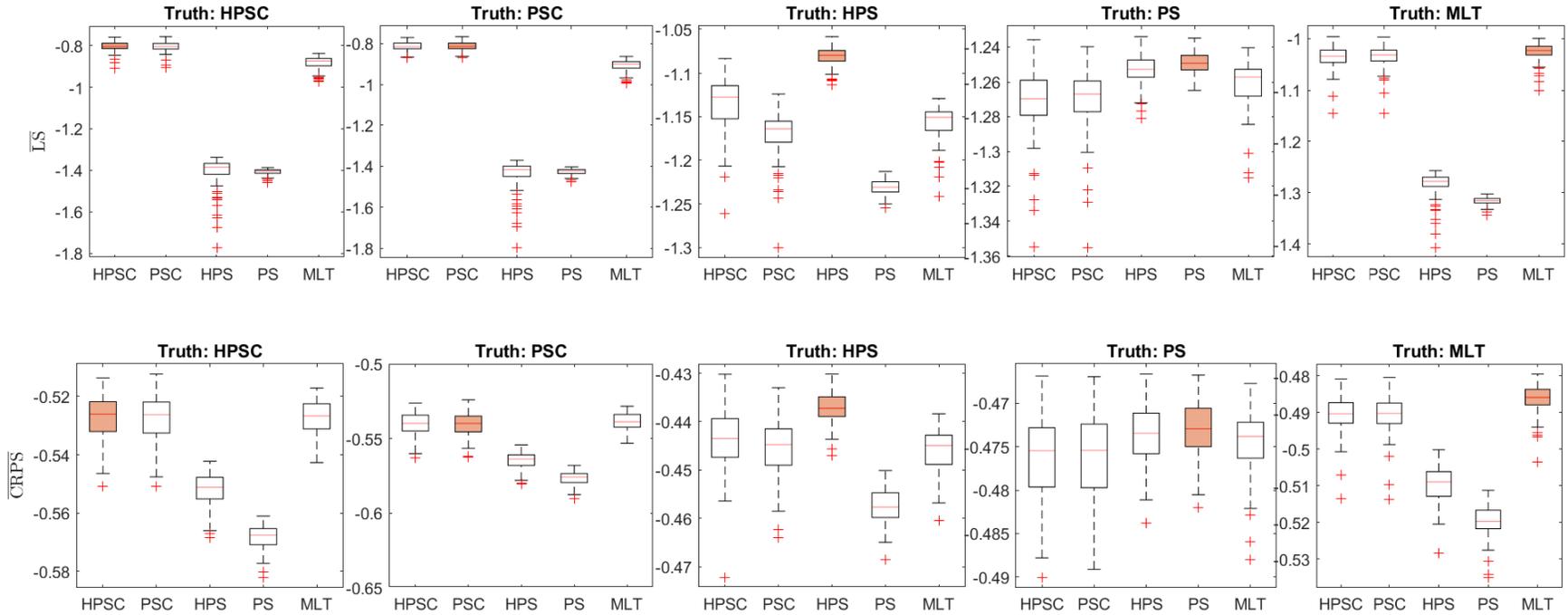}
	\end{center}
The top panels report the mean logarithmic score ($\overline{\mbox{LS}}$), and the bottom panels the mean CRPS ($\overline{\mbox{CRPS}}$). These are out-of-sample density forecasting metrics averaged over observations in a 101st replicate. Results are orientated so that higher values correspond to greater accuracy.
The columns give results for replicates simulated from each of the DGPs obtained from fitting the five distributional regression models to the original data. 
In each panel boxplots of the metrics for each the 100 replicates are given, with one boxplot for each of the five methods.
The shaded boxplots are for the cases where the method matches the DGP used to generate the data, which will typically be most accurate.
	\label{fig:simgeyser}
\end{sidewaysfigure}

\begin{sidewaysfigure}[ht]
	\caption{Simulation results for the DGPs constructed from fits to the {\bf Rents} dataset.}
	\begin{center}
	\includegraphics[width=0.9\textwidth,angle=0]{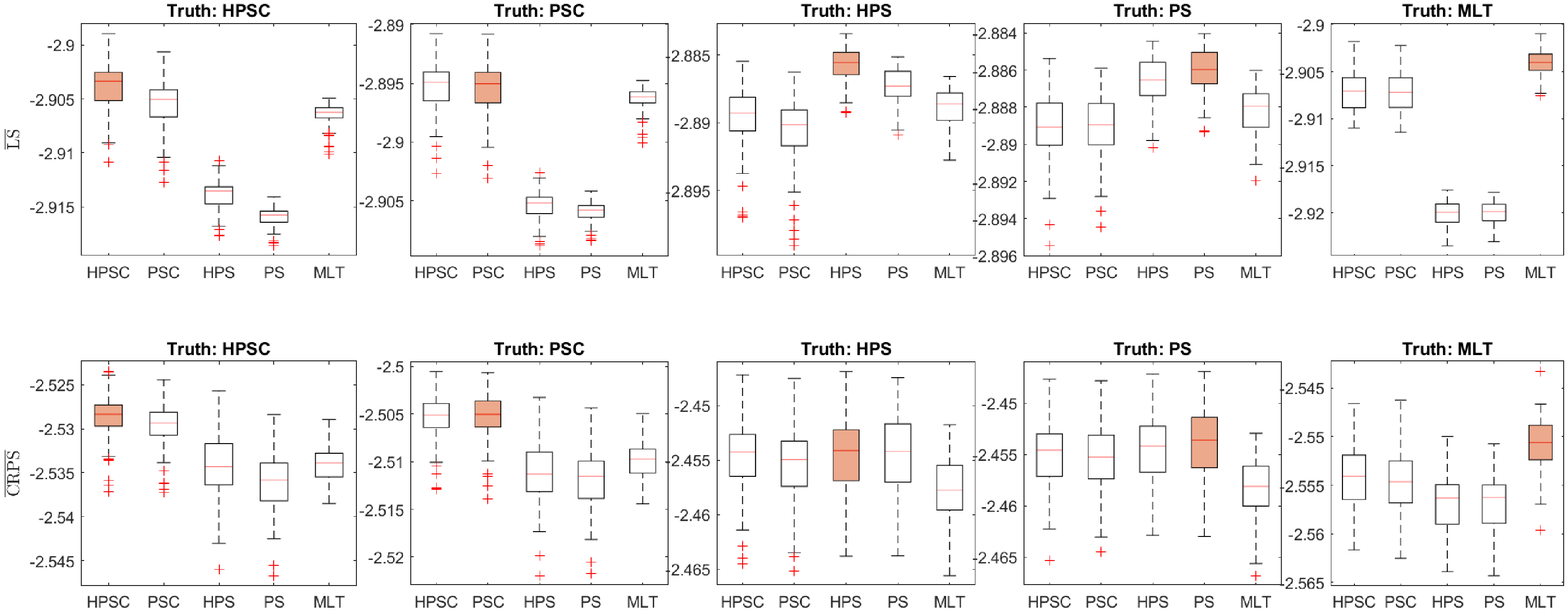}
	\end{center}
The top panels report the mean logarithmic score ($\overline{\mbox{LS}}$), and the bottom panels the mean CRPS ($\overline{\mbox{CRPS}}$). These are out-of-sample density forecasting metrics averaged over observations in a 101st replicate. Results are orientated so that higher values correspond to greater accuracy.
The columns give results for replicates simulated from each of the DGPs obtained from fitting the five distributional regression models to the original data. 
In each panel boxplots of the metrics for each the 100 replicates are given, with one boxplot for each of the five methods.
The shaded boxplots are for the cases where the method matches the DGP used to generate the data, which will typically be most accurate.
	\label{fig:simrents}
\end{sidewaysfigure}

\begin{sidewaysfigure}[ht]
	\caption{Simulation results for the DGPs constructed from fits to the {\bf Amazon} dataset.}
	\begin{center}
		\includegraphics[width=0.9\textwidth,angle=0]{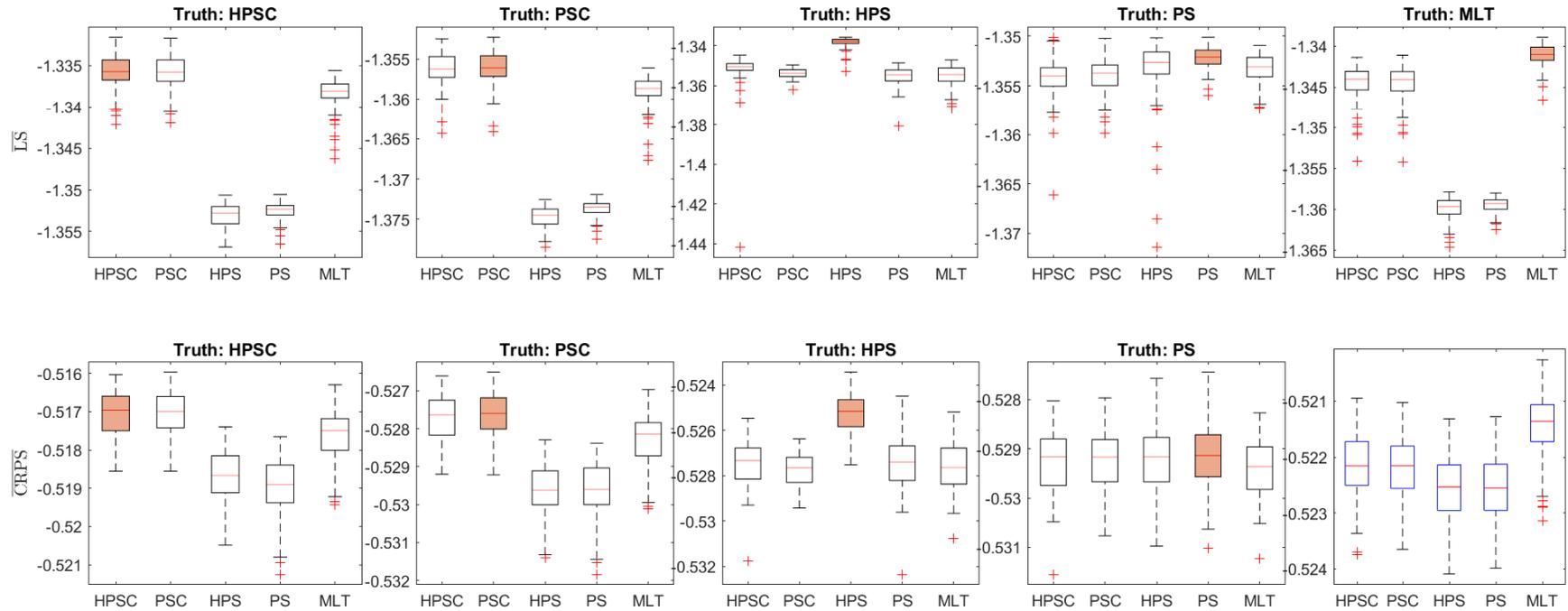}
	\end{center}
The top panels report the mean logarithmic score ($\overline{\mbox{LS}}$), and the bottom panels the mean CRPS ($\overline{\mbox{CRPS}}$). These are out-of-sample density forecasting metrics averaged over observations in a 101st replicate. Results are orientated so that higher values correspond to greater accuracy.
The columns give results for replicates simulated from each of the DGPs obtained from fitting the five distributional regression models to the original data. 
In each panel boxplots of the metrics for each the 100 replicates are given, with one boxplot for each of the five methods.
The shaded boxplots are for the cases where the method matches the DGP used to generate the data, which will typically be most accurate.
	\label{fig:simamazon}
\end{sidewaysfigure}

The results for the 5 DGPs constructed from the Incomes dataset are given in the main paper,
while  those for the 15 DGPs constructed from the Geyser, Rents and Amazon datasets are given in 
Figs.~\ref{fig:simgeyser}, \ref{fig:simrents} and \ref{fig:simamazon}. The top panels of these figures
give the $\overline{\mbox{LS}}$ results, and the bottom panels the corresponding $\overline{\mbox{CRPS}}$ results. Each panel corresponds
to a different DGP and contains five boxplots, one for each method fit to the replicates. Each 
boxplot  is constructed from the 100 density forecasting metrics arising
from the 100 replicates. Generally speaking, one would expect refitting the same method to as
used to construct the DGP, to provide the highest accuracy density forecasts. These boxplots are shaded. Therefore, our focus is on the `next best' performing method. And here the HPSC method
performs particularly well, being the `next best' in more cases than the other approaches. 
In particular, HPSC either equals or out-performs the MLT benchmark method for all DGPs and metrics, except for the PS DGP.

Last, we note here how we simulate from the fitted models to construct the
replicates. For the PS and HPS  we extract the estimated mean (and variance) functions 
for the response $Y$, and simulate from the resulting normal distributions. For the MLT 
we extract the estimated transformation function, and use the \texttt{simulate.mlt} in the R package
`mlt'. For the PSC and HPSC we extract the estimates of the mean function $\tilde m$ (and variance
function $g$) for the pseudo-response $\tilde Z$, along with the estimated margin $\hat{F}_Y$. Simulation
is then based on generating values from the conditionally Gaussian pseudo-response model, and
applying the transformation $\hat{F}_Y^{-1}\circ \Phi$ to these values.  

\clearpage
\newpage

\vspace{-15pt}
\section{Details and Derivations for Section~5}
\vspace{-10pt}
To implement Algorithm~2 a similar strategy as described in Section~\ref{app:subsec:vb} can be used. In particular, the gradients for $\betavec$ and $\alphavec$ only change in the design matrices $B(\xvec),V(\wvec)$ (see Section 5.1 of the main paper for the specification) and the prior precision matrices $P_{\beta},P_{\alpha}$. The latter depend on the prior choice which we discuss now.

\vspace{-12pt}
\subsection{Specification of the Horseshoe Prior for the Copula Parameters}
\vspace{-7pt}
For the parameters $\lbrace \thetavec_{\beta},\thetavec_{\alpha}\rbrace$ we employ the horseshoe prior such that 
\[
\thetavec_{\beta}=\lbrace \lambda_{\beta,1},\ldots,\lambda_{\beta,p_1},\tau_{\beta}\rbrace\quad\thetavec_{\alpha}=\lbrace \lambda_{\alpha,1},\ldots,\lambda_{\alpha,p_2},\tau_{\alpha}\rbrace,
\]
($p_1=240$, $p_2=96$ in our example), 
with prior distributions
\[
\beta_j|\lambda_j\sim\ND(0,\lambda_{\beta,j}^2), \quad \lambda_{\beta,j}|\tau_{\beta}\sim\mbox{C}^+(0,\tau_{\beta}),\quad\tau_{\beta}\sim\mbox{C}^+(0,1),\quad j=1,\ldots p_1
\]
and similar for $\alphavec$. Here, $\mbox{C}^+(\cdot)$ denotes a half Cauchy distribution. As a result, we obtain
\[
P_{\beta}(\thetavec_{\beta})=\diag(\lambda_{\beta,1}^2,\ldots,\lambda_{\beta,p_1}^2)^{-1},\quad P_{\alpha}(\thetavec_{\alpha})=\diag(\lambda_{\alpha,1}^2,\ldots,\lambda_{\alpha,p_2}^2)^{-1},
\]
such that $S(\xvec,\wvec,\alphavec,\lambdavec_{\beta})=\diag(s_{1},\ldots,s_{n})$ and
\[
s_i=\lbrack\exp(\vvec_i'\alphavec+\bvec_i'\diag(\lambda_{\beta,1}^2,\ldots,\lambda_{\beta,p_1}^2)\bvec_i)\rbrack^{-1/2}.
\]

\vspace{-12pt}
\subsection{Log Posterior Distribution}
\vspace{-7pt}
The joint log-posterior distribution $\log(h(\varthetavec))=l_{\varthetavec}$ of $$\varthetavec=\lbrace\betavec,\alphavec,\log(\lambda_{\beta,1}^2),\ldots,\log(\lambda_{\beta,p_1}^2),\log(\tau_{\beta}),\log(\lambda_{\alpha,1}^2),\ldots,\log(\lambda_{\alpha,p_2}^2),\log(\tau_{\alpha})\rbrace$$ is proportional to 
\begin{equation*}\begin{aligned}
l_{\varthetavec}&\propto-\frac{1}{2}\sum_{i=1}^n(\log(s_i^2))-\frac{1}{2}\sum_{i=1}^n(\log(\sigma_i^2))-\frac{1}{2}(\zvec-SB\betavec)'(S\Sigma S)^{-1}(\zvec-SB\betavec)  \\%likelihood part
  & \quad+\frac{1}{2}\sum_{j=1}^{p_1}(\log(\lambda_{\beta,j}^2)) -\frac{1}{2}\sum_{j=1}^{p_1}\frac{\beta_j^2}{\lambda_{\beta,j}^2}  %prior beta
    +\quad\frac{1}{2}\sum_{j=1}^{p_2}(\log(\lambda_{\alpha,j}^2)) -\frac{1}{2}\sum_{j=1}^{p_2}\frac{\alpha_j^2}{\lambda_{\alpha,j}^2}\\ %prior alpha
    &\quad-(p_1-1)\log(\tau_{\beta})-\sum_{j=1}^{p_1}\log\left(1+\frac{\lambda_{\beta,j}^22}{\tau_{\beta}^2}\right) %prior llambda2s
    -(p_2-1)\log(\tau_{\alpha})-\sum_{j=1}^{p_2}\log\left(1+\frac{\lambda_{\alpha,j}^22}{\tau_{\alpha}^2}\right)\\%prior llambda2V
    &\quad -\log(1+\tau_{\beta}^2)-\log(1+\tau_{\alpha}^2), %prior ltau
\end{aligned}\end{equation*}
and where we transform $\lambda_{\cdot,j}^2$ and $\tau_{\cdot}$ to the log-scale for convenience.

\vspace{-12pt}
\subsection{Gradients of $\log(\lambda_{\beta,j}^2)$}
\vspace{-7pt}
\begin{equation*}\begin{aligned}
\nabla_{\log(\lambda_{\beta,j}^2)}l_{\varthetavec}&=
\frac{1}{2}\frac{\beta_j^2}{\lambda_{\beta,j}^2}-\frac{\lambda_{\beta,j}^2/\tau_{\beta}^2}{1+\lambda_{\beta,j}^2/\tau_{\beta}^2}
             +\frac{1}{2}+\frac{1}{2}\sum_{i=1}^n b_{ij}^2\lambda_{\beta,j}^2 s_i^2- \frac{1}{2}\sum_{i=1}^n \frac{z_i^2 b_{ij}^2\lambda_{\beta,j}^2}{\sigma_i^2} + \frac{1}{2}\betavec'B'(W\Sigma^{-1}S)\zvec,
\end{aligned}\end{equation*}
where $b_{ij}$ is the $j$-th element of $\bvec_i$ and $W=\diag(w_1,\ldots,w_n)$, $w_i=b_{ij}^2\lambda_{\beta,j}^2$.
\vspace{-12pt}
\subsection{Gradient of $\log(\tau_{\beta})$}
\vspace{-7pt}

\begin{equation*}\begin{aligned}
\nabla_{\log(\tau_{\beta})}l_{\varthetavec}&=
-(p_1-1)+2\sum_{j=1}^{p_1}\frac{\lambda_{\beta,j}^2}{\tau_{\beta}^2}\left(1+\frac{\lambda_{\beta,j}^2}{\tau_{\beta}^2}\right)^{-1}-2\frac{\tau_{\beta}^2}{1+\tau_{\beta}^2}.
\end{aligned}\end{equation*}

\vspace{-12pt}
\subsection{Gradients of $\log(\lambda_{\alpha,j}^2)$}
\vspace{-7pt}

\begin{equation*}\begin{aligned}
\nabla_{\log(\lambda_{\alpha,j}^2)}l_{\varthetavec}&=
\frac{1}{2}\frac{\alpha^2}{\lambda_{\alpha,j}^2}-\frac{\lambda_{\alpha,j}^2/\tau_{\alpha}^2}{1+\lambda_{\alpha,j}^2/\tau_{\alpha}^2}
             +\frac{1}{2}.
\end{aligned}\end{equation*}

\vspace{-12pt}
\subsection{Gradient of $\log(\tau_{\alpha})$}
\vspace{-7pt}

\begin{equation*}\begin{aligned}
\nabla_{\log(\tau_{\alpha})}l_{\varthetavec}&=
-(p_2-1)+2\sum_{j=1}^{p_2}\frac{\lambda_{\alpha,j}^2}{\tau_{\alpha}^2}\left(1+\frac{\lambda_{\alpha,j}^2}{\tau_{\alpha}^2}\right)^{-1}-2\frac{\tau_{\alpha}^2}{1+\tau_{\alpha}^2}.
\end{aligned}\end{equation*}
\clearpage
\newpage

\vspace{-15pt}
\section{Additional Figures for Section~5}
\vspace{-10pt}

\begin{figure}[htbp]
\caption{Quantile-Quantile (QQ) plots of the residuals of three distributional regression models fit to the electricity price data}
\begin{center}
\includegraphics[width=0.9\textwidth,angle=0]{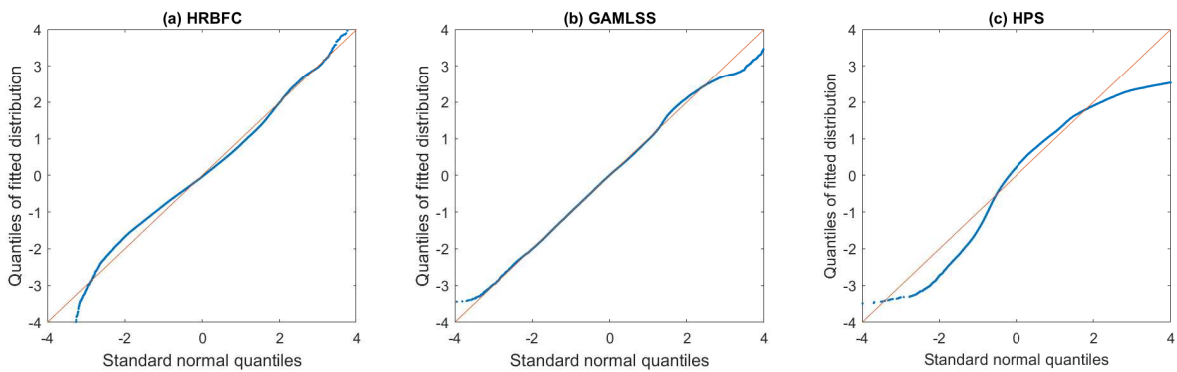}
\end{center}
The residuals are obtained using the  mean of $Y=\log(\mbox{Price}+101)$ for all $87,648$ observations, computed
from the (a) HRBFC regression copula, (b) GAMLSS and (c)~HPS models.\label{fig:fitted:electricity}
\end{figure}

\begin{figure}[htbp]
\caption{Plot of lower bound ${\cal L}(\lambdavec)$ against step number in the SGA algorithm for the
electricity price data}
\begin{center}
\includegraphics[width=0.475\textwidth,angle=0]{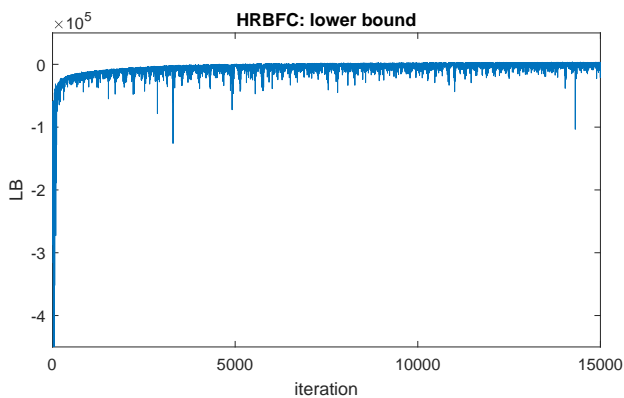}
\end{center}
We note that due to the large sample size and large number of basis terms,
this example is the slowest to compute. A total of 1000 steps of the SGA algorithm
takes approximately 180 minutes to execute on a contemporary laptop, with code written
in Matlab.
\label{fig:LB:electricity}
\end{figure}

\begin{figure}[p]
\caption{Predictive distributions of the logarithm of electricity prices from the {\bf HRBFC regression copula} model.}
\begin{center}
\includegraphics[width=0.95\textwidth,angle=0]{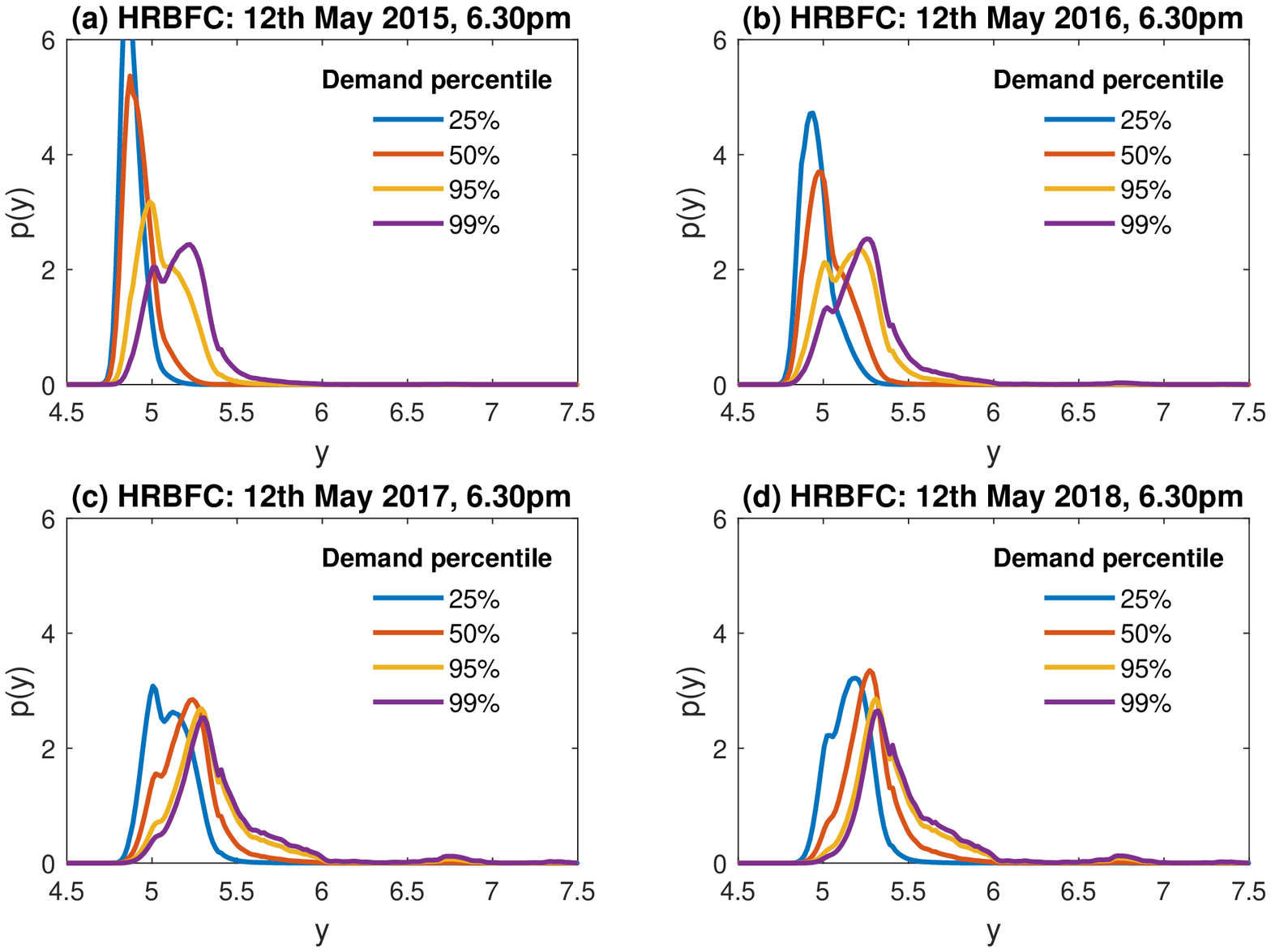}
\end{center}
The four panels provide predictions for 18:30 on 12 May in the years (a)~2015 (b)~2016, (c)~2017 and (d)~2018.
In each panel, the predictive densities are constructed at four levels of demand corresponding to 
the 0.25, 0.5, 0.95 and 0.99 percentiles of demand at 18:30. Note the accentuation of the upper tail of 
price from 2015 to 2018. \label{fig:dens:dy:electricity}
\end{figure}

\begin{figure}[p]
\caption{Predictive distributions of the logarithm of electricity prices from the {\bf GAMLSS} model.}
\begin{center}
\includegraphics[width=0.95\textwidth,angle=0]{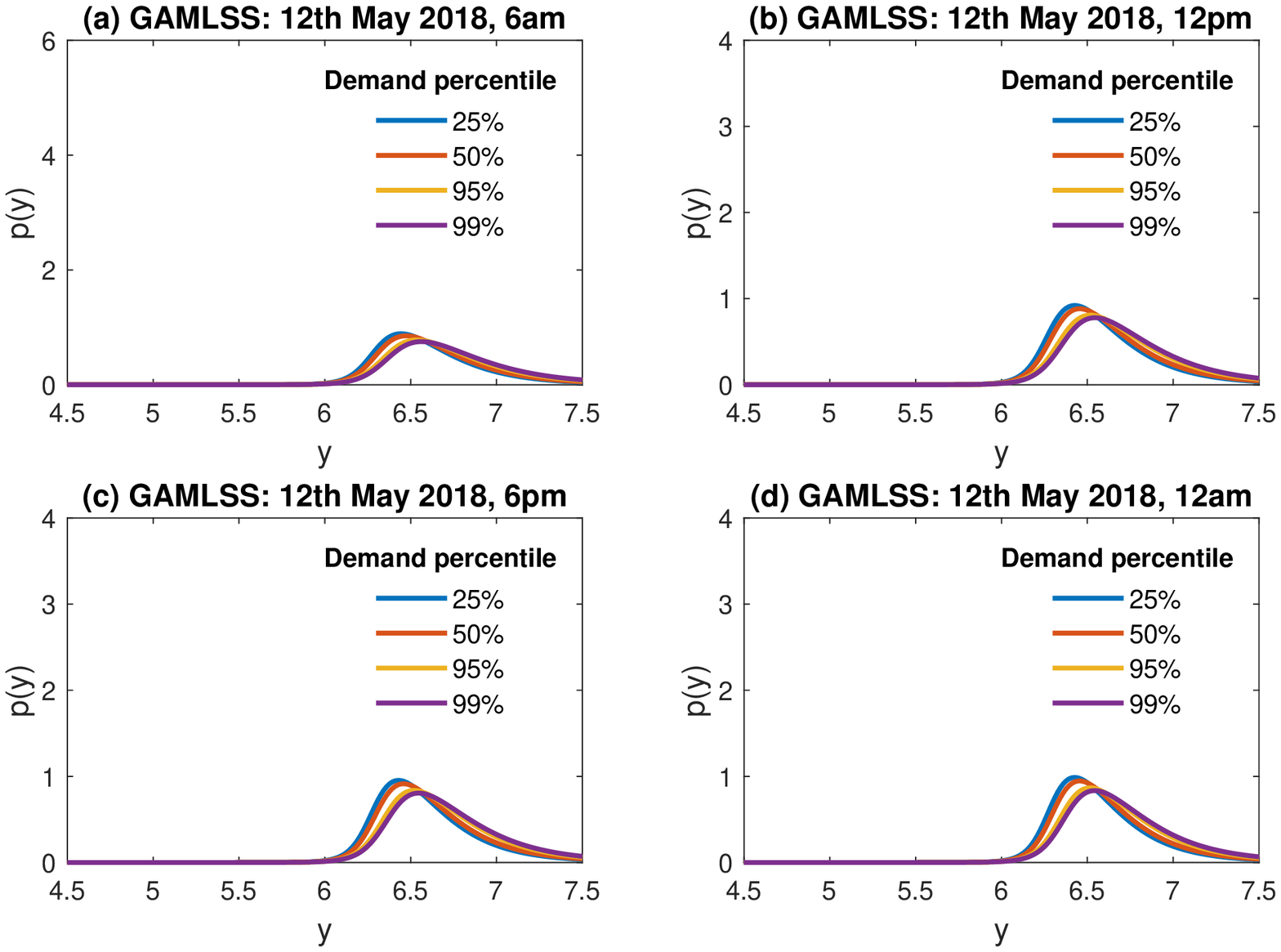}
\end{center}
The four panels provide predictions for 12 May 2018 at~(a) 06:00, (b)~12:00, (c)~18:00 and (d)~24:00.
In each panel, the predictive densities are constructed at four levels of demand corresponding to 
the 0.25, 0.5, 0.95 and 0.99 percentiles of demand at each time of day. \label{fig:dens:gamlss:tod:electricity}
\end{figure}

\begin{figure}[p]
\caption{Predictive distributions of the logarithm of electricity prices from the {\bf HPS} model.}
\begin{center}
\includegraphics[width=0.95\textwidth,angle=0]{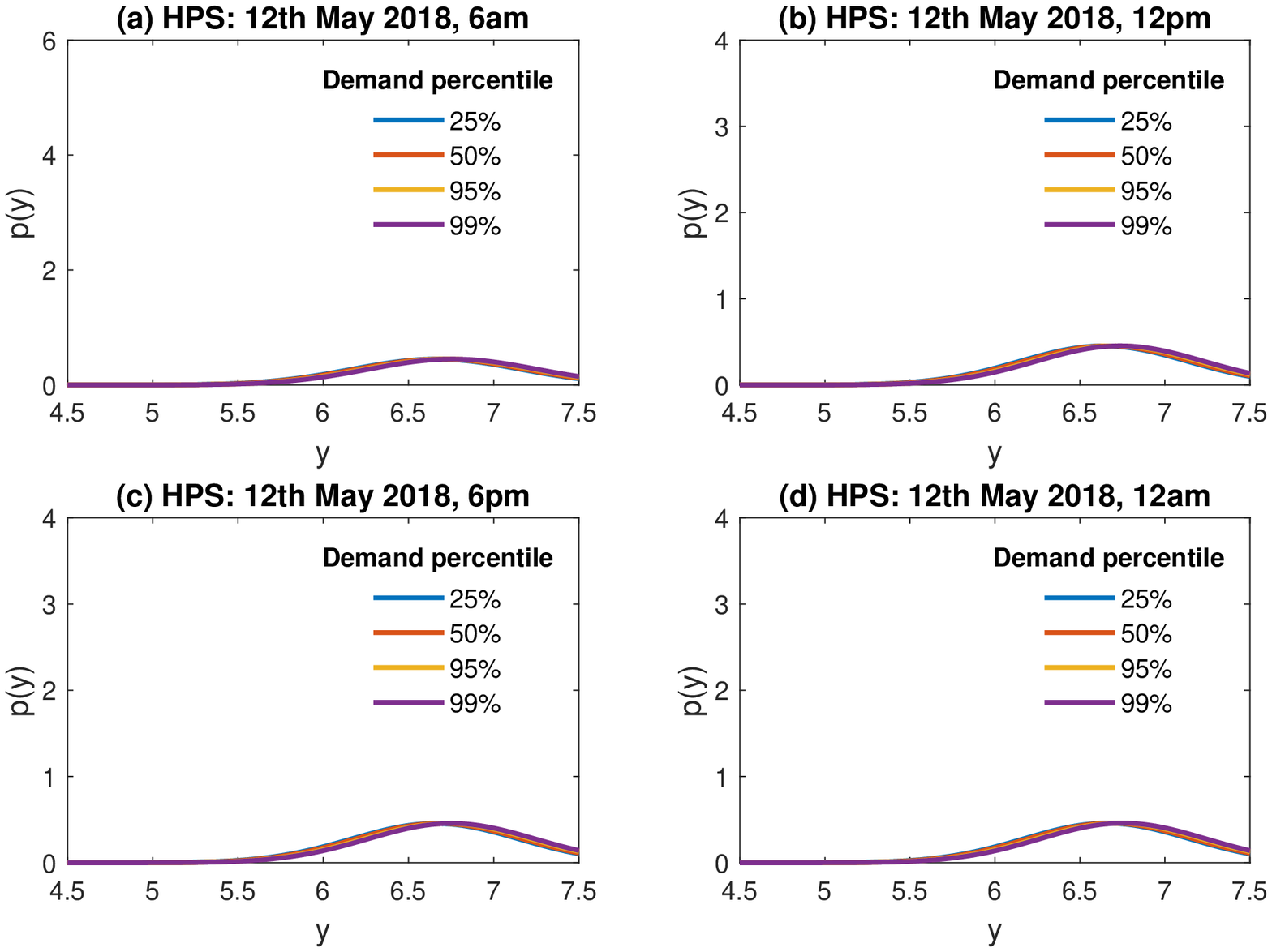}
\end{center}
The four panels provide predictions for 12 May 2018 at~(a) 06:00, (b)~12:00, (c)~18:00 and (d)~24:00.
In each panel, the predictive densities are constructed at four levels of demand corresponding to 
the 0.25, 0.5, 0.95 and 0.99 percentiles of demand at each time of day. \label{fig:dens:gamlss:dy:electricity}
\end{figure}

\begin{figure}[p]
\caption{Predictive distributions of the logarithm of electricity prices from the {\bf GAMLSS} model.}
\begin{center}
\includegraphics[width=0.95\textwidth,angle=0]{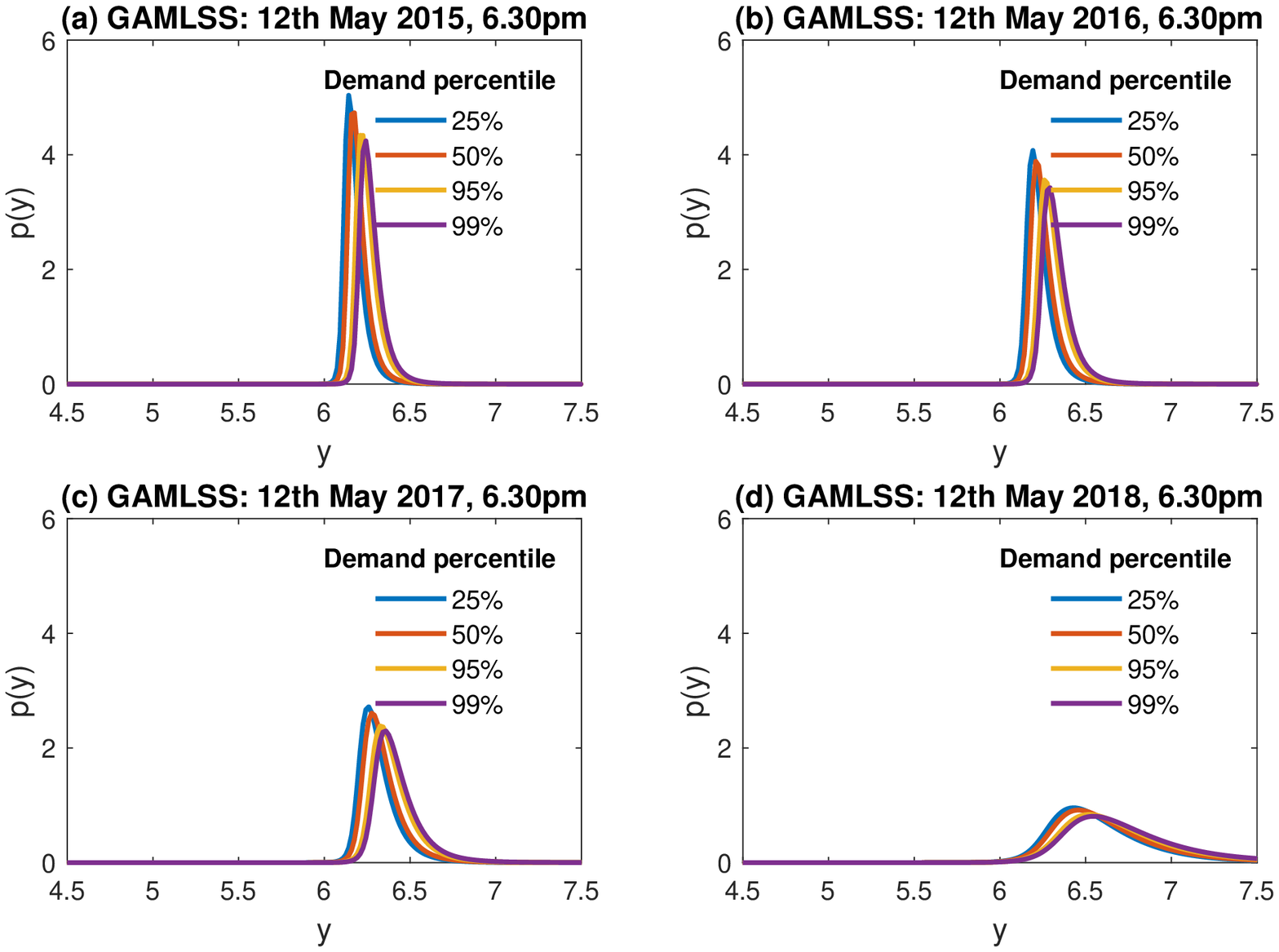}
\end{center}
The four panels provide predictions for 18:30 on 12 May in the years (a)~2015 (b)~2016, (c)~2017 and (d)~2018.
In each panel, the predictive densities are constructed at four levels of demand corresponding to 
the 0.25, 0.5, 0.95 and 0.99 percentiles of demand at 18:30. Note the accentuation of the upper tail of 
price from 2015 to 2018. \label{fig:dens:hps:tod:electricity}
\end{figure}

\begin{figure}[p]
\caption{Predictive distributions of the logarithm of electricity prices from the {\bf HPS} model.}
\begin{center}
\includegraphics[width=0.95\textwidth,angle=0]{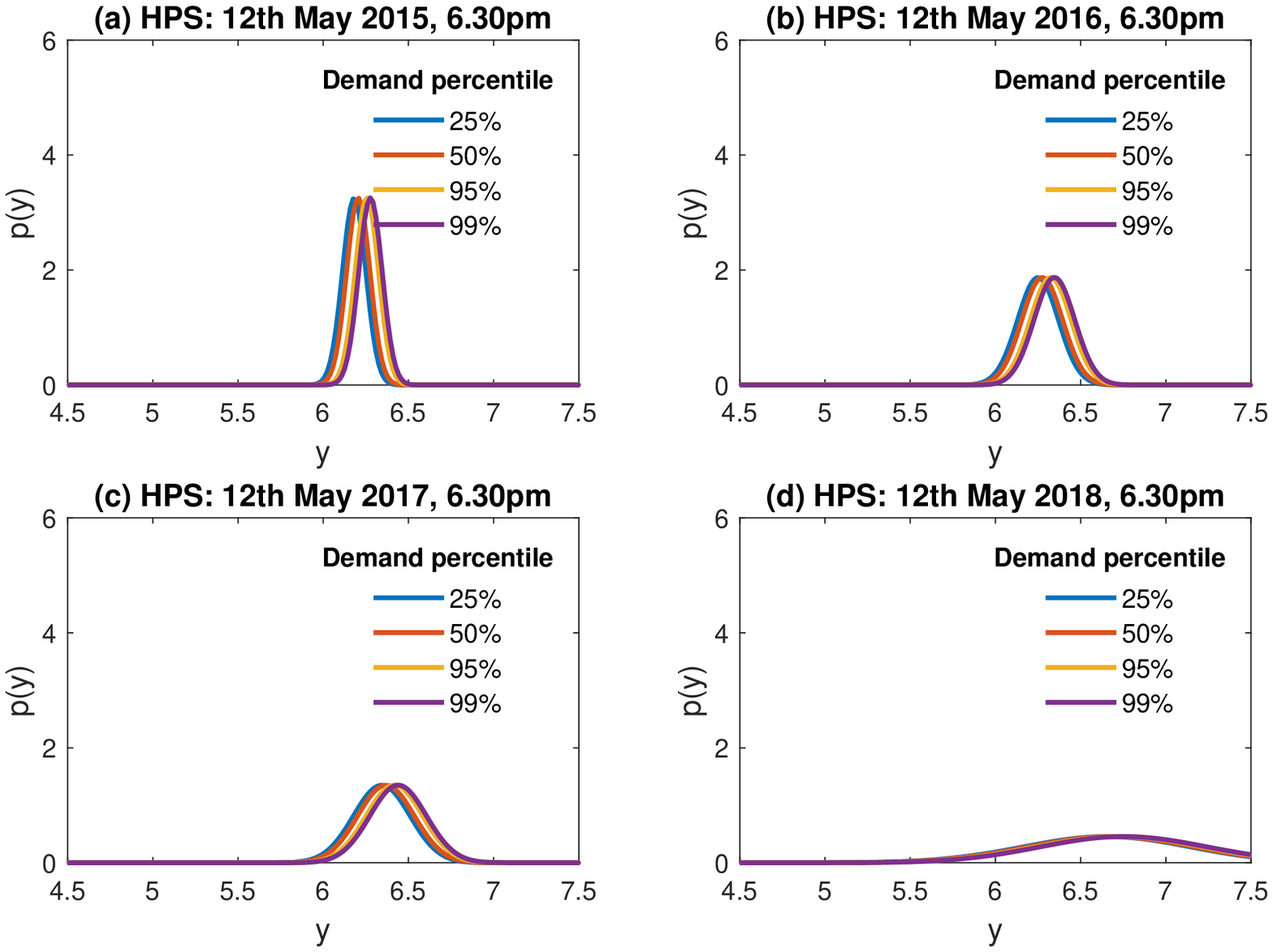}
\end{center}
The four panels provide predictions for 18:30 on 12 May in the years (a)~2015 (b)~2016, (c)~2017 and (d)~2018.
In each panel, the predictive densities are constructed at four levels of demand corresponding to 
the 0.25, 0.5, 0.95 and 0.99 percentiles of demand at 18:30. Note the increase in variance from 2015 to 2018.\label{fig:dens:hps:dy:electricity}
\end{figure}

\end{document}